\documentclass[12pt]{article}

\usepackage{enumitem}
\usepackage{epsfig}

\usepackage{graphicx}
\usepackage{axodraw2}
\usepackage{float}
\usepackage{cite}
\usepackage{amsfonts}
\usepackage{amssymb}
\usepackage{amsmath}
\usepackage{xcolor}

\def\square{\kern1pt\vbox{\hrule height 1.2pt
\hbox{\vrule width 1.2pt\hskip 3pt
\vbox{\vskip 6pt}\hskip 3pt\vrule width 0.6pt}
\hrule height 0.6pt}\kern1pt}
\def\ltwid{\mathrel{\raise.3ex\hbox{$<$\kern-.75em\lower1ex\hbox{$\sim$}}}}
\def\gtwid{\mathrel{\raise.3ex\hbox{$>$\kern-.75em\lower1ex\hbox{$\sim$}}}}

\begin{document}

\begin{titlepage}
\begin{flushright}
CCTP-2024-6 \\ UFIFT-QG-24-03
\end{flushright}

\vspace{0cm}

\begin{center}
\bf{Alternate Computation of Gravitational Effects 
from a Single Loop of Inflationary Scalars}
\end{center}

\vspace{0cm}

\begin{center}
S. P. Miao$^{\ast}$
\end{center}
\begin{center}
\it{Department of Physics, National Cheng Kung University, \\
Tainan City 70101, TAIWAN }
\end{center}

\vspace{0cm}

\begin{center}
N. C. Tsamis$^{\dagger}$
\end{center}
\begin{center}
\it{Institute of Theoretical \& Computational Physics, and \\
Department of Physics, University of Crete \\
GR-710 03 Heraklion, HELLAS}
\end{center}

\vspace{0cm}

\begin{center}
R. P. Woodard$^{\#}$
\end{center}
\begin{center}
\it{Department of Physics, University of Florida \\
Gainesville, FL 32611, UNITED STATES}
\end{center}

\vspace{0cm}

\begin{center}
ABSTRACT
\end{center}
\hspace{0.3cm} 
We present a new computation of the renormalized graviton 
self-energy induced by a loop of massless, minimally coupled 
scalars on de Sitter background. Our result takes account 
of the need to include a finite renormalization of the 
cosmological constant, which was not included in the first
analysis. We also avoid preconceptions concerning structure 
functions and instead express the result as a linear combination 
of 21 tensor differential operators. By using our result to 
quantum-correct the linearized effective field equation we 
derive logarithmic corrections to both the electric components 
of the Weyl tensor for gravitational radiation and to the two 
potentials which quantify the gravitational response to a 
static point mass.

\vspace{0cm}

\begin{flushleft}
\end{flushleft}

\vspace{0cm}

\begin{flushleft}
$^{\ast}$ {\it e-mail:} spmiao5@phys.ncku.edu.tw \\
$^{\dagger}$ {\it e-mail:} tsamis@physics.uoc.gr \\
$^{\#}$ {\it e-mail:} woodard@phys.ufl.edu
\end{flushleft}

\end{titlepage}

\section{Prologue}

A wide variety of observational evidence points to the very 
early universe having experienced a phase of accelerated 
expansion, or inflation \cite{infl}. Cosmological spacetimes
are described by the scale factor $a(t)$ and its two first
time derivatives $H(t)$ (Hubble parameter) and $\epsilon(t)$
($1st$ slow-roll parameter):
\begin{equation}
ds^2 = -dt^2 + a^2(t) d{\mathbf{x}} \cdot d{\mathbf{x}} 
\quad , \quad
H(t) \equiv \frac{\dot a}{a}
\quad , \quad
\epsilon(t) \equiv -\frac{\dot H}{H^2}
\;\; . \label{cosmo}
\end{equation}
Inflation is characterized by the positivity of both derivatives
of $a(t)$: \\
$H(t) > 0 \;\; \& \;\; 0 \leq \epsilon(t) < 1$. The standard 
inflationary cosmology is that of the maximally symmetric 
de Sitter spacetime:
\footnote{Hellenic indices take on spacetime values
while Latin indices take on space values. Our metric
tensor $g_{\mu\nu}$ has spacelike signature
$( - \, + \, + \, +)$ and our curvature tensor equals
$R^{\alpha}_{~ \beta \mu \nu} \equiv 
\Gamma^{\alpha}_{~ \nu \beta , \mu} +
\Gamma^{\alpha}_{~ \mu \rho} \, 
\Gamma^{\rho}_{~ \nu \beta} -
(\mu \leftrightarrow \nu)$.
Co-moving time is denoted by $t$ and conformal time
by $\eta$.}
\begin{eqnarray}
ds^2 &\!\! = \!\!& 
-dt^2 + a^2(t) d{\mathbf{x}} \cdot d{\mathbf{x}} 
= a^2(\eta)
\Big[ -d\eta^2 + a^2(\eta) d{\mathbf{x}} \cdot d{\mathbf{x}} \Big]
\;\; , \label{dS1} \\
a(t) &\!\! = \!\!& e^{Ht} = -\frac{1}{H \eta} = a(\eta)
\;\; . \label{dS2}
\end{eqnarray}

During inflation, quantum physics implies the production of 
real particles out of the vacuum as long as they are effectively 
massless, possess classically non-conformally invariant free
Lagrangians, and have adequately large wavelength. The carrier 
of the gravitational force, the graviton, is such a particle and 
inflationary evolution eventually will produce a dense ensemble 
of infrared gravitons \cite{Grishchuk:1974ny,Starobinsky:1979ty}. 

Can the universally attractive nature of the gravitational 
interaction alter cosmological parameters, kinematical 
parameters and long-range forces? There are perturbative 
indications for such changes from many loop computations 
which show a time dependence of powers of $\ln a(t)$ 
\cite{Onemli:2002hr,Prokopec:2002uw,Miao:2006pn,Miao:2006gj,
Kahya:2006hc,Prokopec:2008gw,Glavan:2013jca,Wang:2014tza,
Tan:2021lza,Tan:2022xpn}. Because the dimensionless coupling 
constant of quantum gravity is $(G H^2)^{-1}$, at some time 
the secular increase by powers of $\ln a(t)$ will overwhelm 
$(G H^2)^{-1}$ causing perturbation theory to break.

It is always a formidable affair to decipher the dynamics
of a theory after the perturbative analysis has become
invalid. While it is easy to state what is needed - a
re-summation technique for these leading logarithms - it 
is very hard to realize it. Usually one tries to first 
understand what happens in an analogous situation in a 
simpler theory that retains the essential interaction 
structure of gravity. The latter feature is satisfied by 
non-linear $\sigma$-models since they possess the same 
kind of derivative interactions as gravity; they lack the 
tensor structure and gauge fixing dependence of quantum 
gravity. Of these, it is obviously the gauge issue that 
is the strongest simplification because a true physical 
effect is by definition independent of the gauge fixing 
functional. 

Recently a particular non-linear $\sigma$-model - the $AB$ 
model - has been perturbatively analyzed and the required 
re-summation techniques have been indicated: 
curvature-dependent variants of the stochastic technique
\cite{Starobinsky:1994bd,Tsamis:2005hd} {\it and} of the 
renormalization group \cite{Miao:2021gic,Woodard:2023rqo,
Litos:2023nvj}.

Yet another theory that could be similarly analyzed before
facing the full quantum gravity, is that of a massless 
minimally coupled (MMC) scalar in a de Sitter spacetime
which we describe in Section 2. The result for the 1-loop 
MMC correction to the 2-point gravitational function is 
found in Section 3. Its effect on the gravitational mode 
functions and gravitational force are displayed in Sections 
4 and 5 respectively.

\section{The MMC theory}

The dynamics of a MMC scalar in a de Sitter background 
are defined by:
\begin{equation}
\mathcal{L} = \frac{[R - (D\!-\!2) \Lambda] 
\sqrt{-g}}{16 \pi G}
- \frac12 \partial_{\mu} \phi \, \partial_{\nu} \phi \, 
g^{\mu\nu} \sqrt{-g} 
\quad , \quad
\Lambda \equiv (D-1) H^2
\;\; , \label{L}
\end{equation}
leading to the following gravitational field equations:
\begin{equation}
R_{\mu\nu} - \frac12 R \, g_{\mu\nu} 
+ \frac12 (D\!-\!2) \Lambda g_{\mu\nu}
= 8 \pi G \Bigl\{ \partial_{\mu} \phi \, \partial_{\nu} \phi 
- \frac12 g_{\mu\nu} g^{\rho\sigma} 
\partial_{\rho} \phi \, \partial_{\sigma} \phi \Bigr\} 
\;\; . \label{eom}
\end{equation}
The graviton field $h_{\mu\nu}(x)$ is defined by 
conformally rescaling the full metric with the 
scale factor:
\begin{equation}
g_{\mu\nu} \equiv a^2 \, \widetilde{g}_{\mu\nu} 
\equiv a^2 \left( \eta_{\mu\nu} + \kappa h_{\mu\nu} \right) 
\qquad , \qquad 
\kappa^2 \equiv 16 \pi G 
\;\; . \label{hmn}
\end{equation}
Here $\kappa$ is the loop-counting parameter of quantum 
gravity. Our notation throughout is that indices are raised 
and lowered with the Minkowski metric, for instance, 
$h^{\mu\nu} \equiv \eta^{\mu\rho} \eta^{\nu\sigma} 
h_{\rho\sigma}$ and $\partial^{\mu} \equiv \eta^{\mu\rho} 
\partial_{\rho}$. Furthermore, parenthesized indices are 
symmetrized.
\\ [9pt]
$\bullet$ {\it The MMC Model: Counterterms}

Our model (\ref{L}) is not renormalizable, but the divergences
of any theory can be removed by BPHZ (Bogoliubov, Parasiuk
\cite{Bogoliubov:1957gp}, Hepp \cite{Hepp:1966eg} and Zimmermann
\cite{Zimmermann:1968mu,Zimmermann:1969jj} counterterms. Those
of our model (\ref{L}) were the first ever studied using 
dimensional regularization \cite{tHooft:1974toh}. At 1-loop 
order they consist of Eddington and Weyl terms:
\begin{equation}
\Delta \mathcal{L} = c_1 R^2 \sqrt{-g} 
+ c_2 \, C^{\alpha\beta\gamma\delta}
C_{\alpha\beta\gamma\delta} \sqrt{-g} 
\;\; , \label{DL}
\end{equation}
where the coefficients are \cite{Park:2011ww}:
\begin{eqnarray}
c_1 &\!\!\! = \!\!\!& 
\frac{\mu^{D-4} \Gamma(\frac{D}2)}{2^8 \pi^{\frac{D}{2}}}
\frac{(D\!-\!2)}{(D\!-\!1)^2 (D\!-\!3) (D\!-\!4)} 
\;\; , \label{c1def} \\
c_2 &\!\!\! = \!\!\!& 
\frac{\mu^{D-4} \Gamma(\frac{D}2)}{2^8 \pi^{\frac{D}{2}}}
\frac{2}{(D\!+\!1) (D\!-\!1) (D\!-\!3)^2 (D\!-\!4)} 
\;\; . \label{c2def}
\end{eqnarray}

We can decompose $R^2$ into three pieces as follows:
\begin{equation}
R^2 = \Bigl( R - D \Lambda \Bigr)^{\!\! 2} 
+ 2 D \Lambda \Bigl[ R - (D\!-\!2) \Lambda \Bigr]
+ D (D\!-\!4) \Lambda^2 
\;\; , \label{breakup}
\end{equation}
so that the Eddington counterterm becomes the sum of three 
contributions:
\begin{eqnarray}
\Delta \mathcal{L}_{1a} &\!\!\! \equiv \!\!\!& 
c_1 \Bigl( R - D \Lambda \Bigr)^{\!\! 2} \sqrt{-g} 
\;\; , \label{DL1a} \\
\Delta \mathcal{L}_{1b} &\!\!\! \equiv \!\!\!& 
2 D c_1 \Lambda \Bigl[ R - (D\!-\!2) \Lambda \Bigr] \sqrt{-g} 
\;\; , \label{DL1b} \\
\Delta \mathcal{L}_{1c} &\!\!\! \equiv \!\!\!& 
D (D\!-\! 4) c_1 \Lambda^2 \sqrt{-g}
\;\; . \label{DL1c}
\end{eqnarray}

{\bf -} There is also a finite renormalization of the 
cosmological constant which is necessary to make the
graviton self-energy conserved \cite{Tsamis:2023fri}:
\begin{equation}
\Delta \mathcal{L}_3 = 
c_3 \sqrt{-g} = c_3 \, a^D \sqrt{-\widetilde{g}} 
\;\; . \label{DLcc}
\end{equation}
\\ [9pt]
\noindent
$\bullet$ {\it The MMC Model: Conservation}

Stress-energy conservation has been discussed in 
detail in \cite{Tsamis:2023fri}. The starting point 
of the analysis is the Ward identity which follows 
from stress-energy conservation for a matter loop 
contribution to the graviton self-energy: 
\begin{equation}
\mathcal{W}^{\mu}_{~\alpha\beta} \times (-i) \left[ 
\mbox{}^{\alpha\beta} \Sigma^{\rho\sigma}_{\rm total} 
\right](x;x') = 0 
\;\; , \label{conserve}
\end{equation} 
with the Ward operator defined thusly:
\begin{equation}
\mathcal{W}^{\mu}_{~\alpha\beta} \equiv 
\delta^{\mu}_{~(\alpha} \partial_{\beta)}
+ a H \delta^{\mu}_{~0} \, \eta_{\alpha\beta} 
\;\; . \label{ward}
\end{equation}
When the scalar obeys its equation of motion the 
Ward operator annihilates the graviton variation:
\begin{equation}
\mathcal{W}^{\mu}_{~\alpha\beta} \times 
\frac{i \delta S[\phi,0]}{\delta h_{\alpha\beta}(x)} 
= \frac{\kappa}{2} \, \partial^{\mu} \phi(x) \!\times\!
\frac{i \delta S[\phi,0]}{\delta \phi(x)} 
\;\; . \label{wardproperty}
\end{equation} 

{\bf -} It can be shown \cite{Tsamis:2023fri} that in 
de Sitter spacetime and before renormalization we have: 
\begin{eqnarray}
\mathcal{W}^{\mu}_{~\alpha\beta} \times 
(-i) \Bigl[ \mbox{}^{\alpha\beta} \Sigma^{\rho\sigma}_{\rm prim}
\Bigr](x;x') 
&\!\!\! = \!\!\!&
\frac{i \xi \kappa^2}{2} \Biggl\{ 
-\eta^{\mu (\rho} \partial^{\sigma)}
\Bigl[ a^D \delta^D(x \!-\! x') \Bigr]
\nonumber \\
& & \hspace{-4cm}
+ \frac12 a^{D-2} \eta^{\rho\sigma} \partial^{\mu} 
\Bigl[ a^2 \delta^D(x\!-\!x') \Bigr] \Biggr\} 
\;\; , \qquad \label{obstdS}
\end{eqnarray}
where $\, \xi \equiv \frac{(D\!-\!2) D\!-\!1)}{2 D} H^2 k$. 
This equation exhibits the obstacle to achieving conservation 
albeit when only the primitive form of the 1-loop graviton 
self-energy is taken into account.

{\bf -} Upon renormalizing the 1-loop graviton self-energy 
the contributions from the counterterms (\ref{DL}) are 
considered. We employ the decomposition of the Eddington 
counterterm into three pieces (\ref{DL1a}-\ref{DL1c}) and 
have sequentially computed the action of the Ward operator 
on the various pieces. It turns out \cite{Tsamis:2023fri} 
that except for the counterterm (\ref{DL1c}) the Ward operator 
annihilates their contribution to the graviton self-energy:
\begin{eqnarray}
\mathcal{W}^{\mu}_{~\alpha\beta} \times (-i)\left[ 
\mbox{}^{\alpha\beta} \Sigma^{\rho\sigma}_2 \right](x;x') 
& \!=\! & 
\mathcal{W}^{\mu}_{~\alpha\beta} \times (-i)\left[ 
\mbox{}^{\alpha\beta} \Sigma^{\rho\sigma}_{1a} \right](x;x') 
\nonumber \\
& \mbox{} & \hspace{-5.3cm}
= \;
\mathcal{W}^{\mu}_{~\alpha\beta} \times (-i)\left[ 
\mbox{}^{\alpha\beta} \Sigma^{\rho\sigma}_{1b} \right](x;x') 
\;=\; 0
\;\; , \label{2 1a 1b}
\end{eqnarray} 

The only non-zero contributions come from: 
\\ [3pt]
{\it (i)} the cosmological constant counterterm (\ref{DLcc}) 
with coefficient $c_3$, and 
\\ [3pt]
{\it (ii)} the cosmological constant like counterterm (\ref{DL1c}) 
emanating from the Eddington decomposition (\ref{breakup}) with 
coefficient $\, \gamma \equiv D (D \!-\!1)^2 (D \!-\! 4) H^4 c_1$. 
\\ [3pt]
The results are \cite{Tsamis:2023fri}:
\begin{eqnarray}
\mathcal{W}^{\mu}_{~\alpha\beta} \times 
-i\left[ \mbox{}^{\alpha\beta} \Sigma^{\rho\sigma}_{3+1c} 
\right](x;x')
&\!\!\! = \!\!\!&
\frac{i (c_3 + \gamma) \kappa^2}{2} \Biggl\{\! 
-\eta^{\mu (\rho} \partial^{\sigma)} 
\Bigl[ a^D \delta^D(x \!-\! x') \Bigr]
\nonumber \\
& & \hspace{-4cm} 
+ \frac12 a^{D-2} \eta^{\rho\sigma} \partial^{\mu} 
\Bigl[ a^2 \delta^D(x\!-\!x') \Bigr] \! \Biggr\} 
\;\; , \qquad \label{obst 3 1c}
\end{eqnarray}
and because the functional form of the objection 
(\ref{obst 3 1c}) is identical to that of the primitive 
objection (\ref{obstdS}), it is possible to arrange 
their coefficients so that they cancel against each 
other:
\begin{equation}
\xi = - (c_3 + \gamma)
\;\; , \label{cancel}
\end{equation}
and achieve the desired conservation. Substituting
into (\ref{cancel}) the values of $k$, $\xi$
and $\gamma$ we see that this corresponds to a
finite renormalization of the cosmological constant
in $D\!=\!4$:
\begin{equation}
c_3 = 
-\frac{\mu^{D-4} H^4}{2^7 \pi^{\frac{D}2}} 
\frac{(D\!-\!2) \Gamma(\frac{D}2 \!+\! 1)}{D\!-\!3} 
- \frac{H^D}{(4\pi)^{\frac{D}2}}
\frac{(D\!-\!2) \Gamma(D)}{4 \Gamma(\frac{D}2 \!+\! 1)} 
\; \longrightarrow \;
-\frac{H^4}{8 \pi^2} 
\;\; . \label{c3}
\end{equation}

\section{The 1-loop QFT results: Self-Energy}

As discussed previously, the first step in our project 
involving (\ref{L}) is its perturbative contribution to 
the 1-loop graviton self-energy given by:
\begin{eqnarray}
-i \Bigl[\mbox{}^{\mu\nu} \Sigma^{\rho\sigma}\Bigr](x;x') 
& = & 
\Bigl\langle \Omega \, \Bigl\vert \, T^*\Biggl[ 
\frac{i\delta S[\varphi,g]}{\delta h_{\mu\nu}(x)}_{\varphi\varphi} 
\!\!\! \times \, 
\frac{i\delta S[\varphi,g]}{\delta h_{\rho\sigma}(x')}_{\varphi\varphi} 
\nonumber \\
& & \hspace{-0.1cm} 
+ \frac{i\delta^2 S[\varphi,g]}
{\delta h_{\mu\nu}(x) \delta h_{\rho\sigma}(x')}_{\varphi\varphi} 
+ \frac{i \delta^2 \Delta S[g]}
{\delta h_{\mu\nu}(x) \delta h_{\rho\sigma}(x')}_{1} 
\Biggr] \Bigr\vert \, \Omega \Bigr\rangle 
\;\; , \qquad \label{sigma}
\end{eqnarray}
where the subscripts indicate how many weak fields contribute, 
the $T^*$ symbol stands for time-ordering with any derivatives 
taken outside, and $\Delta S[g]$ denotes the 1-loop counterterm 
action. 
\footnote{The Feynman diagrams corresponding to the 3 terms 
in (\ref{sigma}) can be seen in Fig.~1-3.}
The 1st and 2nd variations required for (\ref{sigma}) are:
\footnote{The variation with respect to the graviton is related
to the variation with respect to the metric as
$\, \frac{\delta}{\delta h_{\mu\nu}(x)} = \kappa a^2 
\frac{\delta}{\delta g_{\mu\nu}(x)}$ due to (\ref{hmn}).}
\begin{eqnarray}
& & \hspace{-1.1cm}
\frac{i\delta S[\varphi,g]}{\delta h_{\mu\nu}(x)}_{\varphi\varphi} 
=
\frac{i \kappa}{2} a^{D-2} \Bigl[ \partial^{\mu} \varphi \; 
\partial^{\nu} \varphi 
- \frac12 \eta^{\mu\nu} \partial^{\alpha} \varphi \;
\partial_{\alpha} \varphi \Bigr] 
\;\; , \label{var1} \\
& & \hspace{-1.1cm} 
\frac{i\delta^2 S[\varphi,g]}
{\delta h_{\mu\nu}(x) \delta h_{\rho\sigma}(x')}_{\varphi\varphi} 
= 
\frac{\kappa^2}{2} a^{D-2} \Biggl[
- \eta^{\mu (\rho} \partial^{\sigma)} \varphi \; \partial^{\nu} \varphi 
- \eta^{\nu (\rho} \partial^{\sigma)} \varphi \; \partial^{\mu} \varphi
\nonumber \\ 
& &
+ \frac12 \eta^{\mu\nu} \partial^{\rho} \varphi \; \partial^{\sigma} \varphi
+ \frac12 \eta^{\rho\sigma} \partial^{\mu} \varphi \; \partial^{\nu} \varphi
+ \frac12 \Bigl( \eta^{\mu (\rho} \eta^{\sigma) \nu}
- \frac12 \eta^{\mu\nu} \eta^{\rho\sigma} \Bigr) \; 
\partial^{\alpha} \varphi \; \partial_{\alpha} \varphi 
\Biggr] 
\nonumber \\
& &
\times \, i\delta^D(x \!-\! x') 
\;\; . \label{var2}
\end{eqnarray}
\noindent
$\bullet$ {\it The 4-point Contribution}

The 4-point contribution (Fig.~2) is the expectation value 
of (\ref{var2}):
\begin{eqnarray}
-i \Bigl[\mbox{}^{\mu\nu} \Sigma_4^{\rho\sigma}\Bigr](x;x') 
&\!\! = \!\!& 
\frac{\kappa^2}{2} a^{D-2} \Biggl[
- \eta^{\mu (\rho} \partial^{\prime \sigma)} \partial^{\nu} 
\, i\Delta(x;x') 
- \eta^{\nu (\rho} \partial^{\prime \sigma)} \partial^{\mu} 
\, i\Delta(x;x') 
\nonumber \\
& & \hspace{-0.7cm} 
+ \frac12 \eta^{\mu\nu} \partial^{\rho} \partial^{\prime \sigma} 
\, i\Delta(x;x') 
+ \frac12 \eta^{\rho\sigma} \partial^{\mu} \partial^{\prime\nu} 
\, i\Delta(x;x') 
\nonumber \\
& & \hspace{-0.7cm} 
+ \frac12 \Bigl( \eta^{\mu (\rho} \eta^{\sigma) \nu} 
- \frac12 \eta^{\mu\nu} \eta^{\rho\sigma} \Bigr) 
\partial^{\alpha} \partial'_{\alpha} \, i\Delta(x;x') 
\Biggr] i\delta^D(x \!-\! x') 
\;\; , \label{4ptA}
\end{eqnarray}
which simplifies to:
\footnote{Due to the identity:
$\partial_{\alpha} \partial'_{\beta} \, i\Delta(x;x') 
\Bigl\vert_{x' = x} = 
- \Bigl( \frac{D\!-\!1}{D}\Bigr) k H^2 g_{\alpha\beta}
\quad , \quad 
k \equiv \frac{H^{D-2}}{(4\pi)^{\frac{D}2}} 
\frac{\Gamma(D\!-\!1)}{\Gamma(\frac{D}2)}$.}
\begin{equation}
-i \Bigl[\mbox{}^{\mu\nu} \Sigma_4^{\rho\sigma}\Bigr](x;x') = 
\frac{(D\!-\!1) (D\!-\!4)}{4 D} \kappa^2 k H^2 a^D \Bigl( 
\frac12 \eta^{\mu\nu} \eta^{\rho\sigma} 
- \eta^{\mu (\rho} \eta^{\sigma) \nu} \Bigr) 
i \delta^D(x \!-\! x') 
\;\; , \label{4ptB}
\end{equation}
and hence vanishes in $D \!=\! 4$ spacetime dimensions.

\newpage
\noindent
$\bullet$ {\it The 3-point Contribution}

The expectation value of (\ref{var1}) is the 3-point 
contribution to the 1-loop self-energy (Fig.~1):
\begin{eqnarray}
 -i \Bigl[\mbox{}^{\mu\nu} \Sigma_3^{\rho\sigma}\Bigr](x;x') 
&\!\! = \!\!&
\Bigl(\frac{i \kappa}{2}\Bigr)^2 (a a')^{D-2} \Biggl\{ 
2 \partial^{\mu} \partial^{\prime (\rho} \, i\Delta(x;x') \;
\partial^{\prime \sigma)} \partial^{\nu} \, i\Delta(x;x') 
\nonumber \\
& & \hspace{-3.1cm} 
- \eta^{\mu\nu} \partial^{\alpha} \partial^{\prime \rho} \,
i\Delta(x;x') \;
\partial_{\alpha} \partial^{\prime \sigma} \, 
i\Delta(x;x') 
- \eta^{\rho\sigma} \partial^{\mu} \partial^{\prime \beta} \,
i\Delta(x;x') \;
\partial^{\nu} \partial'_{\beta} i\Delta(x;x') 
\nonumber \\
& & \hspace{-3.1cm} 
+ \frac12 \eta^{\mu\nu} \eta^{\rho\sigma} \partial^{\alpha}
\partial^{\prime \beta} \, i\Delta(x;x') \;
\partial_{\alpha} \partial'_{\beta} \, i\Delta(x;x') \Biggr\} 
\nonumber \\
& & \hspace{-3.1cm} 
\equiv -i\Sigma_{3i} \, -i\Sigma_{3ii} \, -i\Sigma_{3iii} 
\, -i\Sigma_{3iv} 
\;\; . \label{3pt}
\end{eqnarray}
Notice that the first term $-i\Sigma_{3i}$, besides being 
the most difficult, recovers all three remaining terms by 
suitable contractions:
\begin{eqnarray}
\lefteqn{ -i \Bigl[\mbox{}^{\mu\nu} \Sigma_3^{\rho\sigma}\Bigr](x;x') 
=
\Bigl[ \delta^{\mu}_{~\alpha} \delta^{\nu}_{~\beta} 
- \frac12 \eta^{\mu\nu} \eta_{\alpha\beta} \Bigr] 
\Bigl[ \delta^{\rho}_{~\gamma} \delta^{\sigma}_{~\delta}
- \frac12 \eta^{\rho\sigma} \eta_{\gamma\delta} \Bigr] } 
\nonumber \\
& & \hspace{2.5cm} 
\times \, 2 \Bigl(\frac{i \kappa}{2}\Bigr)^2 (a a')^{D-2} \,
\partial^{\alpha} \partial^{\prime (\gamma} i\Delta(x;x') \, 
\partial^{\prime \delta)} \partial^{\beta} i\Delta(x;x') 
\;\; . \qquad \label{3pt2} 
\end{eqnarray}

All terms are quartically divergent, whereas the product of 
two undifferentiated propagators is logarithmically divergent. 
In view of the two derivatives on each propagator, we must 
therefore retain three terms in the expansion of each propagator
\cite{Onemli:2002hr,Onemli:2004mb}:
\begin{eqnarray}
\lefteqn{ i\Delta(x;x') = 
\frac1{4 \pi^{\frac{D}2}} \Biggl\{ 
\frac{2 \Gamma(\frac{D}2)}{D \!-\! 2} 
\frac1{(a a' \Delta x^2)^{\frac{D}2 - 1}} }
\nonumber \\
& & \hspace{1cm} 
+ \frac{\Gamma(\frac{D}2 \!+\! 1)}{2 (D \!-\! 4)} 
\frac{H^2}{(a a' \Delta x^2)^{\frac{D}2 -2}} 
+ \frac{\Gamma(\frac{D}2 \!+\! 2)}{16 (D \!-\! 6)} 
\frac{H^4}{(a a' \Delta x^2)^{\frac{D}2 -3}} + \dots \Biggr\} 
\nonumber \\
& & \hspace{1cm} 
+ k \, \Biggl\{
-\pi {\rm cot}\Bigl( \frac{D\pi}{2}\Bigr) 
+ \ln(a a') 
+ \Bigl( \frac{D \!-\!1}{2 D}\Bigr) H^2 a a' \Delta x^2 + \dots \Biggr\} 
\;\; . \qquad \label{propexp}
\end{eqnarray}
Coordinate differences are indicated throughout by a $\Delta$:
\begin{equation}
\Delta x^{\mu} \equiv (x \!-\! x')^{\mu} 
\quad , \quad 
\Delta \eta \equiv \eta \!-\! \eta' 
\quad , \quad 
\Delta r \equiv \Vert \vec{x} \!-\! \vec{x}'
\Vert \; . \label{differences}
\end{equation}
Taking two derivatives of the propagator gives:
\begin{eqnarray}
\lefteqn{ \partial^{\mu} \partial^{\prime \rho} i\Delta(x;x') 
= 
\frac{\delta^{\mu}_{~0} \delta^{\rho}_{~0} i\delta^D(x \!-\! x')}{a^{D-2}} 
+ 
\frac{\Gamma(\frac{D}2)}{2 \pi^{\frac{D}2} (a a')^{\frac{D}2 -1}} \Biggl\{ 
\frac{\eta^{\mu\rho}}{\Delta x^D} 
- \frac{D \Delta x^{\mu} \Delta x^{\rho}}{\Delta x^{D+2}} } 
\nonumber \\
& & \hspace{-0.5cm} 
+ \frac{(D\!-\!2) [a H \delta^{\mu}_{~0} \Delta x^{\rho} 
\!-\! \Delta x^{\mu} a' H \delta^{\rho}_{~0}]}{2 \Delta x^D} 
+ \frac{(D\!-\!2) a a' H^2 \delta^{\mu}_{~0} \delta^{\rho}_{~0}}
{4 \Delta x^{D-2}} 
+ \frac{D}{8} a a' H^2 
\nonumber \\
& & \hspace{0cm} 
\times \Biggl[ \frac{\eta^{\mu\rho}}{\Delta x^{D-2}} 
- \frac{(D\!-\!2) \Delta x^{\mu} \Delta x^{\rho}}{\Delta x^{D}} 
+ \frac{(D\!-\!4) [a H \delta^{\mu}_{~0} \Delta x^{\rho} 
\!-\! \Delta x^{\mu} a' H \delta^{\rho}_{~0}]}
{2 \Delta x^{D-2}} 
\nonumber \\
& & \hspace{-0.5cm} 
+ \frac{(D\!-\!4) a a' H^2 \delta^{\mu}_{~0} \delta^{\rho}_{~0}}
{4 \Delta x^{D-4}} \Biggr] 
+ \frac{D (D \!+\! 2)}{128} a^2 {a'}^2 H^4 \Biggl[ 
\frac{\eta^{\mu\rho}}{\Delta x^{D-4}} 
- \frac{(D\!-\!4) \Delta x^{\mu} \Delta x^{\rho}}{\Delta x^{D-2}} 
\nonumber \\
& & \hspace{0cm} 
+ \frac{(D\!-\!6) [a H \delta^{\mu}_{~0} \Delta x^{\rho} 
\!-\! \Delta x^{\mu} a' H \delta^{\rho}_{~0}]}{2 \Delta x^{D-4}} 
+ \frac{(D\!-\!6) a a' H^2 \delta^{\mu}_{~0} \delta^{\rho}_{~0}}
{4 \Delta x^{D-6}} \Biggr] + \dots \Biggr\} 
\nonumber \\
& & \hspace{-0.5cm} 
+ \, k \, \Biggl\{ 0 - \Bigl( \frac{D\!-\!1}{D} \Bigr) a a' H^2 
\Bigl[ \eta^{\mu\rho} - a H \delta^{\mu}_{~0} \Delta x^{\rho} 
+ a' H \Delta x^{\mu} \delta^{\rho}_{~0} 
\nonumber \\
& & \hspace{4cm} 
- \frac12 a a' H^2 \Delta x^2 \delta^{\mu}_{~0} \delta^{\rho}_{~0}
\Bigr] + \dots \Biggr\}
\;\; . \qquad \label{difpropexp}
\end{eqnarray}

Now the product of the two doubly-differentiated propagators 
in (\ref{3pt2}) consists of two local terms plus a product of 
two nonlocal terms:
\begin{eqnarray}
\lefteqn{ 2 \Bigl(\frac{i \kappa}{2}\Bigr)^2 (a a')^{D-2} \,
\partial^{\mu} \partial^{\prime (\rho} i\Delta(x;x') \,
\partial^{\prime \sigma)} \partial^{\nu} i\Delta(x;x') = } 
\nonumber \\
& & \hspace{-0.3cm} 
+ \Bigl(\frac{D\!-\!1}{D} \Bigr) \kappa^2 k H^2 a^D 
\delta^{(\mu}_{~~0} \eta^{\nu) (\rho} \delta^{\sigma)}_{~~0} \,
i\delta^D(x \!-\! x') 
\nonumber \\
& & \hspace{-0.3cm} 
-\frac{\kappa^2 \Gamma^2(\frac{D}2)}{8 \pi^D} 
\Biggl\{ \frac{\eta^{\mu (\rho}}{\Delta x^D} 
- \frac{D \Delta x^{\mu} \Delta x^{(\rho}}{\Delta x^{D+2}} 
+ \dots \Biggr\} 
\Biggl\{ \frac{\eta^{\sigma) \nu}}{\Delta x^D} 
- \frac{D \Delta x^{\sigma)} \Delta x^{\nu}}{\Delta x^{D+2}} 
+ \dots \Biggr\}
\qquad \label{product}
\end{eqnarray}
The first 8 terms in each of the curly bracketed expressions
can make a non-zero contribution and are shown in Table~\ref{Vi}.
Because of symmetrization there are a total of 36 independent 
products of these 8 terms. 

We shall not present the very complicated procedure of reducing
the aforementioned products to isolate the divergences and the
non-local finite parts they contain. It should be sufficient 
to incorporate the reuslts in the analysis that follows.
\\ [9pt]
$\bullet$ {\it The 3-point Contribution: Tensor Basis}

A covariant generalization of \cite{Tan:2021ibs} provides
an appropriate tensor basis for the primitive divergent 
contributions:
\begin{equation}
-i\Bigl[ \mbox{}^{\mu\nu} \Sigma^{\rho\sigma}_{\rm prim} \Bigr](x;x') 
= 
\mathcal{K} \times \sum_{i=1}^{21} T^{i}(x;x') 
\times \Bigl[ \mbox{}^{\mu\nu} D_{i}^{\rho\sigma} \Bigr] 
\times i\delta^D(x \!-\! x') 
\;\; ,
\label{genform1}
\end{equation}
where the divergent prefactor $\mathcal{K}$ is:
\begin{equation}
\mathcal{K} \equiv \frac{\kappa^2 \Gamma^2(\frac{D}2)}{8 \pi^D} 
\times K = 
\frac{\kappa^2 \mu^{D-4} \Gamma(\frac{D}2)}{2 \pi^{\frac{D}2} 
(D\!-\!3) (D\!-\!4)} 
\;\; . \label{Kdef}
\end{equation}
The tensor differential operators $[\mbox{}^{\mu\nu} D_{i}^{\rho\sigma}]$ 
are listed in the table below:

\begin{table}[H]
\setlength{\tabcolsep}{8pt}
\def\arraystretch{1.5}
\centering
\begin{tabular}{|@{\hskip 1mm }c@{\hskip 1mm }||c||c|c||c|c|}
\hline
$i$ & $[\mbox{}^{\mu\nu} D^{\rho\sigma}_i]$ & $i$ & 
$[\mbox{}^{\mu\nu} D^{\rho\sigma}_i]$
& $i$ & $[\mbox{}^{\mu\nu} D^{\rho\sigma}_i]$ \\
\hline\hline
1 & $\eta^{\mu\nu} \eta^{\rho\sigma}$ & 8 & 
$\partial^{\mu} \partial^{\nu} \eta^{\rho\sigma}$ & 15 & 
$\delta^{(\mu}_{~~0} \partial^{\nu)} 
\delta^{\rho}_{~0} \delta^{\sigma}_{~0}$ \\
\hline
2 & $\eta^{\mu (\rho} \eta^{\sigma) \nu}$ & 9 & 
$\delta^{(\mu}_{~~0} \eta^{\nu) (\rho}
\delta^{\sigma)}_{~~0}$ & 16 & 
$\delta^{\mu}_{~0} \delta^{\nu}_{~0} \partial^{\rho}
\partial^{\sigma}$ \\
\hline
3 & $\eta^{\mu\nu} \delta^{\rho}_{~0} \delta^{\sigma}_{~0}$ & 10 & 
$\delta^{(\mu}_{~~0} \eta^{\nu) (\rho} \partial^{\sigma)}$ & 17 & 
$\partial^{\mu} \partial^{\nu} \delta^{\rho}_{~0}
\delta^{\sigma}_{~0}$ \\
\hline
4 & $\delta^{\mu}_{~0} \delta^{\nu}_{~0} \eta^{\rho\sigma}$ & 11 & 
$\partial^{(\mu} \eta^{\nu) (\rho} \delta^{\sigma)}_{~~0}$ & 18 & 
$\delta^{(\mu}_{~~0} \partial^{\nu)}
\delta^{(\rho}_{~~0} \partial^{\sigma)}$ \\
\hline
5 & $\eta^{\mu\nu} \delta^{(\rho}_{~~0} \partial^{\sigma)}$ & 12 & 
$\partial^{(\mu} \eta^{\nu)(\rho} \partial^{\sigma)}$ & 19 & 
$\delta^{(\mu}_{~~0} \partial^{\nu)}
\partial^{\rho} \partial^{\sigma}$ \\
\hline
6 & $\delta^{(\mu}_{~~0} \partial^{\nu)} \eta^{\rho\sigma}$ & 13 & 
$\delta^{\mu}_{~0} \delta^{\nu}_{~0} \delta^{\rho}_{~0} 
\delta^{\sigma}_{~0}$ & 20 & 
$\partial^{\mu} \partial^{\nu} 
\delta^{(\rho}_{~~0} \partial^{\sigma)}$ \\
\hline
7 & $\eta^{\mu\nu} \partial^{\rho} \partial^{\sigma}$ & 14 & 
$\delta^{\mu}_{~0} \delta^{\nu}_{~0} \delta^{(\rho}_{~~0} 
\partial^{\sigma)}$ & 21 & 
$\partial^{\mu} \partial^{\nu} \partial^{\rho} \partial^{\sigma}$ \\
\hline
\end{tabular}
\caption{\footnotesize 
The 21 basis tensors used in expression (\ref{genform1}).
The pairs $(3,4)$, $(5,6)$, $(7,8)$, $(10,11)$, $(14,15)$, 
$(16,17)$ and $(19,20)$ are related by reflection.}
\label{Tbasis}
\end{table}

The same tensor basis is appropriate for the counterterm
contributions (Fig.~3):
\begin{equation}
-i\Bigl[ \mbox{}^{\mu\nu} \Sigma^{\rho\sigma}_{\rm count} \Bigr](x;x') 
= 
\mathcal{K} \times \sum_{i=1}^{21} \Delta T^{i}(x;x') 
\times \Bigl[ \mbox{}^{\mu\nu} D_{i}^{\rho\sigma} \Bigr] 
\times i\delta^D(x \!-\! x') 
\;\; ,
\label{genform2}
\end{equation}

\noindent
$\bullet$ {\it The 3-point Contribution: Primitive Divergences}

The primitive divergences coming only from the first term 
$\, -i\Sigma_{3i} \,$ provide contributions of the same form 
as (\ref{genform1}), but with different coefficients that we 
shall call $\mathcal{T}^i(x;x')$. Including the three trace 
terms according to (\ref{3pt2}) gives complicated relations for 
those $T^{i\,}$'s whose $[\mbox{}^{\mu\nu} D^{\rho\sigma}]$'s 
contain either $\eta^{\mu\nu}$ or $\eta^{\rho\sigma}$. Hence, 
a subset of the $T^{i\,}$'s will be equal to the 
$\mathcal{T}^{i\,}$'s: 
\begin{equation}
T^2 = \mathcal{T}^2
\qquad , \qquad
T^i = \mathcal{T}^i \quad \forall i \geq 9
\;\; , \label{T=T}
\end{equation}
while the remaining $T^{i\,}$'s will be linear combinations
of the $\mathcal{T}^{i\,}$'s. Of these, $T^1$ satisfies: 
\begin{eqnarray}
T^1 &\!\!\! = \!\!\!&
\frac{(D\!-\!2)^2}{4} \mathcal{T}^1 
+ \frac{(D\!-\!4)}{4} \mathcal{T}^2 
- \frac{(D\!-\!2)}{4} (\mathcal{T}^3 + \mathcal{T}^4) 
+ \frac{(D\!-\!2)}{4} (\mathcal{T}^5 + \mathcal{T}^6) \partial_0 
\nonumber \\
& & \hspace{0cm} 
+ \frac{(D\!-\!2)}{4} (\mathcal{T}^7 + \mathcal{T}^8) \partial^2
- \frac14 \mathcal{T}^9 
+ \frac14 (\mathcal{T}^{10} + \mathcal{T}^{11}) \partial_0
+ \frac14 \mathcal{T}^{12} \partial^2 
+ \frac14 \mathcal{T}^{13} 
\nonumber \\
& & \hspace{0cm} 
- \frac14 (\mathcal{T}^{14} + \mathcal{T}^{15} ) \partial_0
- \frac14 (\mathcal{T}^{16} + \mathcal{T}^{17}) \partial^2 
+ \frac14 \mathcal{T}^{18} \partial^2_0 
\nonumber \\
& & \hspace{0cm} 
+ \frac14 (\mathcal{T}^{19} + \mathcal{T}^{20}) \partial^2 \partial_0 
+ \frac14 \mathcal{T}^{21} \partial^4 
\;\; , \label{T1def}
\end{eqnarray}
while the remaining ones are simpler and are presented 
in Table~\ref{Ti-to-Ti}.

The $\mathcal{T}^i(x;x')$ have been laboriously computed 
and are presented in Table~\ref{Tinitial}. For the cases
$i=2$ and $i \geq 9$ they are identical with $T^i(x;x')$. 
The most complicated case when $T^i \neq \mathcal{T}^i$ is
(\ref{T1def}):
\footnote{Note that differences of scale factors combine 
to give $\; a - a' = a a' H \Delta \eta$. 
\newline
The resulting factors of $\Delta \eta$ acting on 
$\delta^D(x - x')$ can be reduced using:
$\Delta \eta \partial_0 \longrightarrow -1 \;\; ,
\newline
\Delta \eta^2 \partial^2 \longrightarrow -2 \;\; , \;\;
\Delta \eta \partial_0 \partial^2 \longrightarrow 
-\partial^2 + 2 \partial_0^2 \;$.} 
\begin{eqnarray}
T^1(x;x') &\!\!\!= \!\!\!&
\frac{(D^2 \!-\! 2D \!-\! 2)\partial^4}
{32 (D\!+\!1) (D\!-\!1) (D\!-\!2)} 
+ \frac{(3 D^3 \!-\! 18 D^2 \!+\! 24 D \!-\! 16) a a' H^2 \partial^2}
{512 (D\!-\!1)} 
\quad \nonumber \\
& & \hspace{-1.7cm} 
- \frac{(D\!-\!2) (D\!-\! 3) a a' H^2 \partial_0^2}{64 (D\!-\!1)} 
+ \frac{(D \!-\! 2) (D \!-\! 4) (D^4 \!-\! 48 D \!+\! 64) a^2 {a'}^2 H^4}
{4096 (D\!-\!1)} 
\;\; . \label{T1final}
\end{eqnarray}
The remaining cases for which $T^i \neq \mathcal{T}^i$ are 
given in Table~\ref{Tfinal}.
\\ [9pt]
\noindent
$\bullet$ {\it The 3-point Contribution: Weyl Counterterm}

A simple computation shows that the contribution of the 
Weyl counterterm to the graviton self-energy equals:
\begin{equation}
\frac{\delta^2 \, i\Delta S_2}
{\delta h_{\mu\nu}(x) \delta h_{\rho\sigma}(x')} 
\Bigr\vert_{h_{\mu\nu} = 0} = 
2 \kappa^2 c_2 \, \mathcal{C}^{\alpha\beta\gamma\delta\mu\nu}
\Bigl[ a^{D-4} \mathcal{C}_{\alpha\beta\gamma\delta}^{~~~~~\rho\sigma} 
i \delta^D(x\!-\!x') \Bigr] 
\;\; , \label{S2contA}
\end{equation}
where the tensor differential operator
$\mathcal{C}_{\alpha\beta\gamma\delta}^{~~~~~\mu\nu}$ is defined 
via the linearized, conformally rescaled Weyl tensor thusly:
\begin{equation}
\widetilde{C}_{\alpha\beta\gamma\delta} = 
\mathcal{C}_{\alpha\beta\gamma\delta}^{~~~~~\mu\nu} \!\times\! 
\kappa h_{\mu\nu} + O(\kappa^2 h^2) 
\;\; . \label{Cdef}
\end{equation}
Its explicit form is:
\begin{eqnarray}
\mathcal{C}_{\alpha\beta\gamma\delta}^{~~~~~\mu\nu} 
&\!\!\! = \!\!\!&
\mathcal{D}_{\alpha\beta\gamma\delta}^{~~~~~\mu\nu}
- \frac1{D\!-\!2} \Bigl[ \eta_{\alpha\gamma} 
\mathcal{D}_{\beta\delta}^{~~\mu\nu}
- \eta_{\gamma\beta} \mathcal{D}_{\delta\alpha}^{~~\mu\nu}  
+ \eta_{\beta\delta} \mathcal{D}_{\alpha\gamma}^{~~\mu\nu}
- \eta_{\delta\alpha} \mathcal{D}_{\gamma\beta}^{~~\mu\nu} \Bigr]
\nonumber \\
& & \hspace{2.9cm}
+ \frac{(\eta_{\alpha\gamma} \eta_{\beta\delta} \!-\! 
\eta_{\alpha\delta} \eta_{\beta\gamma}) \mathcal{D}^{\mu\nu} }
{(D\!-\!1) (D\!-\!2)} 
\;\; , \label{Cexplicit}
\end{eqnarray}
where the various derivatives are:
\begin{eqnarray}
\mathcal{D}_{\alpha\beta\gamma\delta}^{~~~~~\mu\nu} 
&\!\!\! \equiv \!\!\!&
-\frac12 \Bigl[ \delta^{(\mu}_{~~\alpha} \delta^{\nu)}_{~~\gamma} \, 
\partial_{\beta} \partial_{\delta}
- \delta^{(\mu}_{~~\gamma} \delta^{\nu)}_{~~\beta} \,
\partial_{\delta} \partial_{\alpha}
+ \delta^{(\mu}_{~~\beta} \delta^{\nu)}_{~~\delta} \, 
\partial_{\alpha} \partial_{\gamma}
\nonumber \\
& & \hspace{6.3cm}
- \delta^{(\mu}_{~~\delta} \delta^{\nu)}_{~~\alpha} \, 
\partial_{\gamma} \partial_{\beta} \Bigr] 
\;\; , \label{D4def} \\
\mathcal{D}_{\beta\delta}^{~~\mu\nu} 
&\!\!\! \equiv \!\!\!& 
\eta^{\alpha\gamma} \mathcal{D}_{\alpha\beta\gamma\delta}^{~~~~~\mu\nu} 
= 
-\frac12 \Bigl[ \eta^{\mu\nu} \partial_{\beta} \partial_{\delta} 
- 2 \partial^{(\mu} \delta^{\nu)}_{~~(\beta} \partial_{\delta)} 
+ \delta^{(\mu}_{~~\beta} \delta^{\nu)}_{~~\delta)} \partial^2 \Bigr]
\;\; , \qquad \label{D2def} \\
\mathcal{D}^{\mu\nu} 
&\!\!\! \equiv \!\!\!& 
\eta^{\beta\delta} \mathcal{D}_{\beta\delta}^{~~\mu\nu} 
= 
\partial^{\mu} \partial^{\nu} - \eta^{\mu\nu} \partial^2
= \Pi^{\mu\nu} 
\;\; . \label{D0def}
\end{eqnarray}
Consequently, the explicit form of (\ref{S2contA}) becomes:
\footnote{
Taking also into account the usual expansion of the measure 
factor, \\
$\, a^{D-4} = 1 + (D\!-\!4) \ln(a) + O[(D\!-\!4)^2]$.}
\begin{eqnarray}
\frac{\delta^2 \, i\Delta S_2}
{\delta h_{\mu\nu}(x) \delta h_{\rho\sigma}(x')} 
\Bigr\vert_{h_{\mu\nu} = 0} 
&\!\!\! = \!\!\!& 
2 \kappa^2 c_2 \Bigl( \frac{D\!-\!3}{D\!-\!2} \Bigr)
\Bigl[ \Pi^{\mu (\rho} \Pi^{\sigma) \nu} 
- \frac{\Pi^{\mu\nu} \Pi^{\rho\sigma}}{D\!-\!1} \Bigr] 
i \delta^D(x \!-\! x')  
\qquad \nonumber \\
& & \hspace{-2.7cm} 
+ \frac{\kappa^2}{2^6 \!\cdot\!3 \!\cdot\! 5 \!\cdot\! \pi^2} \,
\mathcal{C}^{\alpha\beta\gamma\delta\mu\nu} 
\Bigl[ \ln(a) \, \mathcal{C}_{\alpha\beta\gamma\delta}^{~~~~~\rho\sigma} 
\, i \delta^4(x\!-\!x') \Bigr] 
+ O(D\!-\!4) 
\;\; ,  \qquad \label{S2contB}
\end{eqnarray}
with the prefactor of the divergence equaling:
\begin{equation}
2 \kappa^2 \, c_2 \Bigl( \frac{D\!-\!3}{D\!-\!2}\Bigr) 
= 
\mathcal{K} \times
\frac1{32 (D\!+\!1) (D\!-\!1) (D\!-\!2)} 
\;\; . \label{Weylpre}
\end{equation}

The divergent contribution to each $T^i(x;x')$ is given 
in Table~\ref{WeylDiv}.
\\ [9pt]
\noindent
$\bullet$ {\it The 3-point Contribution: Eddington Counterterm}

In (\ref{breakup}), the Eddington counterterm $R^2$ was 
decomposed into three pieces. We shall analyze them in
reverse order.

{\bf -} The last of these is the counterterm (\ref{DL1c}) 
which is the same as a cosmological constant and finite:
\begin{equation}
\frac{\delta^2 \, i\Delta S_{1c}}
{\delta h_{\mu\nu}(x) \delta h_{\rho\sigma}(x')} 
\Bigr\vert_{h_{\mu\nu} = 0} 
= 
\frac{\kappa^2 H^4}{64 \pi^2} 
\Bigl[ -\eta^{\mu (\rho} \eta^{\sigma) \nu} 
+ \frac12 \eta^{\mu\nu} \eta^{\rho\sigma} \Bigr] 
a^4 \, i \delta^4(x \!-\! x') 
\;\; . \label{S1ccont}
\end{equation}

{\bf -} The second of these is the Einstein counterterm 
(\ref{DL1b}) whose contribution to the graviton
self-energy is:
\begin{eqnarray}
\frac{\delta^2 \, i\Delta S_{1b}}
{\delta h_{\mu\nu}(x) \delta h_{\rho\sigma}(x')} 
\Bigr\vert_{h_{\mu\nu} = 0} 
&\!\!\! = \!\!\!&
D (D\!-\!1) c_1 \kappa^2 H^2 \Biggl\{ 
\Bigl[ \eta^{\mu (\rho} \eta^{\sigma) \nu} 
- \eta^{\mu\nu} \eta^{\rho\sigma} \Bigr]
\mathcal{D} \, i \delta^D(x \!-\! x') 
\nonumber \\
& & \hspace{-2.9cm} 
+ \Bigl[ 2 \partial^{\prime (\mu} \eta^{\nu) (\rho} \partial^{\sigma)} 
+ \eta^{\mu\nu} \partial^{\rho} \partial^{\sigma} 
+ \partial^{\prime \mu} \partial^{\prime \nu} \eta^{\rho\sigma} \Bigr] 
\Bigl[ a^{D-2} \, i \delta^D(x \!-\! x')
\Bigr] \Biggr\} 
\;\; , \label{S1bcont} 
\end{eqnarray}
where $\mathcal{D} \equiv \partial^{\mu} a^{D-2} \partial_{\mu}$.
By employing the delta function to express the scale factor 
$a^{D-2}$ as half primed and half unprimed so that:
\begin{equation}
a^{D-2} \quad \longrightarrow \quad
(a a')^{\frac{D}2 -1} =
a a' \Big( 1 - (\tfrac{D}2 \!-\! 2) \ln(a a') 
+ O[(D \!-\! 4)^2] \Big) 
\;\; , \label{Einsteinscale}
\end{equation}
by moving all scale factors to the left, for example:
\begin{equation}
\mathcal{D} \, i\delta^D(x \!-\! x') = 
(a a')^{\frac{D}2 - 1} \Bigl[ 
\partial^2 +\tfrac{D}2 ( \tfrac{D}2 \!-\! 1) a a' H^2 \Bigr] 
i \delta^D(x \!-\! x') 
\;\; , \label{examp1}
\end{equation}
by extracting a factor of $\mathcal{K}$ from the 
multiplicative prefactor:  
\begin{equation}
D (D\!-\!1) c_1 \kappa^2 H^2 = 
\mathcal{K} \times \frac{D (D\!-\!2) H^2}{128 (D\!-\!1)}
\;\; , \label{Einsteinpre}
\end{equation}
we derive the divergences shown in Table~\ref{EinsteinDiv}.

{\bf -} Finally, we consider the divergences associated with 
$\Delta \mathcal{L}_{1a}$ of (\ref{DL1a}). We follow the 
same procedure as for the Weyl countertem and express 
$R - D \Lambda$ as a tensor differential operator acting 
on a single graviton:
\begin{equation}
R - D \Lambda = 
\frac1{a^2} \!\times\! \overline{\mathcal{F}}^{\mu\nu} \!\times\! 
\kappa h_{\mu\nu} + O(\kappa^2 h^2) 
\;\; , \label{Fdef}
\end{equation} 
where the tensor differential operator 
$\overline{\mathcal{F}}^{\mu\nu}$ is:
\begin{eqnarray}
\overline{\mathcal{F}}^{\mu\nu} 
&\!\!\! \equiv \!\!\!& 
\partial^{\mu} \partial^{\nu} - \eta^{\mu\nu}
\Bigl[ \partial^2 - (D\!-\!1) a H \partial_0 \Bigr] 
- 2 (D\!-\!1) a H \delta^{(\mu}_{~~0} \partial^{\nu)}
\nonumber \\
& & \hspace{1cm}
+ D (D\!-\!1) a^2 H^2 \delta^{\mu}_{~0} \delta^{\nu}_{~0} 
\;\; .
\end{eqnarray}
The operator we actually need is obtained by partially 
integrating $\overline{\mathcal{F}}^{\mu\nu}$ to obtain:
\begin{eqnarray}
\mathcal{F}^{\mu\nu} 
&\!\!\! \equiv \!\!\!& 
\partial^{\mu} \partial^{\nu} 
- \eta^{\mu\nu} \Big[ \partial^2 + (D\!-\!1) a H \partial_0 
+ (D\!-\!1) a^2 H^2 \Big]
+ 2 (D\!-\! 1) a H \delta^{(\mu}_{~~0} \partial^{\nu)} 
\nonumber \\
& & \hspace{0.3cm} 
+ (D\!-\!2) (D\!-\!1) a^2 H^2 \delta^{\mu}_{~0} \delta^{\nu}_{~0} 
\;\; , \label{Fexplicit}
\end{eqnarray}
The resulting second variation of $S_{1a}$ is:
\begin{eqnarray}
\frac{\delta^2 \, i\Delta S_{1a}}
{\delta h_{\mu\nu}(x) \delta h_{\rho\sigma}(x')} 
\Bigr\vert_{h_{\mu\nu} = 0} 
&\!\!\! = \!\!\!&
2 \kappa^2 c_1 \, \mathcal{F}^{\mu\nu} 
\Bigl[ a^{D-4} \mathcal{F}^{\rho\sigma} i \delta^D(x \!-\! x') \Bigr] 
\\
& & \hspace{-4.4cm} = 
2 \kappa^2 c_1 \mathcal{F}^{\mu\nu} \mathcal{F}^{\rho\sigma} 
i \delta^D(x \!-\! x') 
+ \frac{\kappa^2}{2^6 \!\cdot\! 3^2 \!\cdot\! \pi^2} \,
\mathcal{F}^{\mu\nu} 
\Bigl[ \ln(a) \, \mathcal{F}^{\rho\sigma} i \delta^4(x \!-\! x') \Bigr]
+ \dots \qquad \label{S1avariation}
\end{eqnarray}
The multiplicative prefactor is:
\begin{equation}
2 \kappa^2 c_1 = \mathcal{K} \times \frac{(D\!-\!2)}{64 (D\!-\!1)^2} 
\;\; . \label{1aprefactor}
\end{equation}
Table~\ref{EddingtonDiv} gives the divergences contributed by $S_{1a}$.
\\ [9pt]
\noindent
$\bullet$ {\it The 3-point Contribution: Primitive + Counterterm Results}

The primitive divergences of the 1-loop graviton 
self-energy that have just been evaluated, can be 
summarized in Table~\ref{Tprimitive} which lists 
the basis coefficients $T^i(x;x')$.

Similarly, the counterterm divergences of the 1-loop graviton 
self-energy that have just been evaluated, can be summarized 
in Table~\ref{cterms} which lists the basis coefficients 
$\Delta T^i(x;x')$.

By adding Tables~\ref{Tprimitive} and \ref{cterms} we arrive
at the residuals shown in Table~\ref{residual}. A few remarks
are in order:
\\ [5pt]
{\bf -} The residuals actually vanish for tensor factors 
$[\mbox{}^{\mu\nu} D_i^{\rho\sigma}]$ with $13 \leq i \leq 21$. 
\\ [5pt]
{\bf -} For $i = 1$, $i = 2$ and $i = 12$ the residual 
vanishes in $D=4$.
\\ [5pt]
{\bf -} For $3 \leq i \leq 11$ there is a more complicated 
cancellation scheme based on the observation:
\begin{equation}
\Delta \eta \partial^{\mu} = 
\partial^{\mu} \Delta \eta + \delta^{\mu}_{~0}
\;\; . \label{commute}
\end{equation}
{\bf -} Sometimes (for instance, $i = 7$ and $i = 8$) this 
must be done twice before the $\Delta \eta$ acts on the 
$i \delta^D(x-x')$ and vanishes. The clusters of tensor 
factors which cancel in this way are:
\\ [3pt]
1. The case of $i = 10$ and $i = 11$, which combine to 
cancel $i = 9$.
\\ [3pt]
2. The case of $i = 8$, which contributes to $i = 6$ to 
produce a $\Delta \eta$ term that cancels $i = 4$.
\\ [3pt]
3. The case of $i = 7$, which contributes to $i = 5$ to 
produce a $\Delta \eta$ term that cancels $i = 3$.

The final result is displayed in Table~\ref{finite} with 
each tensor factor being proportional to at least one factor 
of $(D-4)$. The final step, reported in Table~\ref{D4limit},
multiplies by the factor of $\, \tfrac{1}{D-4} \,$ in 
$\mathcal{K}$:
\begin{equation}
\mathcal{K} = \frac{\kappa^2}{2 \pi^2} \times \frac1{D-4} 
+ O[(D\!-\! 4)]
\;\; , \label{divergence}
\end{equation}
and then takes the limit of $D=4$.
\\ [9pt]
$\bullet$ {\it The 3-point Contribution: Finite Local Contributions}

There are six sources of finite, local contributions to the graviton
1-loop self-energy:
\\ [5pt]
{\bf -} The local terms arising from the action of two derivatives
on the scalar propagator which is given in equation (\ref{difpropexp}).
Upon substituting the local contribution from (\ref{difpropexp}) into
equation (\ref{3pt2}) and setting $D=4$ gives:
\begin{equation}
\frac{3 \kappa^2 H^4}{32 \pi^2} \Bigl\{ 
\delta^{(\mu}_{~~0} \eta^{\nu)(\rho} \delta^{\sigma)}_{~~0} 
- \frac12 \delta^{\mu}_{~0} \delta^{\nu}_{~0} \eta^{\rho\sigma}
- \frac12 \eta^{\mu\nu} \delta^{\rho}_{~0} \delta^{\sigma}_{~0} 
- \eta^{\mu\nu} \eta^{\rho\sigma} \Bigr\} a^4 i \delta^4(x \!-\! x') 
\;\; . \label{source1}
\end{equation}
{\bf -} The local terms arising from adding the primitive 
divergences to the counterterms; they can be found in
Table~\ref{D4limit}:
\begin{eqnarray}
\lefteqn{
\frac{a a' \kappa^2 H^2}{96 \pi^2} \Biggl\{
-\Bigl( \frac54 \partial^2 + 2 a a' H^2 \Bigr) 
\eta^{\mu\nu} \eta^{\rho\sigma} 
+ \frac12 \partial^2 \eta^{\mu (\rho} \eta^{\sigma) \nu} 
- a' H \eta^{\mu\nu} \delta^{(\rho}_{~~0} \partial^{\sigma)} } 
\nonumber \\
& & \hspace{1.1cm} 
+ a \delta^{(\mu}_{~~0} \partial^{\nu)} \eta^{\rho\sigma}
+ \eta^{\mu\nu} \partial^{\rho} \partial^{\sigma} 
+ \partial^{\mu} \partial^{\nu} \eta^{\rho\sigma} 
- 3 a a' H^2 \delta^{(\mu}_{~~0} \eta^{\nu) (\rho} \delta^{\sigma)}_{~~0} 
\nonumber \\
& & \hspace{1.1cm} 
- a' H \delta^{(\mu}_{~~0} \eta^{\nu) (\rho} \partial^{\sigma)}
+ a H \partial^{(\mu} \eta^{\nu) (\rho} \delta^{\sigma)}_{~~0} 
- \partial^{(\mu} \eta^{\nu) (\rho} \partial^{\sigma)} 
\Biggr\} \, i \delta^4(x \!-\! x') 
\; . \qquad \label{source2}
\end{eqnarray}
{\bf -} The local logarithm terms we found in equation (\ref{S2contB}) 
from the Weyl counterterm:
\begin{equation}
\frac{\kappa^2}{2^6 \!\cdot\! 3 \!\cdot\! 5 \!\cdot\! \pi^2} \,
\mathcal{C}^{\alpha\beta\gamma\delta\mu\nu} 
\Bigl[ \ln(a) \, \mathcal{C}_{\alpha\beta\gamma\delta}^{~~~~~\rho\sigma} 
\, i \delta^4(x \!-\! x') \Bigr] 
\;\; . \label{source4} 
\end{equation}
{\bf -} The local terms we found in equation (\ref{S1ccont}) from 
the finite renormalization of the cosmological constant:
\begin{equation}
\frac{\kappa^2 H^4 a^4}{64 \pi^2} 
\Bigl[ -\eta^{\mu (\rho} \eta^{\sigma) \nu} 
+ \frac12 \eta^{\mu\nu} \eta^{\rho\sigma} \Bigr] 
i \delta^4(x \!-\! x') 
\;\; . \label{source3}
\end{equation}
{\bf -} The local logarithm terms we found in equation 
(\ref{S1bcont}) from the Einstein counterterm:
\begin{eqnarray}
\lefteqn{ 
\frac{\kappa^2 H^2}{192 \pi^2} \, a a' \ln(a a') \Biggl\{
\Bigl[ \eta^{\mu (\rho} \eta^{\sigma) \nu} \!-\! 
\eta^{\mu\nu} \eta^{\rho\sigma} \Bigr] 
\Bigl[ \partial^2 \!+\! 2 a a' H^2 \Bigr] } 
\nonumber \\
& & \hspace{-0.3cm} 
+ 2 \Bigl[ - \partial^{(\mu} \eta^{\nu) (\rho} \partial^{\sigma)}
\!-\! a' H \delta^{(\mu}_{~~0} \eta^{\nu) (\rho} \partial^{\sigma)} 
\!+\! a H \partial^{(\mu} \eta^{\nu) (\rho} \delta^{\sigma)}_{~~0} 
\!+\! a a' H^2 \delta^{(\mu}_{~~0} \eta^{\nu) (\rho} \delta^{\sigma)}_{~~0} 
\Bigr] 
\nonumber \\
& & \hspace{-0.3cm} 
+ \eta^{\mu\nu} \Bigl[ \partial^{\rho} \partial^{\sigma} 
\!-\! 2 a H \delta^{(\rho}_{~~0} \partial^{\sigma)} 
\!+\! 2 a^2 H^2 \delta^{\rho}_{~0} \delta^{\sigma}_{~0} \Bigr] 
\nonumber \\
& & \hspace{-0.3cm} 
+ \eta^{\rho\sigma} \Bigl[ \partial^{\mu} \partial^{\nu} 
\!-\! 2 a H \delta^{(\mu}_{~~0} \partial^{\nu)} 
\!+\! 2 a^2 H^2 \delta^{\mu}_{~0} \delta^{\nu}_{~0} \Bigr] 
\Biggr\} \, i \delta^4(x \!-\! x') 
\;\; . \label{source6}
\end{eqnarray}
{\bf -} The local logarithm terms we found in equation 
(\ref{S1avariation}) from the Eddington counterterm:
\begin{equation}
\frac{\kappa^2}{2^6 \!\cdot\! 3^2 \!\cdot\! \pi^2} \, 
\mathcal{F}^{\mu\nu}
\Bigl[ \ln(a) \, \mathcal{F}^{\rho\sigma} i \delta^4(x \!-\! x') \Bigr] 
\;\; . \label{source5}
\end{equation}

\vspace{0.3cm}
\noindent
$\bullet$ {\it The 3-point Contribution: Finite Non-local Contributions}

The finite, non-local contributions are of the form:
\begin{eqnarray}
-i\Bigl[ \mbox{}^{\mu\nu} \Sigma^{\rho\sigma}_{\rm nonloc} \Bigr](x;x') 
&\!\!\! = \!\!\!& 
-\frac{\kappa^2}{8 \pi^4} \!\times\! \sum_{i=1}^{21} 
\Biggl\{ T^{i}_{A}(a,a',\partial) 
\!\times\! \Bigl[ \mbox{}^{\mu\nu} D_{i}^{\rho\sigma} \Bigr] 
\!\times\! \ln(\mu^2 \Delta x^2) 
\nonumber \\
& & \hspace{0.1cm} 
+ T^{i}_{B}(a,a',\partial) 
\!\times\! \Bigl [\mbox{}^{\mu\nu} D_{i}^{\rho\sigma} \Bigr] 
\!\times\! \partial^2 \Bigl[\frac{\ln(\mu^2 \Delta x^2)}{\Delta x^2} 
\Bigr] \Biggr\} 
\;\; , \qquad \label{genform3}
\end{eqnarray}
where the tensor differential operators in (\ref{genform3}) 
have been defined before in Table~\ref{Tbasis}. As before, 
the finite non-local contributions coming from the first 
term $\, -i\Sigma_{3i}$ in (\ref{3pt}) have coefficients
$\mathcal{T}^i$. Including the three trace terms 
$\, -i\Sigma_{3ii, \, 3iii, \, 3iv} \,$ gives the full
coefficients $T^i$. The relations between these two sets 
of coefficients were displayed in (\ref{T=T}) and in
Table~\ref{Ti-to-Ti}; here, since we only condider finite 
contributions we can set $D \!=\! 4$ in the latter relations.

Our final results for $\mathcal{T}^i_A$ are given in 
Table~\ref{TauA}, with the trace terms included to give
$T^i_A$ in Table~\ref{TA}. Similarly, Table~\ref{TauB}
gives $\mathcal{T}^i_B$, and Table~\ref{TB} presents
$T^i_B$.
\\ [9pt]
$\bullet$ {\it The 3-point Contribution: Summary}

We conclude this Section by summarizing the various
contributions to the 1-loop renormalized graviton
self-energy. We organize these according to the 
tensor operators they contain.
\\ [5pt]
{\bf -} First we gather the contributions coming from
expressing the linearized Weyl tensor and Ricci scalar
as 2nd order tensor differential operators contracted 
into the graviton field - (\ref{Cdef}) and (\ref{Fdef}).
The resulting contributions are (\ref{source4}) and 
(\ref{source5}):
\begin{eqnarray}
-i\Bigl[ \mbox{}^{\mu\nu} \Sigma^{\rho\sigma}_{\rm ren} \Bigr](x;x')
&\!\!\! = \!\!\!&
\frac{\kappa^2}{2^6 \!\cdot\!3 \!\cdot\! 5 \!\cdot\! \pi^2} \,
\Bigl[ \ln(a) \, \mathcal{C}_{\alpha\beta\gamma\delta}^{~~~~~\rho\sigma} 
\, i \delta^4(x\!-\!x') \Bigr]
\nonumber \\
& & \hspace{0cm} 
+ \frac{\kappa^2}{2^6 \!\cdot\! 3^2 \!\cdot\! \pi^2} \,
\mathcal{F}^{\mu\nu} 
\Bigl[ \ln(a) \, \mathcal{F}^{\rho\sigma} i \delta^4(x \!-\! x') \Bigr]
\;\; . \qquad \label{renloc}
\end{eqnarray}
\\ [-9pt]
{\bf -} Next we gather the contributions to the 1-loop
graviton self-energy involving the 21 tensor differential 
operators of Table~\ref{Tbasis}. There are local 
contributions that can be expressed as (\ref{genform1}):
\begin{equation}
-i \Bigl[\mbox{}^{\mu\nu} \Sigma^{\rho\sigma}_{\rm loc}\Bigr](x;x') = 
\frac{\kappa^2 H^2}{192 \pi^2} \sum_{i=1}^{21} T^i(a,a',\partial) \times
\Bigl[\mbox{}^{\mu\nu} D_i^{\rho\sigma}\Bigr] \, i\delta^4(x \!-\! x') 
\; \; , \label{local}
\end{equation}
and are shown in Table~\ref{localTs}; as well as non-local 
contributions that can be expressed as (\ref{genform3}):
\begin{eqnarray}
-i\Bigl[ \mbox{}^{\mu\nu} \Sigma^{\rho\sigma}_{\rm nonloc} \Bigr](x;x') 
&\!\!\! = \!\!\!& 
-\frac{\kappa^2}{8 \pi^4} \!\times\! \sum_{i=1}^{21} 
\Biggl\{ T^{i}_{A}(a,a',\partial) 
\!\times\! \Bigl[ \mbox{}^{\mu\nu} D_{i}^{\rho\sigma} \Bigr] 
\!\times \ln(\mu^2 \Delta x^2) 
\nonumber \\
& & \hspace{0.1cm} 
+ T^{i}_{B}(a,a',\partial) 
\!\times\! \Bigl [\mbox{}^{\mu\nu} D_{i}^{\rho\sigma} \Bigr] 
\!\times \partial^2 \Bigl[\frac{\ln(\mu^2 \Delta x^2)}{\Delta x^2} 
\Bigr] \Biggr\} 
\;\; , \qquad \label{nonlocal}
\end{eqnarray}
with the explicit results for the coefficients
$T^i_{A} (a, a', \partial)$, $T^i_{B} (a, a', \partial)$ 
shown in Tables~\ref{TA},\ref{TB} respectively.

We conclude with a description of the origin of the various
entries in Table~\ref{localTs}. In the 1st column under the 
name ``Residuals" we have included the contributions of 
(\ref{source2}) from the renormalization residuals of 
Table~\ref{D4limit}, in the 2nd column under the name 
``$\ln(a a') \times \,$Einstein" the logarithmic 
contributions of (\ref{source6}) coming from the Einstein 
term, in the 3rd column under the name $``\Delta \Lambda"$ 
the contributions of (\ref{source3}) from the cosmological 
constant like term and of (\ref{DLcc}) with $c_3$ given 
by (\ref{c3}) from the cosmological counterterm, and in 
the 4th column under the name ``Marginal" the contributions 
of (\ref{source1}) emanating from the action of the two 
derivatives on the scalar propagator.

\begin{table}[H]
\setlength{\tabcolsep}{8pt}
\def\arraystretch{1.5}
\centering
\begin{tabular}{|@{\hskip 1mm }c@{\hskip 1mm }||c||c||c||c||}
\hline
$i$ & Residuals 
    & $\ln(a a') \times \,$Einstein 
	& $\Delta \Lambda$ 
	& Marginal \\
\hline\hline
1 & $\!\!-\frac52 a a' \partial^2 -4 a^2 {a'}^2 H^2\!\!$ 
  & $\!\!-a a' \partial^2 - 2 a^2 {a'}^2 H^2\!\!$ 
  & $\!\!-\frac92 a^2 {a'}^2 H^2\!\!$ 
  & $-3 a^2 {a'}^2 H^2$ \\
\hline
2 & $a a' \partial^2$ 
  & $a a' \partial^2 + 2 a^2 {a'}^2 H^2$ 
  & $\!\!9 a^2 {a'}^2 H^2\!\!$ 
  & $-6 a^2 {a'}^2 H^2$ \\
\hline
3 & $0$ & $2 a^3 a' H^2$ & $0$ & $0$ \\
\hline
4 & $0$ & $2 a {a'}^3 H^2$ & $0$ & $0$ \\
\hline
5 & $-2 a {a'}^2 H$ & $-2 a^2 a' H$ & $0$ & $0$ \\
\hline
6 & $2 a^2 a' H$ & $2 a {a'}^2 H$ & $0$ & $0$ \\
\hline
7 & $2 a a'$ & $a a'$ & $0$ & $0$ \\
\hline
8 & $2 a a'$ & $a a'$ & $0$ & $0$ \\
\hline
9 & $-6 a^2 {a'}^2 H^2$ & $2 a^2 {a'}^2 H^2$ & $0$ & $0$ \\
\hline
10 & $-2 a {a'}^2 H$ & $-2 a {a'}^2 H$ & $0$ & $0$ \\
\hline
11 & $2 a^2 a'^2 H$ & $2 a^2 a' H$ & $0$ & $0$ \\
\hline
12 & $-2 a a'$ & $-2 a a'$ & $0$ & $0$ \\
\hline
\end{tabular}
\caption{\footnotesize 
Local contributions to each $T^i(a,a',\partial)$ from the
various sources.}
\label{localTs}
\end{table}

\section{Solving the Effective Field Equations}

The linearized effective field equation for the graviton
field $h_{\mu\nu}(x)$ (\ref{hmn}) is:
\begin{equation}
\mathcal{D}^{\mu\nu\rho\sigma} \kappa h_{\rho\sigma}(x) 
= 8 \pi G T^{\mu\nu}(x)
+ \int \!\! d^4x' \, [\mbox{}^{\mu\nu} \Sigma^{\rho\sigma}](x;x') 
\, \kappa h_{\rho\sigma}(x') 
\;\; , \label{EFE}
\end{equation}
where $[\mbox{}^{\mu\nu} \Sigma^{\rho\sigma}](x;x')$ is 
the graviton self-energy in the $''in\!-\!in''$ formalism 
\cite{Schwinger:1960qe,Mahanthappa:1962ex,Bakshi:1962dv,
Bakshi:1963bn,Keldysh:1964ud,Chou:1984es,Jordan:1986ug,
Calzetta:1986ey,Ford:2004wc}, $T^{\mu\nu}(x)$ is minus 
the variation of the matter action with respect to 
$h_{\mu\nu}(x)$ and $\mathcal{D}^{\mu\nu\rho\sigma}$ 
is the Lichnerowicz operator on de Sitter background:
\begin{eqnarray}
\lefteqn{\mathcal{D}^{\mu\nu\rho\sigma} h_{\rho\sigma} = 
\frac12 a^2 \Bigl[ \partial^2 h^{\mu\nu} 
- \eta^{\mu\nu} \partial^2 h 
+ \eta^{\mu\nu} \partial^{\rho} \partial^{\sigma} h_{\rho\sigma} 
+ \partial^{\mu} \partial^{\nu} h 
- 2 \partial^{\rho} \partial^{(\mu} h^{\nu)}_{~~\rho} \Bigr] } 
\nonumber \\
& & \hspace{0.5cm} 
+ H a^3 \Bigl[ \eta^{\mu\nu} \partial_0 h 
- \partial_0 h^{\mu\nu}
- 2 \eta^{\mu\nu} \partial^{\rho} h_{\rho 0} 
+ 2 \partial^{(\mu} h^{\nu)}_{~~0} \Bigr]
+ 3 H^2 a^4 \eta^{\mu\nu} h_{00} 
\;\; . \qquad \label{Lichnerowicz} 
\end{eqnarray}
The purpose of this Section is to use the 1-loop scalar 
contribution to the graviton self-energy to solve (\ref{EFE}) 
for 1-loop corrections to plane wave gravitational radiation 
and for the gravitational response to a static point mass.
The Section begins by giving the $''in\!-\!in''$ form of 
the 1-loop scalar contribution to 
$[\mbox{}^{\mu\nu} \Sigma^{\rho\sigma}](x;x')$. 
Then, we explain generally how equation (\ref{EFE}) can be 
solved for 1-loop corrections to $h_{\mu\nu}(x)$. Setting 
$T^{\mu\nu}(x) = 0$ gives dynamical gravitons, and setting 
$T^{\mu\nu}(x) = -\delta^{\mu}_{~0} \delta^{\nu}_{~0} M a 
\, \delta^3(\vec{x})$ gives the Newtonian potential.

\subsection{The Graviton Self-Energy}

After making the simple conversion from the in-out formalism
of Section 3 to the in-in formalism \cite{Ford:2004wc} the 1-loop 
scalar contribution to the graviton self-energy 
$[\mbox{}^{\mu\nu} \Sigma_1^{\rho\sigma}](x;x')$ can be written:
\begin{eqnarray}
\lefteqn{ [\mbox{}^{\mu\nu} \Sigma_1^{\rho\sigma}](x;x') 
= 
- \frac{\kappa^2 \mathcal{C}^{\alpha\beta\gamma\delta\mu\nu}}
{960\pi^2} 
\Bigl[ \ln(a) {\mathcal{C}'}_{\alpha\beta\gamma\delta}^{~~~~~\rho\sigma} 
\delta^4(\Delta x) \Bigr] } 
\nonumber \\
& & \hspace{-0.3cm}
- \frac{\kappa^2 \mathcal{F}^{\mu\nu}}{576 \pi^2} 
\Bigl[ \ln(a) {\mathcal{F}'}^{\rho\sigma} \delta^4(\Delta x) \Bigr] 
- \frac{\kappa^2 H^2}{192 \pi^2} \sum_{i=1}^{21} 
\widehat{T}^i(a,a',\partial) [\mbox{}^{\mu\nu} D^{\rho\sigma}_{i}] \delta^4(\Delta x) 
\nonumber \\
& & \hspace{-0.3cm} 
+ \frac{\kappa^2 H^2}{384 \pi^3} \sum_{i=1}^{21} \Biggl\{ 
\widehat{T}^i_{A}(a,a',\partial) [\mbox{}^{\mu\nu} D^{\rho\sigma}_{i}]
\times \theta(\Delta \eta \!-\! \Delta r)
\nonumber \\
& & \hspace{0.3cm}
+ \widehat{T}^i_{B}(a,a',\partial) \, [\mbox{}^{\mu\nu} D^{\rho\sigma}_{i}] 
\, \partial^4 \Bigl[ \theta(\Delta \eta \!-\! \Delta r) 
\Bigl( \ln[\mu^2 (\Delta \eta^2 \!-\! \Delta r^2)] 
\!-\! 1 \Bigr) \Bigr] \! \Biggr\} 
\;\; . \qquad \label{SigmaSK}
\end{eqnarray}

\noindent 
The notation in expression (\ref{SigmaSK}) requires explanation:
\\
- The tensor differential operators $[\mbox{}^{\mu\nu} D_i^{\rho\sigma}]$ 
are given in Table~\ref{Tbasis}. \\
- The coefficient functions $\widehat{T}^i(a,a',\partial)$ and 
$\widehat{T}^i_{A}(a,a',\partial)$ are listed in Table~\ref{TTA}, 
while the $\widehat{T}^i_{B}(a,a',\partial)$ are listed in 
Table~\ref{TB2}. \\
- The tensor differential operators $\mathcal{F}^{\mu\nu}$ and 
$\mathcal{C}_{\alpha\beta\gamma\delta}^{~~~~~\mu\nu}$ were defined 
in (\ref{Cdef}) and (\ref{Fdef}-\ref{Fexplicit}) respectively by 
expanding the Weyl tensor and Ricci scalar in powers of the graviton 
field $h_{\mu\nu}$.

\begin{table}[H]
\setlength{\tabcolsep}{8pt}
\def\arraystretch{1.5}
\centering
\begin{tabular}{|@{\hskip 1mm }c@{\hskip 1mm }||c||c|}
\hline
$i$ & $\widehat{T}^i(a,a',\partial)$ & $\widehat{T}_A^i(a,a',\partial)$ \\
\hline\hline
1 & $-a a' \ln(a a') [\partial^2 + 2 a a' H^2]$ & 
$-\frac14 a^2 {a'}^2 H^2 \partial^2 \partial_0^2$ \\
& $-3 a a' \partial^2 - a a' \partial_0^2 - 6 a^2 {a'}^2 H^2$ & $$ \\
\hline
2 & $a a' \ln(a a') [\partial^2 + 2 a a' H^2]$ & $0$ \\
& $+ a a' \partial^2 + 3 a^2 {a'}^2 H^2$ & \\
\hline
3 & $2 a^3 a' H^2 [\ln(a a') + \frac92]$ & $0$ \\
\hline
4 & $2 a {a'}^3 H^2 [\ln(a a') + \frac92]$ & $0$ \\
\hline
5 & $-2 a^2 a' H [\ln(a a') + 4]$ & 
$\frac12 a^2 {a'}^2 H^2 \partial_0 \partial^2$ \\
\hline
6 & $2 a {a'}^2 H [\ln(a a') + 4]$ & 
$\frac12 a^2 {a'}^2 H^2 \partial_0 \partial^2$ \\
\hline
7 & $a a' [\ln(a a') + 3]$ & 
$- \frac{1}{2} a^2 a' H \partial_0 \partial^2 
+ \frac{1}{2} a^2 {a'}^2 H^2 \partial^2$ \\
\hline
8 & $a a' [\ln(a a') + 3]$ & 
$\frac{1}{2} a {a'}^2 H \partial_0 \partial^2 
+ \frac{1}{2} a^2 {a'}^2 H^2 \partial^2$ \\
\hline
9 & $2 a^2 {a'}^2 H^2 [\ln(a a') + 1]$ & $0$ \\
\hline
10 & $-2 a {a'}^2 H [\ln(a a') + 2]$ & $0$ \\
\hline 
11 & $2 a^2 a' H [\ln(a a') + 2]$ & $0$ \\
\hline
12 & $-2 a a' [\ln(a a') + 1]$ & $-a^2 {a'}^2 H^2 \partial^2$ \\
\hline
18 & $0$ & $-3a^2 {a'}^2 H^2 \partial^2$ \\
\hline
19 & $0$ & $a^2 a' H \partial^2$ \\
\hline
20 & $0$ & $-a {a'}^2 H \partial^2$ \\
\hline
21 & $0$ & $\frac{1}{2} a a' \partial^2 - a^2 {a'}^2 H^2$ \\
\hline
\end{tabular}
\caption{\footnotesize Coefficients $\widehat{T}^i(a,a',\partial)$ and 
$\widehat{T}^i_{A}(a,a',\partial)$ which appear in expression (\ref{SigmaSK}).}
\label{TTA}
\end{table}

\begin{table}[H]
\setlength{\tabcolsep}{8pt}
\def\arraystretch{1.5}
\centering
\begin{tabular}{|@{\hskip 1mm }c@{\hskip 1mm }||c||c||c|}
\hline
$i$ & $\widehat{T}_B^i(a,a',\partial)$ & $i$ & $\widehat{T}_B^i(a,a',\partial)$ \\
\hline\hline 
1  & $-\frac{3 \partial^4}{80 H^2} + \frac{a a'\partial^2}{8} 
   + \frac{a a' \partial_0^2}{8}$ & 
12 & $\frac{\partial^2}{40 H^2} + \frac{a a'}{2}$ \\
\hline
2 & $-\frac{\partial^4}{80 H^2} - \frac{a a' \partial^2}{4} 
  - \frac{a^2 {a'}^2 H^2}{2}$ & 
13 & $-\frac{3 a^2 {a'}^2 H^2}{2}$ \\
\hline
3 & $\frac{{a'}^2 \partial^2}{4} + \frac{3 a {a'}^2 H \partial_0}{4}
  + \frac{a^2 {a'}^2 H^2}{2}$ & 
14 & $\frac{3 a^2 a' H}{2}$ \\
\hline
4 & $\frac{a^2 \partial^2}{4} - \frac{3 a^2 a' H \partial_0}{4}
  + \frac{a^2 {a'}^2 H^2}{2}$ & 
15 & $-\frac{3 a {a'}^2 H}{2}$ \\
\hline
5 & $-\frac{a' \partial^2}{4 H} - \frac{3 a a' \partial_0}{4} 
  + \frac{a {a'}^2 H}{4}$ & 
16 & $-\frac{a^2}{4}$ \\
\hline
6 & $\frac{a \partial^2}{4 H} - \frac{3 a a' \partial_0}{4} 
  - \frac{a^2 a' H}{4}$ & 
17 & $-\frac{{a'}^2}{4}$ \\
\hline
7 & $\frac{3 \partial^2}{80 H^2} + \frac{a \partial_0}{8 H} 
  - \frac{a^2}{8}$ &
18 & $\frac{3 a a'}{2}$ \\
\hline
8 & $\frac{3 \partial^2}{80 H^2} - \frac{a' \partial_0}{8 H} 
  - \frac{{a'}^2}{8}$ &
19 & $-\frac{a}{4 H}$ \\
\hline
9 & $-\frac{3 a^2 {a'}^2 H^2}{2}$ &
20 & $\frac{a'}{4 H}$ \\
\hline
10 & $\frac{a^2 a' H}{2}$ &
21 & $-\frac1{20 H^2}$ \\
\hline
11 & $-\frac{a {a'}^2 H}{2}$ & & $$ \\ 
\hline
\end{tabular}
\caption{\footnotesize Coefficient functions $\widehat{T}^i_{B}(a,a',\partial)$ 
which appear in expression (\ref{SigmaSK}).}
\label{TB2}
\end{table}

The terms in expression (\ref{SigmaSK}) involving 
summation have the generic form of coefficient functions 
of $(a,a',\partial)$) multiplying tensor differential 
operators $[\mbox{}^{\mu\nu} D_i^{\rho\sigma}]$, all
acting on three different functions of $(x-x')^{\mu}$:
\begin{eqnarray}
&\mbox{}&
\delta^4(x \!-\! x') 
\qquad , \qquad 
\theta(\Delta \eta \!-\! \Delta r)  
\nonumber \\
&\mbox{}&
f_B(x;x') \equiv
\partial^4 \Bigl[ \theta(\Delta \eta \!-\! \Delta r) 
\Bigl( \ln[\mu^2 (\Delta \eta^2 \!-\! \Delta r^2)] \!-\! 1\Bigr) 
\Bigr] \;\; . 
\label{3funcs}
\end{eqnarray}
Important relations convert the three functions into one another:
\begin{eqnarray}
\partial^4 \, \theta(\Delta \eta \!-\! \Delta r) 
&\!\!\! = \!\!\!& 
8\pi \delta^4(x \!-\! x') 
\;\; , \qquad \label{f2tof1} \\
\Delta \eta \, f_B(x;x') 
&\!\!\! = \!\!\!& 
-2 \partial_0 \partial^2 \, \theta(\Delta \eta \!-\! \Delta r) 
\;\; . \qquad \label{f3tof2}
\end{eqnarray}
The middle function also obeys the relation:
\begin{equation}
(\Delta \eta \partial^2 \!+\! 2 \partial_0) 
\; \theta(\Delta \eta \!-\! \Delta r) = 0 
\;\; . \label{f2tozero}
\end{equation}
   
\subsection{Perturbative Solution}

We possess only the single scalar loop contribution 
to the graviton self-energy so it is only possible 
to solve equation (\ref{EFE}) perturbatively:
\begin{equation}
h_{\mu\nu} = h^{(0)}_{\mu\nu} + \kappa^2 h^{(1)}_{\mu\nu} 
+ \kappa^4 h^{(2)}_{\mu\nu} + \ldots 
\label{hexpand}
\end{equation}
We consider the stress tensor to be 0th order 
so $h^{(0)}_{\mu\nu}$ obeys the equation:
\begin{equation}
\mathcal{D}^{\mu\nu\rho\sigma} \kappa h^{(0)}_{\rho\sigma}(x) 
= 
8 \pi G \, T^{\mu\nu}(x) 
\;\; . \label{0thorder}
\end{equation}
The 1-loop correction we seek obeys:
\begin{equation}
\mathcal{D}^{\mu\nu\rho\sigma} \, \kappa^3 h^{(1)}_{\rho\sigma}(x) 
= 
\int \! d^4x' \, [\mbox{}^{\mu\nu} \Sigma_1^{\rho\sigma}](x;x') 
\, \kappa h^{(0)}_{\rho\sigma}(x') 
\;\; . \label{1storder}
\end{equation}

The $D=4$ contributions from the first two terms of 
$[\mbox{}^{\mu\nu} \Sigma^{\rho\sigma}_1](x;x')$ in 
expression (\ref{SigmaSK}) can be written in terms of 
linearized curvatures. Relation (\ref{Cdef}) implies 
that the first term is:
\begin{eqnarray}
\lefteqn{-\frac{\kappa^2 \mathcal{C}^{\alpha\beta\gamma\delta\mu\nu}}
{960\pi^2} 
\Biggl[ \ln(a) \! \int \!\! d^4x' \, 
\mathcal{C}_{\alpha\beta\gamma\delta}^{~~~~~\rho\sigma} 
\, \delta^4(x \!-\! x') \times 
\kappa h^{(0)}_{\rho\sigma}(x') \Biggr] }
\nonumber \\
& & \hspace{1.3cm} 
= -\frac{\kappa^2 \, \mathcal{C}^{\alpha\beta\gamma\delta\mu\nu}}
{960\pi^2} \Bigl[ \ln(a) \, C^{(0)}_{\alpha\beta\gamma\delta} \Bigr] 
= 
\frac{\kappa^2 \partial_{\rho} \partial_{\sigma}}{480 \pi^2} 
\Bigl[ \ln(a) \, C^{(0)\, \rho\mu\sigma\nu} \Bigr] 
\;\; , \label{Cterm}
\end{eqnarray}
and relation (\ref{Fdef}) implies a similar form for the 
second term:
\begin{equation}
-\frac{\kappa^2 \mathcal{F}^{\mu\nu}}{576 \pi^2} 
\Biggl[ \ln(a) \! \int \! d^4x' \, 
\mathcal{F}^{\rho\sigma} \delta^4(x - x') 
\times \kappa h^{(0)}_{\rho\sigma}(x') \Biggr] 
= 
- \frac{\kappa^2 \mathcal{F}^{\mu\nu}}{576 \pi^2} 
\Bigl[ \ln(a) \, a^2 R^{(0)} \Bigr]
\; . \label{Fterm} 
\end{equation}
 
\subsection{Dynamical Gravitons}

Dynamical gravitons are characterized by their 3-momenta 
$\vec{k}$ and polarization $\lambda$. The graviton field 
for a dynamical graviton takes the form:
\begin{equation}
\kappa h_{\mu\nu}(x) = \epsilon_{\mu\nu}(\vec{k},\lambda) 
e^{i \vec{k} \cdot \vec{x}} u(t,k) 
\;\; , \label{graviton}
\end{equation}
where $u(t,k)$ is the graviton mode function and the 
polarization tensors take the same form in cosmology 
that they do in flat space. In particular, their temporal 
components vanish, they are transverse and traceless:
\begin{equation}
\epsilon_{0\mu}(\vec{k},\lambda) = 0 
\qquad , \qquad 
k_i \epsilon_{i\mu}(\vec{k}, \lambda) = 0 
\qquad , \qquad 
\epsilon_{ii}(\vec{k},\lambda) = 0 
\;\; . \label{poltens}
\end{equation}
The action of the Lichnerowicz operator (\ref{Lichnerowicz}) 
on such a field is:
\begin{equation}
\mathcal{D}^{\mu\nu\rho\sigma} h_{\rho\sigma}(x) = 
\epsilon^{\mu\nu}(\vec{k},\lambda) e^{i \vec{k} \cdot \vec{x}} 
\times \Big( \!\! -\frac{a^2}{2} \Big)
\Bigl[ \partial_0^2 + 2 a H \partial_0 + k^2 \Bigr] u(t,k) 
\;\; . \label{gravLHS}
\end{equation}

The general perturbative expansion (\ref{hexpand}) implies 
a similar expansion for the graviton mode function:
\begin{equation}
u(t,k) = u_0(t,k) + \kappa^2 u_1(t,k) 
+ \kappa^4 u_2(t,k) + \ldots 
\end{equation}
The point of this sub-section is to compute the 1st order 
solution $u_1(t,k)$. The canonically normalized 0th order 
solution is well known:
\begin{equation}
u_0(t,k) = 
\frac{H}{\sqrt{2 k^3}} \Bigl[1 - \frac{i k}{H a}\Bigr] 
\exp\Bigl[ \frac{i k}{H a} \Bigr] = 
\frac{H}{\sqrt{2 k^3}} \, (1 + i k \eta) e^{-i k \eta}
\;\; . \label{u0}
\end{equation}
Its first two conformal time derivatives are:
\begin{equation}
\partial_0 u_0 = \frac{H}{\sqrt{2 k^3}} 
\Bigl[-\frac{k^2}{H a}\Bigr] \exp\Bigl[ \frac{i k}{H a} \Bigr] 
\quad , \quad 
\partial^2_0 u_0 = \frac{H}{\sqrt{2 k^3}} 
\Bigl[ k^2 + \frac{i k^3}{H a} \Bigr] 
\exp\Bigl[ \frac{i k}{H a} \Bigr] 
\;\; . \label{u0ID1}
\end{equation}
Relations (\ref{u0ID1}) imply four useful identities:
\begin{eqnarray}
(\partial_0^2 - k^2) u_0 = -2 i k \partial_0 u_0 
& \qquad , \qquad &
(\partial_0^2 - k^2) u_0 = \tfrac12 (\partial_0 - i k)^2 u_0 
\; , \qquad \label{u0ID2} \\ 
(\partial_0^2 + k^2) u_0 = -2 H a \partial_0 u_0 
& \qquad , \qquad &
(\partial_0^2 + k^2)^2 u_0 = 0 
\; . \label{u0ID3}
\end{eqnarray}

The right hand side of the effective field equation (\ref{EFE}) 
for dynamical gravitons (\ref{graviton}-\ref{poltens}) consists 
of the two local contributions (\ref{Cterm}-\ref{Fterm}) and a 
series of local and non-local terms proportional to the 21 tensor 
differential operators of Table~\ref{Tbasis}. 
\\ [5pt]
{\bf -} {\it Contributions from the $\mathcal{F}$-term 
(\ref{Fterm}):} Because dynamical gravitons are source-free 
solutions, the linearized Ricci scalar vanishes: $R^{(0)} 
= 0$. Hence, there are no such contributions.
\\ [5pt]
{\bf -} {\it Contributions from the $\mathcal{C}$-term 
(\ref{Cterm}):} To evaluate (\ref{Cterm}) we first make 
a $3+1$ expansion of the derivatives:
\begin{eqnarray}
\frac{\kappa^2 \partial_{\rho} \partial_{\sigma}}{480 \pi^2} 
\Bigl[ \ln(a) \, C^{(0)\, \rho\mu\sigma\nu} \Bigr] 
&\!\!\! = \!\!\!&
\frac{\kappa^2 \partial^2_{0}}{480 \pi^2} \Bigl[ 
\ln(a) \, C^{(0)\, 0 \mu 0 \nu} \Bigr] 
- \frac{2 \kappa^2 \partial_0 \partial_i }{480 \pi^2} 
\Bigl[ \ln(a) \, C^{(0)\, 0 (\mu \nu) i} \Bigr]
\nonumber \\
&\mbox{}& 
+ \frac{\kappa^2 \partial_i \partial_j}{480 \pi^2} 
\Bigl[ \ln(a) \, C^{(0)\, i \mu j \nu} \Bigr] 
\;\; . \qquad \label{3+1}
\end{eqnarray}
Now note that, for dynamical gravitons, the non-zero 
components of the linearized Weyl tensor are:
\begin{eqnarray}
C^{(0)}_{0i0j} &\!\!\! = \!\!\!& 
e^{i \vec{k} \cdot \vec{x}} \epsilon_{ij} 
\times (-\tfrac14) (\partial_0^2 - k^2) \, u_0 
\;\; , \label{C0i0j} \\
C^{(0)}_{0ijk} &\!\!\! = \!\!\!& 
e^{i \vec{k} \cdot \vec{x}} \Bigl( 
\epsilon_{ij} k_k - \epsilon_{ik} k_j \Bigr) 
\times \tfrac{i}{2} \partial_0 u_0 
\;\; , \label{C0ijk} \\
C^{(0)}_{ijk\ell} &\!\!\! = \!\!\!& 
e^{i \vec{k} \cdot \vec{x}} \Bigl( 
\epsilon_{ik} \delta_{j\ell} - \epsilon_{kj} \delta_{\ell i} 
+ \epsilon_{j \ell} \delta_{i k} - \epsilon_{\ell i} \delta_{k j} \Bigr) 
\times (-\tfrac14) (\partial_0^2 + k^2) \, u_0 
\qquad \qquad \nonumber \\
& & 
+ e^{i \vec{k} \cdot \vec{x}} \Bigl( 
\epsilon_{ik} k_j k_{\ell} - \epsilon_{k j} k_{\ell} k_i 
+ \epsilon_{j \ell} k_i k_k - \epsilon_{\ell i} k_k k_j \Bigr) 
\times \tfrac12 u_0 
\;\; . \label{Cijkl}
\end{eqnarray}
Substituting (\ref{C0i0j}-\ref{Cijkl}) into (\ref{3+1}) 
and exploiting the properties (\ref{poltens}) of the 
polarization tensor and the identities (\ref{u0ID2}) 
obeyed by the mode function implies:
\begin{eqnarray}
\frac{\kappa^2 \partial_{\rho} \partial_{\sigma}}{480 \pi^2} 
\Bigl[ \ln(a) \, C^{(0)\, \rho\mu\sigma\nu} \Bigr] 
&\!\!\! = \!\!\!& 
\frac{\kappa^2 e^{i \vec{k} \cdot \vec{x}} \epsilon^{\mu\nu}}
{480 \pi^2} 
\Bigl\{ -\tfrac18 ( \partial_0 \!+\! i k)^2 
\Bigl[ \ln(a) (\partial_0 \!-\! i k)^2 u_0 \Bigr] \Bigr\}
\qquad \nonumber \\
&\!\!\! = \!\!\! & 
\frac{\kappa^2 e^{i \vec{k} \cdot \vec{x}} \epsilon^{\mu\nu}}
{480 \pi^2} 
\!\times\! (-\tfrac{i}{2}) k H^2 a^2 \partial_0 u_0 
\;\; . \label{Cresult}
\end{eqnarray}
\noindent
{\bf -} {\it Contributions from the ``Summation'' terms:} 
The remaining terms on the right hand side of equation 
(\ref{EFE}) all involve sums of the tensor differential 
operators $[\mbox{}^{\mu\nu} D_i^{\rho\sigma}]$ of 
Table~\ref{Tbasis}. These can be partially integrated to 
act on the factor of $h_{\rho\sigma}(x')$, whereupon most 
of the $[\mbox{}^{\mu\nu} D_i^{\rho\sigma}]$ give zero because 
they access a temporal component, or a divergence, or a 
trace. Only the case of $[\mbox{}^{\mu\nu} D_2^{\rho\sigma}] = 
\eta^{\mu (\rho} \eta^{\sigma)\nu}$ contributes, and a further 
simplification is that the coefficient function
$\widehat{T}^2_{A}(a,a',\partial)$ vanishes. Each surviving term has 
a common factor of $e^{i \vec{k} \cdot \vec{x}} \epsilon^{\mu\nu}$ 
which can be canceled out from equations (\ref{gravLHS}) and 
(\ref{Cresult}) to give a scalar equation for $u_1(t,k)$:
\begin{eqnarray}
\lefteqn{-\frac{a^2}{2} \Bigl[ \partial_0^2 
\!+\! 2 a H \partial_0 \!+\! k^2 \Bigr] \kappa^2 u_1 
\; = \;
\frac{\kappa^2 H^2 a^2}{480 \pi^2} \!\times\! (-i)k \partial_0 u_0 } 
\nonumber \\
& & \hspace{1.5cm} 
- \frac{\kappa^2 H^2}{192 \pi^2} \! 
\int \! d^4x' \, T^2(a,a',\partial) \, \delta^4(x \!-\! x') 
\times e^{-i \vec{k} \cdot \Delta \vec{x}} u_0(t',k) 
\nonumber \\
& & \hspace{1.5cm} 
+ \frac{\kappa^2 H^2}{384 \pi^3} \! 
\int \! d^4x' \, T^2_{B}(a,a',\partial) \, f_B(x;x') 
\times e^{-i \vec{k} \cdot \Delta \vec{x}} u_0(t',k) 
\;\; . \qquad \label{modeeqn}
\end{eqnarray}

{\it (i)} For the term in (\ref{modeeqn}) involving the 
coefficient $\widehat{T}^2(a,a',\partial)$, it is possible to obtain 
an exact result:
\begin{equation}
\widehat{T}^2(a,a',\partial) = 
a a' \ln(a a') \Bigl[ \partial^2 + 2 a a' H^2 \Bigr] 
+ a a' \partial^2 + 3 a^2 {a'}^2 H^2 
\;\; . \label{T2}
\end{equation}
Performing the delta function integral, and using relations 
(\ref{u0ID2}-\ref{u0ID3}) gives:
\begin{eqnarray}
\lefteqn{ - \frac{\kappa^2 H^2}{192 \pi^2} \! 
\int \! d^4x' \, \widehat{T}^2(a,a',\partial) \, \delta^4(x \!-\! x') 
\times e^{-i \vec{k} \cdot \Delta \vec{x}} u_0(t',k) } 
\nonumber \\
& & \hspace{0cm} 
= -\frac{\kappa^2 H^2}{192 \pi^2} \Bigl\{
-a \ln(a) (\partial_0^2 \!+\! k^2) [a u_0] + 4 a^4 \ln(a) H^2 u_0 
\nonumber \\
& & \hspace{2.2cm} 
- a (\partial_0^2 \!+\! k^2) [a \ln(a) u_0] 
- a (\partial_0^2 \!+\! k^2) [a u_0] + 3 a^4 H^2 u_0 \Bigr\}
\qquad \\
& & \hspace{0cm} 
= \frac{\kappa^2 H^2}{96 \pi^2} \Bigl\{ a^4 H^2 u_0 + a^3
H \partial_0 u_0 \Bigr\} 
\;\; . \label{deltaterm}
\end{eqnarray}
Expression (\ref{u0}) implies that $u_0(t,k)$
approaches a constant at late times: 
\begin{equation}
\lim_{t \rightarrow \infty} u_0(t,k) 
= \frac{H}{\sqrt{2 k^3}} 
\equiv u_{\infty} 
\;\; . \label{u0late}
\end{equation}
Expression (\ref{u0ID1}) implies that $\partial_0 u_0$ falls 
off like $1/a$, so (\ref{deltaterm}) goes like $a^4$.
\\ [5pt]
\mbox{} \hspace{0.3cm}
{\it (ii)} For the non-local part of (\ref{modeeqn}) which 
involves $\widehat{T}^2_{B}(a,a',\partial)$:
\begin{equation}
\widehat{T}^2_{B}(a,a',\partial) = -\frac{\partial^4}{80 H^2} 
- \tfrac14 a a' \partial^2 - \tfrac12 a^2 {a'}^2 H^2 
\;\; , \label{T2B}
\end{equation}
upon substituting (\ref{T2B}) in the final term of 
(\ref{modeeqn}), reflecting the derivatives and partially 
integrating to act on $u_0(t',k)$ we get:
\begin{eqnarray}
\lefteqn{ \frac{\kappa^2 H^2}{384 \pi^3} 
\! \int \! d^4x' \, f_B(x;x') 
\times e^{-i \vec{k} \cdot \Delta \vec{x}} } 
\nonumber \\
& & \hspace{1.5cm} 
\times \Bigl\{ -\frac{({\partial'}_0^2 \!+\! k^2)^2}{80 H^2}
- \tfrac14 a ({\partial'}_0^2 \!+\! k^2) a' 
- \tfrac12 a^2 {a'}^2 H^2 \Bigr\} \,
u_0(t',k) \;\; . 
\qquad \label{TBstep1} 
\end{eqnarray}
The mode function identities (\ref{u0ID3}) serve to eliminate 
the derivatives on the 2nd line of (\ref{TBstep1}):
\begin{equation}
\Bigl\{ -\frac{({\partial'}_0^2 \!+\! k^2)^2}{80 H^2}
- \tfrac14 a ({\partial'}_0^2 \!+\! k^2) a' 
- \tfrac12 a^2 {a'}^2 H^2 \Bigr\} \, u_0(t', k)
=
-\tfrac12 a^2 {a'}^3 H^3 \Delta \eta \; u_0(t',k) 
\; . \label{TBstep2}
\end{equation}
At this stage we can exploit relation (\ref{f3tof2}) to 
simplify the $\widehat{T}^i_B$ contribution to:
\begin{eqnarray}
\lefteqn{ \frac{\kappa^2 H^2}{384 \pi^3} 
\! \int \! d^4x' \, \partial_0 \partial^2 \Bigl[ 
\theta(\Delta \eta \!-\! \Delta r) \Bigr] 
\times e^{-i \vec{k} \cdot \Delta \vec{x}} 
\times a^2 {a'}^3 H^3 u_0(t',k) }
\nonumber \\
& & \hspace{0.7cm} 
= -\frac{\kappa^2 H^2}{96 \pi^2} \, 
a^2 H^3 \partial_0 (\partial_0^2 \!+\! k^2) 
\! \int_{\eta_i}^{\eta} \!\! d\eta' \, {a'}^3 u_0(t',k) 
\! \int_{0}^{\Delta \eta} \!\! dr \, r^2 \, \frac{\sin(kr)}{kr}
\qquad \\
& & \hspace{0.7cm} 
= -\frac{\kappa^2 H^2}{96 \pi^2} \, a^2 H^3 
\! \int_{\eta_i}^{\eta} \!\! d\eta' \, {a'}^3 u_0(t',k) 
\cos(k \Delta \eta) 
\;\; . \label{TBstep3}
\end{eqnarray}
The integrand in (\ref{TBstep3}) is a total derivative:
\begin{equation}
{a'}^3 u_0(t',k) \cos(k \Delta \eta) = 
\frac{u_{\infty}}{4 H} \, 
\frac{\partial}{\partial \eta'} \Bigl\{ 
{a'}^2 e^{i k \eta - 2 i k \eta'} 
+ \Bigl[ {a'}^2 - \frac{2 i k a'}{H} \Bigr] e^{-i k \eta} \Bigr\} 
\;\; . \label{TBint}
\end{equation}
If we keep only the upper limit contribution the result is:
\begin{equation}
-\frac{\kappa^2 H^2}{96 \pi^2} \, a^2 H^3 
\! \int_{\eta_i}^{\eta} \!\! d\eta' \, {a'}^3 u_0(t',k) 
\cos(k \Delta \eta) 
\; \longrightarrow \; 
-\frac{\kappa^2 H^2}{96 \pi^2} 
\times a^4 H^2 u_0(t,k) 
\;\; . \label{TBstep4}
\end{equation}
\noindent
{\bf -} {\it The Total Contribution:} 
Combining expressions (\ref{Cresult}), (\ref{deltaterm}) 
and (\ref{TBstep4}) gives:
\begin{eqnarray}
\lefteqn{-\frac{a^2}{2} \Bigl[ 
\partial_0^2 \!+\! 2 a H \partial_0 \!+\! k^2 \Bigr] \kappa^2 u_1 } 
\nonumber \\
& & \hspace{0.3cm} 
= \frac{\kappa^2 H^2}{96 \pi^2} \Bigl\{ 
-\tfrac{i k}{10 a H} \!\times\! a^3 H \partial_0 u_0 
+ a^4 H^2 u_0 + a^3 H \partial_0 u_0 - a^4 H^2 u_0 \Bigr\} 
\qquad \\
& & \hspace{0.3cm} 
= -\frac{a^2}{2} \!\times\! \frac{\kappa^2 H^2 k^2}{48 \pi^2} 
\Bigl\{ 1 + \Bigl[ 1 \!-\! \tfrac1{10} \Bigr] \frac{ik}{a H} 
+ \ldots \Bigr\}
\;\; . \label{newmodeeqn}
\end{eqnarray}
Equation (\ref{newmodeeqn}) is easy to solve at late times:
\begin{equation}
\kappa^2 u_1(t,k) 
\; \longrightarrow \;
\frac{\kappa^2 H^2}{48 \pi^2} \!\times\! u_{\infty} \Bigl\{
1 -\tfrac{3 i}{10} ( \tfrac{k}{a H})^3 \ln(a) + \ldots \Bigr\} 
\;\; . \label{u1late}
\end{equation}
The fact that there is no growing contribution agrees with 
previous analyses \cite{Park:2011kg,Leonard:2014zua}. It is 
useful to combine the 1-loop correction (\ref{u1late}) with 
the tree order result and express both in terms of the 
``electric'' components of the Weyl tensor:
\begin{equation}
C_{0i0j}(t,\vec{x}) 
\; \longrightarrow \; 
C^{(0)}_{0i0j}(t,\vec{x}) \Bigl\{
1 - \tfrac{3 \kappa^2 H^2}{160 \pi^2} \, \ln(a) 
+ O(\kappa^4)\Bigr\} 
\;\; \label{radiation}
\end{equation}
The functional form, although not the sign, resembles the 
logarithmic enhancement induced by inflationary gravitons 
in the electric field strength of electromagnetic radiation 
\cite{Wang:2014tza}.

\subsection{Response to A Point Mass}\label{response}

The symmetries of cosmology are homogeneity and isotropy. 
Four components of the metric are scalar under these 
symmetries, of which any two can be gauged away. We 
choose the two non-zero scalar potentials to be:
\begin{equation}
\kappa h_{00} = 2 \Psi(t,r) 
\qquad , \qquad 
\kappa h_{ij} = -2 \Phi(t,r) \delta_{ij} 
\;\; , \label{scalarpots}
\end{equation}
so that the invariant element in conformal coordinates is: 
\begin{equation}
ds^2 = -a^2 (1 - 2 \Psi) d\eta^2 
+ a^2 (1 - 2 \Phi) d\vec{x} \!\cdot\! d\vec{x}
\;\; . \label{PsiPhi}
\end{equation}

In the spacetime geometry (\ref{scalarpots}-\ref{PsiPhi}) 
the left hand side of the effective field equation 
(\ref{EFE}):
\begin{equation}
E^{\mu\nu} \equiv
\mathcal{D}^{\mu\nu\rho\sigma} \kappa h_{\rho\sigma}
\;\; , \label{EFElhs} 
\end{equation}
can be $3+1$ decomposed to give:
\begin{eqnarray}
E^{00} &\!\!\! = \!\!\!& 
a^2 \Bigl[ -6 a^2 H^2 \Psi 
+ (-2 \nabla^2 \!+\! 6 a H \partial_0) \Phi \Bigr] 
\;\; , \qquad \label{E00} \\
E^{0i} &\!\!\! = \!\!\!& 
a^2 \partial^i \Bigl[ -2 a H \Psi + 2 \partial_0 \Phi \Bigr] 
\;\; , \qquad \label{E0i} \\
E^{ij} &\!\!\! = \!\!\!& 
a^2 \partial^i \partial^j \Bigl[ -\Psi - \Phi \Bigr]
+ a^2 \delta^{ij} \Bigl\{ 
(\nabla^2 \!+\! 2 a H \partial_0 \!+\! 6 a^2 H^2) \Psi 
\qquad \nonumber \\
& & \hspace{0.3cm} 
+ (\nabla^2 \!-\! 4 a H \partial_0 \!-\! 2 \partial_0^2) \Phi \Bigr\} 
\;\; . \label{Eij}
\end{eqnarray}
Relations (\ref{E00}-\ref{Eij}) suggest that we identify 
four scalar components:
\begin{equation}
E^{00} \equiv \mathcal{E}_1 
\qquad , \qquad E^{0i} 
\equiv \partial^i \mathcal{E}_2
\qquad , \qquad 
E^{ij} \equiv \partial^i \partial^j \mathcal{E}_3 
+ \delta^{ij} \mathcal{E}_4 
\;\; . \label{scalarE}
\end{equation}
Conservation implies two relations between them:
\begin{eqnarray}
&\mbox{}& \hspace{-2.2cm}
\partial_{\nu} E^{\mu\nu} 
+ a H \delta^{\mu}_{~0} E^{\rho}_{~\rho} = 0 
\label{LHScons} \\
&\mbox{}& 
\Longrightarrow \quad \left\{ 
\begin{matrix}
\partial_0 \mathcal{E}_1 + \nabla^2 \mathcal{E}_2 
+ a H (-\mathcal{E}_1 \!+\! \nabla^2 \mathcal{E}_3 
\!+\! 3 \mathcal{E}_4) = 0 \\
\partial_0 \mathcal{E}_2 + \nabla^2 \mathcal{E}_3 + \mathcal{E}_4 = 0 
\end{matrix}
\right\} 
\;\; . \label{Econs}
\end{eqnarray}
Hence, up to integration constants, any two of the four 
components (\ref{scalarE}) determine the other two.

In the geometry (\ref{scalarpots}-\ref{PsiPhi}) 
the right hand side of equation (\ref{EFE}):
\begin{equation}
S^{\mu\nu}(x) \equiv 
8\pi G T^{\mu\nu} + \int \! d^4x' \, 
[\mbox{}^{\mu\nu} \Sigma^{\rho\sigma}](x;x') 
\, \kappa h_{\rho\sigma}(x') 
\;\; , \label{Source}
\end{equation}
must obviously have the same tensor structure (\ref{scalarE}) 
as the left hand side:
\begin{equation}
S^{00} \equiv \mathcal{S}_1 
\qquad , \qquad 
S^{0i} \equiv \partial^i \mathcal{S}_2
\qquad , \qquad 
S^{ij} \equiv \partial^i \partial^j \mathcal{S}_3 
+ \delta^{ij} \mathcal{S}_4 
\;\; . \label{scalarS}
\end{equation}
Therefore we can solve any two of the four scalar equations 
$\mathcal{E}_i = \mathcal{S}_i$. The simplest choice is 
obviously the combination of $i=2$ and $i=3$, which implies 
first order equations for $\Psi$ and $\Phi$:
\begin{equation}
2 a \partial_0 (a \Psi) = 
-\mathcal{S}_2 - 2 (\partial_0 \!-\! 2 a H) \mathcal{S}_3 
\qquad , \qquad 
2 a \partial_0 (a \Phi) = 
\mathcal{S}_2 - 2 a H \mathcal{S}_3 
\;\; . \label{twoeqns}
\end{equation}
However, one must bear in mind that $\mathcal{S}_2$ and 
$\mathcal{S}_3$ alone leave $\Psi$ and $\Phi$ unfixed up 
to a free function $f(r)$:
\begin{equation}
\Delta \Psi(t,r) = \frac{f(r)}{a} = -\Delta \Phi(t,r) 
\;\; . \label{ambiguity}
\end{equation}
Because the $i=4$ equation $\mathcal{E}_4 = \mathcal{S}_4$ 
vanishes for (\ref{ambiguity}), this ambiguity must be 
fixed by appealing to the $i=1$ equation, $\mathcal{E}_1 = 
\mathcal{S}_1$: 
\begin{equation}
\Delta \mathcal{E}_1 = 2 a \nabla^2 f(r) 
\;\; . \label{fixambig}
\end{equation}

For a static point mass $M$ in an expanding universe we find:
\begin{eqnarray}
T^{\mu\nu}(x) & \equiv & 
-2\frac{\delta}{\delta \kappa h_{\mu\nu}(x)} \Biggl\{
-M \int \! d\tau \sqrt{-g_{\alpha\beta}(\chi(\tau)) 
\dot{\chi}^{\alpha}(\tau) \dot{\chi}^{\beta}(\tau)} 
\, \Biggr\}_{\genfrac{}{}{0pt}{2}
{h_{\alpha\beta} \,=\, 0}{\chi^{\mu} = (\tau,\vec{x})} } 
\qquad \\
& = & 
-\delta^{\mu}_{~0} \delta^{\nu}_{~0} M a \delta^3(\vec{x}) 
\;\; . \label{stress}
\end{eqnarray}
This is one of those cases for which the $i=1$ equation 
$\mathcal{E}_1 = \mathcal{S}_1 $ must be employed to 
determine the full 0th order solutions:
\begin{equation}
\Psi_0(t,r) = \frac{G M}{a r} = -\Phi_0(t,r) 
\;\; . \label{0thpots}
\end{equation}
There are three derivatives of $\Psi_0(t,r)$ which shall
be important in the analysis that follows:
\begin{equation}
\partial_0 \Psi_0 = -a H \Psi_0 
\quad , \quad 
\partial_0^2 \Psi_0 = 0 
\quad , \quad 
\nabla^2 \Psi_0 = -\frac{4\pi G M \delta^3(\vec{x})}{a} 
\;\; . \label{Psi0ds}
\end{equation}
\noindent 
{\bf -} {\it Contributions from the $\mathcal{F}$-term 
(\ref{Fterm}) and $\mathcal{C}$-term (\ref{Cterm}):} 
The linearized Ricci scalar and Weyl tensor are:
\begin{eqnarray}
R^{(0)} &\!\!\! = \!\!\!& 
-\frac{2 \nabla^2 \Psi_0}{a^2} 
\;\; , \label{Ricci} \\
C^{(0)}_{0i0j} &\!\!\! = \!\!\!& 
(-\partial_i \partial_j + \tfrac13 \delta_{ij} \nabla^2) \Psi_0 
\;\; , \label{Weyl0i0j} \\
C^{(0)}_{0ijk} &\!\!\! = \!\!\!& 0 
\;\; , \label{Weyl0ijk} \\
C^{(0)}_{ijk\ell} &\!\!\! = \!\!\!& 
\Bigl[ \tfrac23 (\delta_{ik} \delta_{j\ell} 
\!-\! \delta_{i\ell} \delta_{jk}) \nabla^2 
\!-\! \delta_{ik} \partial_j \partial_{\ell} 
\!+\! \delta_{kj} \partial_{\ell} \partial_i 
\!-\! \delta_{j\ell} \partial_i \partial_k 
\!+\! \delta_{\ell i} \partial_k \partial_j \Bigr] \Psi_0
\;\; . \qquad \label{Weylijkl}
\end{eqnarray}
Although the Ricci scalar is proportional to a 
$\delta$-function, the $\delta$-function contributions 
to the Weyl tensor (\ref{Weyl0i0j}-\ref{Weylijkl}) 
all cancel. Substituting into expressions 
(\ref{Cterm}-\ref{Fterm}) and segregating appropriate 
components yields the local contributions to 
$\mathcal{S}_{1-3}$:
\begin{eqnarray}
\mathcal{S}_{1\mathcal{F}\mathcal{C}} 
&\!\!\! = \!\!\!& 
-\frac{\kappa^2 (\nabla^2 \!-\! 3 a H \partial_0 \!+\! 9 a^2 H^2)}
{576 \pi^2} 
\Bigl[ -2 \ln(a) \, \nabla^2 \Psi_0 \Bigr] 
\nonumber \\
& & \hspace{0.3cm} 
+ \frac{\kappa^2 \nabla^2}{480 \pi^2} \Bigl[
-\tfrac23 \ln(a) \, \nabla^2 \Psi_0 \Bigr] 
\;\; , \label{localS1} \\
\mathcal{S}_{2\mathcal{F}\mathcal{C}} 
&\!\!\! = \!\!\!& -\frac{\kappa^2 
(-\partial_0 \!+\! 3 a H)}{576 \pi^2} \Bigl[
-2 \ln(a) \, \nabla^2 \Psi_0 \Bigr] 
+ \frac{\kappa^2 \partial_0}{480 \pi^2} \Bigl[
\tfrac23 \ln(a) \, \nabla^2 \Psi_0 \Bigr] 
\;\; , \qquad \label{localS2} \\
\mathcal{S}_{3\mathcal{F}\mathcal{C}} 
&\!\!\! = \!\!\!& 
-\frac{\kappa^2}{576 \pi^2} \Bigl[
-2 \ln(a) \, \nabla^2 \Psi_0 \Bigr] 
+ \frac{\kappa^2}{480 \pi^2} \Bigl[ 
(\tfrac13 \nabla^2 \!-\! \partial_0^2) \ln(a) \, \Psi_0 \Bigr] 
\;\; . \label{localS3}
\end{eqnarray}
Of these the only contribution that does not vanish 
away from the origin comes from $\mathcal{S}_3$:
\footnote{This is because $\nabla^2 \Psi_0 = 
-\frac{4\pi G M}{a} \delta^3(\vec{x})$.}
\begin{equation}
\mathcal{S}_{3\mathcal{F}\mathcal{C}} 
\Bigl\vert_{\vec{x} \neq 0} = 
\frac{\kappa^2 H^2 a^2}{480 \pi^2} \times \Psi_0(t,r) 
\;\; . \label{largerS3}
\end{equation}

\begin{table}[H]
\setlength{\tabcolsep}{8pt}
\def\arraystretch{1.5}
\centering
\begin{tabular}{|@{\hskip 1mm }c@{\hskip 1mm }||c||c|c||c|c|}
\hline
$i$ & $[\mbox{}^{\mu\nu} D^{\rho\sigma}_i] \!\times\! h_{\rho\sigma}(x')$ & 
$i$ & $[\mbox{}^{\mu\nu} D^{\rho\sigma}_i] \!\times\! h_{\rho\sigma}(x')$ & 
$i$ & $[\mbox{}^{\mu\nu} D^{\rho\sigma}_i] \!\times\! h_{\rho\sigma}(x')$ \\
\hline\hline
1 & $\eta^{\mu\nu} h(x')$ & 
8 & $\partial^{\mu} \partial^{\nu} \!\times\! h(x')$ & 
15 & $\delta^{(\mu}_{~~0} \partial^{\nu)} \!\times\! h_{00}(x')$ \\
\hline
2 & $h^{\mu\nu}(x')$ & 
9 & $\delta^{(\mu}_{~~0} h^{\nu)}_{~~0}(x')$ & 
16 & $\!\!\delta^{\mu}_{~0} \delta^{\nu}_{~0} 
\partial^{\rho} \partial^{\sigma} \!\times\! h_{\rho\sigma}(x')\!\!$ \\
\hline
3 & $\eta^{\mu\nu} h_{00}(x')$ & 
10 & $\delta^{(\mu}_{~~0} \partial_{\rho} \!\times\! h^{\nu)\rho}(x')$ & 
17 & $\partial^{\mu} \partial^{\nu} \!\times\! h_{00}(x')$ \\
\hline
4 & $\delta^{\mu}_{~0} \delta^{\nu}_{~0} h(x')$ & 
11 & $\partial^{(\mu} \!\times\! h^{\nu)}_{~~0}(x')$ & 
18 & $\delta^{(\mu}_{~~0} \partial^{\nu)} \partial^{\rho} 
\times\! h_{\rho 0}(x')$ \\
\hline
5 & $\eta^{\mu\nu} \partial^{\rho} \!\times\! h_{\rho 0}(x')$ & 
12 & $\partial^{(\mu} \partial_{\rho} \!\times\! h^{\nu)\rho}(x')$ & 
19 & $\!\! \delta^{(\mu}_{~~0} \partial^{\nu)} 
\partial^{\rho} \partial^{\sigma} \!\times\! h_{\rho\sigma}(x') \!\!$ \\
\hline
6 & $\delta^{(\mu}_{~~0} \partial^{\nu)} \!\times\! h(x')$ & 
13 & $\delta^{\mu}_{~0} \delta^{\nu}_{~0} h_{00}(x')$ & 
20 & $\partial^{\mu} \partial^{\nu} \partial^{\rho} 
\!\times\! h_{\rho 0}(x')$ \\
\hline
7 & $\eta^{\mu\nu} \partial^{\rho} \partial^{\sigma} 
\!\times\! h_{\rho\sigma}(x')$ & 
14 & $\delta^{\mu}_{~0} \delta^{\nu}_{~0} \partial^{\rho} 
\!\times\! h_{\rho 0}(x')$ & 
21 & $\partial^{\mu} \partial^{\nu} \partial^{\rho} \partial^{\sigma} 
\!\times\! h_{\rho\sigma}(x')$ \\
\hline
\end{tabular}
\caption{\footnotesize Contraction of $h_{\rho\sigma}(x')$ into the 21 tensor 
differential operators given in Table~\ref{Tbasis}, which act on the functions
(\ref{3funcs}).}
\label{Dhmn}
\end{table}

\newpage

\noindent
{\bf -} {\it Contributions from the ``Summation'' terms:}
The other three contributions to the graviton self-energy (\ref{SigmaSK})
involve sums over the 21 tensor differential operators $[\mbox{}^{\mu\nu} 
D_i^{\rho\sigma}]$ listed in Table~\ref{Tbasis}, acting on the three
functions of $(x - x')^{\mu}$ given in expression (\ref{3funcs}), and 
contracted into $\kappa h^{(0)}_{\rho\sigma}(x')$. The results of 
these contractions are presented in Table~\ref{Dhmn}. It remains to 
substitute $\; \kappa h^{(0)}_{00}(x') = 2 \Psi_0(t',r') \;$ and 
$\; \kappa h^{(0)}_{ij}(x') = 2 \Psi_0(t',r') \delta_{ij}$, and then 
identify contributions to each of the three sources $\mathcal{S}_{1-3}$. 
Note the relevant contractions:
\begin{eqnarray}
\kappa h^{(0)}(x') = 4 \Psi_0(t',r') 
& , & 
\partial^{\rho} \!\times\! \kappa h^{(0)}_{\rho 0}(x') 
= -2 \partial_0 \!\times\! \Psi_0(t',r') 
\;\; , \nonumber \\
\partial^{\rho} \partial^{\sigma} 
\!\times\! \kappa h^{(0)}_{\rho\sigma} 
& = & 
2 (\partial_0^2 + \nabla^2) \!\times\! \Psi_0(t',r') 
\;\; . 
\label{contractions}
\end{eqnarray}
Recall that each of the contractions $[\mbox{}^{\mu\nu} D_i^{\rho\sigma}]
\times \kappa h^{(0)}_{\rho\sigma}(x')$ consists of a tensor differential
operator acting on $x^{\mu}$ and  multiplied by the same function 
$\Psi_0(t',r')$. Therefore, we need only keep track of the factor 
$\mathcal{F}^{i}_{1-3}$ which acts on the functions (\ref{3funcs}) 
for each of the three sources. Table~\ref{Twosources} gives these 
factors for the sources $\mathcal{S}_3$ and $\mathcal{S}_2$, while 
Table~\ref{00source} for the source $\mathcal{S}_1$. Because partially 
integrating temporal derivatives would produce unwanted surface terms, 
whereas there are none for spatial derivatives, we have eliminated 
second time derivatives:
\begin{equation}
\partial_0^2 = -\partial^2 + \nabla^2 
\;\; . \label{timetospace}
\end{equation}

\begin{table}[H]
\setlength{\tabcolsep}{8pt}
\def\arraystretch{1.5}
\centering
\begin{tabular}{|@{\hskip 1mm }c@{\hskip 1mm }||c||c|c||c|c|}
\hline
$i$ & $\mathcal{F}^{i}_{3}$ & $i$ & $\mathcal{F}^{i}_{2}$ & $i$ 
& $\mathcal{F}^{i}_{2}$ \\
\hline\hline
8 & $4$ & 6 & $2$ & 17 & $-2 \partial_0$ \\
\hline
12 & $2$ & 8 & $-4 \partial_0$ & 18 & $-\partial_0$ \\
\hline
17 & $2$ & 10 & $1$ & 19 & $-\partial^2 + 2 \nabla^2$ \\
\hline
20 & $-2 \partial_0$ & 11 & $-1$ & 20 & $-2 \partial^2 + 2 \nabla^2$ \\
\hline
21 & $-2 \partial^2 + 4 \nabla^2$ & 15 & $1$ & 21 & 
$2 \partial_0 \partial^2 - 4 \partial_0 \nabla^2$ \\
\hline
\end{tabular}
\caption{\footnotesize The first two columns give the non-zero 
factors contributing to the source $\mathcal{S}_3$. 
The last four columns present the nonzero factors contributing 
to the source $\mathcal{S}_2$.}
\label{Twosources}
\end{table}


\begin{table}[H]
\setlength{\tabcolsep}{8pt}
\def\arraystretch{1.5}
\centering
\begin{tabular}{|@{\hskip 1mm }c@{\hskip 1mm }||c||c|c||c|c|}
\hline
$i$ & $\mathcal{F}_{1}^{i}$ & $i$ & $\mathcal{F}_{1}^{i}$ & $i$ & 
$\mathcal{F}_{1}^{i}$ \\
\hline\hline
1 & $- 4$ & 8 & $-4 \partial^2 + 4 \nabla^2$ & 15 & $-2 \partial_0$ \\
\hline
2 & $2$ & 9 & $-2$ & 16 & $-2 \partial^2 + 4 \nabla^2$ \\
\hline
3 & $-2$ & 10 & $2 \partial_0$ & 17 & $-2 \partial^2 + 2 \nabla^2$ \\
\hline
4 & $4$ & 11 & $2\partial_0$ & 18 & $-2 \partial^2 + 2 \nabla^2$ \\
\hline
5 & $2 \partial_0$ & 12 & $2 \partial^2 - 2 \nabla^2$
& 19 & $2 \partial_0 \partial^2 - 4 \partial_0 \nabla^2$ \\
\hline
6 & $-4 \partial_0$ & 13 & $2$ & 20 & $2 \partial_0 \partial^2 - 
2 \partial_0 \nabla^2$ \\
\hline
7 & $2 \partial^2 -4 \nabla^2$ & 14 & $-2 \partial_0$ & 21 
& $2 \partial^4 - 2 \partial^2 \nabla^2 + 4 \partial_0^2 \nabla^2$ \\
\hline
\end{tabular}
\caption{\footnotesize Source $\mathcal{S}_1$ factors arising from the
contractions $[\mbox{}^{00} D_i^{\rho\sigma}] \!\times\! \kappa 
h^{(0)}_{\rho\sigma}(x')$.}
\label{00source}
\end{table}
\noindent
{\bf -} {\it The Source $\mathcal{S}_3$:}
The simplest source is $\mathcal{S}_3$, which receives 
contributions only from $i=8, 12, 17, 20, 21$. Combining 
information from Table~\ref{TTA} for $\widehat{T}^i(a,a'\partial)$ 
and for $[\mbox{}^{\mu\nu} D_i^{\rho\sigma}] \times 
\kappa h^{(0)}_{\rho\sigma}$ from Table~\ref{Twosources} 
gives:
\begin{eqnarray}
\mathcal{S}_{3T} 
&\!\!\! \equiv \!\!\!& 
-\frac{\kappa^2 H^2}{192 \pi^2} \sum_{i=1}^{21}
\! \int \! d^4x' \, \widehat{T}^i(a,a',\partial) \times \mathcal{F}^i_3 
\times \delta^4(x \!-\! x') \times \Psi_0(t',r') 
\qquad \\
&\!\!\! = \!\!\!& 
-\frac{\kappa^2 H^2 a^2}{192 \pi^2} \! \int \! d^4x' 
\, 8 a a' \, \delta^4(x \!-\! x') \times \Psi_0(t',r') 
\;\; . \label{S3T}
\end{eqnarray}

The same two tables give the initial contribution from 
$\widehat{T}^i_{A}(a,a',\partial)$:
\begin{eqnarray}
\mathcal{S}_{3T_{A}} &\!\!\! \equiv \!\!\!& 
\frac{\kappa^2 H^2}{384 \pi^3} \sum_{i=1}^{21}
\! \int \! d^4x' \, \widehat{T}^i_{A}(a,a',\partial) \times \mathcal{F}^i_3 
\times \theta(\Delta \eta \!-\! \Delta r) 
\times \Psi_0(t',r') 
\\
&\!\!\! = \!\!\!&
\frac{\kappa^2 H^2}{384 \pi^3} \! \int \! d^4x' \Bigl\{
-a a' \partial^4 \!+\! 4 a {a'}^2 H \partial_0 \partial^2 
\!+\! 2 a^2 {a'}^2 H^2 \partial^2 
\nonumber \\
&\mbox{}& \hspace{1.1cm} 
+\Bigl[ 2 a {a'}^2 \partial^2 \!-\! 4 a^2 {a'}^2 H^2 \Bigr]
\nabla^2 \Bigr\} \, \theta(\Delta \eta \!-\! \Delta r) 
\times \Psi_0(t',r')
\;\; , \qquad \label{S3TA}
\end{eqnarray}
as well as the initial contribution from $\widehat{T}^i_{B}(a,a',\partial)$:
\begin{eqnarray}
\mathcal{S}_{3T_{B}} &\!\!\! \equiv \!\!\!&
\frac{\kappa^2 H^2}{384 \pi^3} \sum_{i=1}^{21}
\! \int \! d^4x' \, \widehat{T}^i_{B}(a,a',\partial) \times \mathcal{F}^i_3 
\!\times\! f_B(x;x') \!\times\! \Psi_0(t',r') 
\\
&\mbox{}& \hspace{-1.3cm}
= \frac{\kappa^2 H^2}{384 \pi^3} \! \int \! d^4x' \Bigl\{
\frac{3 \partial^2}{10 H^2} \!-\! \frac{a' \partial_0}{H} 
\!+\! a {a'}^2 H \Delta \eta \!-\! \frac{\nabla^2}{5 H^2} \Bigr\} 
\, f_B(x;x') \times \Psi_0(t',r') 
\qquad \label{S3TB}
\end{eqnarray}
There is some ambiguity in how we describe the three contributions 
(\ref{S3T}), (\ref{S3TA}) and (\ref{S3TB}). For instance, one can 
exploit relation (\ref{f2tof1}) to convert the $-a a' \partial^4$ 
in (\ref{S3TA}) into an additional $+ 4 a a'$ in (\ref{S3T}). 
Also the factor of $+a {a'}^2 H \Delta \eta$ in (\ref{S3TB}) can 
be converted by expression (\ref{f3tof2}) into an additional 
$-2 a {a'}^2 H \partial_0 \partial^2$ in (\ref{S3TA}). When these
simplifications are made the result can be written in terms of
particular cases of the four generic integrals which are defined
and evaluated in the Appendix:
\begin{eqnarray}
\lefteqn{\mathcal{S}_{3T} + \mathcal{S}_{3T_{A}} 
+ \mathcal{S}_{3T_{B}} = }
\nonumber \\
& & 
- \frac{\kappa^2 H^2}{192 \pi^2} \times 12 a^2 \Psi_0 
+ \frac{\kappa^2 H^2}{384 \pi^3} \, \Bigl\{ 
(2 a H \partial_0 \partial^2 \!+\! 2 a^2 H^2 \partial^2) 
\!\times\! I_{A}^2 
\nonumber \\
& & 
+ 2 a \partial^2 \!\times\! I_{A\delta}^{1} 
- 4 a^2 H^2 \!\times\! I_{A\delta}^{2} 
+ \frac{3 \partial^2}{10 H^2} \!\times\! I_{B}^{0} 
- \frac{\partial_0}{H} \!\times\! I_{B}^{1} 
- \frac{1}{5 H^2} \!\times\! I_{B\delta}^{0} \Bigr\} 
\;\; . \qquad \label{S3ints}
\end{eqnarray}
Substituting the relevant results from the Appendix in
(\ref{S3ints}) gives the final contribution from
$\mathcal{S}_3$:
\begin{equation}
\mathcal{S}_{3T} + \mathcal{S}_{3T_{A}} + \mathcal{S}_{3T_{B}} 
= a^2 \Psi_0 \Bigl\{
-\tfrac{\kappa^2}{240 \pi^2 a^2 r^2} 
- \tfrac{\kappa^2 H^2}{48 \pi^2} 
- \tfrac{\kappa^2 H^3 a r}{24 \pi^2} \Bigr\} 
\;\; . \label{S3final}
\end{equation}
\noindent
{\bf -} {\it The Source $\mathcal{S}_2$:}
The contributions to $\mathcal{S}_2$ from $\widehat{T}^i(a,a',\partial)$ 
and $\widehat{T}^i_{A}(a,a',\partial)$ can be obtained from Tables~\ref{TTA} 
and \ref{Twosources}:
\begin{eqnarray}
\lefteqn{\mathcal{S}_{2T} \equiv 
-\frac{\kappa^2 H^2}{192 \pi^2} \sum_{i=1}^{21}
\! \int \! d^4x' \, \widehat{T}^i(a,a',\partial) \times \mathcal{F}^i_2 
\times \delta^4(x \!-\! x') \times \Psi_0(t',r') } 
\\
& & 
= -\frac{\kappa^2 H^2 a^2}{192 \pi^2} \int \! d^4x' \, \Bigl\{ 
\ln(a a') \Bigl[ -a a' \partial_0 
\!-\! 2 a^2 {a'}^2 H^2 \Delta \eta \Bigr]
\!-\! 12 a a' \partial_0  
\nonumber \\
& & \hspace{2.9cm} 
- 4 a^2 a' H \!+\! 12 a {a'}^2 H \Bigr\} \, \delta^4(x \!-\! x') 
\times \Psi_0(t',r') 
\;\; , \qquad \label{S2T} \\
\lefteqn{\mathcal{S}_{2T_{A}} \equiv 
\frac{\kappa^2 H^2}{384 \pi^3} \sum_{i=1}^{21}
\! \int \! d^4x' \, \widehat{T}^i_{A}(a,a',\partial) \times \mathcal{F}^i_2 
\times \theta(\Delta \eta \!-\! \Delta r) \times \Psi_0(t',r') } 
\\
& & \hspace{-0.5cm} 
= \frac{\kappa^2 H^2}{384 \pi^3} \int \! d^4x' \, \Bigl\{
\Bigl[ a a' \partial_0 \!-\! a^2 a' H \!+\! 4 a {a'}^2 H 
\Bigr] \partial^4 
+ \Bigl[ -2 a a' \partial_0 \partial^2 \!+\! 2 a^2 a' H^2 \partial^2 
\nonumber \\
& & \hspace{1.9cm} 
- 4 a {a'}^2 H \partial^2 \!+\! 4 a^2 {a'}^2 H^2 \partial_0 \Bigr]
\nabla^2 \Bigr\} \, \theta(\Delta \eta \!-\! \Delta r) 
\times \Psi_0(t',r')
\;\; . \qquad \label{S2TA}
\end{eqnarray}
Similarly, Tables~\ref{TB2} and \ref{Twosources} give the 
contribution from $\widehat{T}^i_{B}(a,a',\partial)$:
\begin{eqnarray}
\lefteqn{\mathcal{S}_{2T_{B}} \equiv 
\frac{\kappa^2 H^2}{384 \pi^3} \sum_{i=1}^{21}
\! \int \! d^4x' \, \widehat{T}^i_{B}(a,a',\partial) \times \mathcal{F}^i_2 
\times f_B(x;x') \times \Psi_0(t',r') } 
\qquad \\
& & \hspace{-0.6cm} 
= \frac{\kappa^2 H^2}{384 \pi^3} \int \! d^4x' \Bigl\{
-\frac{\partial_0 \partial^2}{4 H^2} \!+\! \Bigl[ 
\frac{3 a}{4 H} \!-\! \frac{a'}{H} \Bigr] \partial^2 
+ \Bigl[ -3 a a' \!+\! {a'}^2 \Bigr] \partial_0 
\!-\! a {a'}^2 H
\qquad \nonumber \\
& & \hspace{1.7cm} 
+ \Bigl[ \frac{\partial_0}{5 H^2} \!-\! \frac{a}{2 H} 
\!+\! \frac{a'}{H} \Bigr] \nabla^2 \Bigr\} 
\times f_B(x;x') \times \Psi_0(t',r') 
\;\; . \qquad \label{S2TB}
\end{eqnarray}
To simplify expressions (\ref{S2T}), (\ref{S2TA}) and (\ref{S2TB})
we use the identities (\ref{f2tof1}-\ref{f3tof2}) and we reduce 
their total to a sum of the integrals evaluated in the Appendix:
\begin{eqnarray}
\mathcal{S}_{2T} \!+\! \mathcal{S}_{2T_{A}} 
\!+\! \mathcal{S}_{2T_{B}} 
&\!\!\! = \!\!\!& 
\frac{\kappa^2 H^2}{384 \pi^3} \, \Big\{ 
2 a^2 H \partial^2 \!\times\! I_{A\delta}^1 
\!-\! (2 a H \partial^2 \!-\! 4 a^2 H^2 \partial_0) 
\! \times \! I_{A\delta}^{2}
\qquad \nonumber \\
&\mbox{}& 
- \Big( \frac{\partial_0 \partial^2}{4 H^2} 
\!+\! \frac{a \partial^2}{4 H} \Big) \!\times\! I_{B}^{0} 
+ \Big( \frac{\partial_0}{5 H^2} \!+\! \frac{a}{2 H} \Big) 
\!\times\! I_{B\delta}^{0} \Big\} 
\;\; . \label{S2ints}
\end{eqnarray}
The final answer is:
\begin{equation}
\mathcal{S}_{2T} \!+\! \mathcal{S}_{2T_{A}} 
\!+\! \mathcal{S}_{2T_{B}} = 
a^3 H \Psi_0 \,
\Bigl\{ -\tfrac{\kappa^2}{80 \pi^2 a^2 r^2} \Bigr\} 
\;\; . \label{S2final}
\end{equation}
\noindent
{\bf -} {\it The Source $\mathcal{S}_1$:}
The factors $\mathcal{F}_1^i$ needed for the $\mathcal{S}_1$ source
can be seen in Table~\ref{00source}.
\footnote{The factor of $\delta^4(x - x')$ in $\mathcal{S}_{1T}$ 
has been used to consolidate factors of $a$ and $a'$, and we have 
suppressed contributions proportional to $\delta^3(\vec{x})$.}
The results are:
\begin{eqnarray}
\lefteqn{\mathcal{S}_{1T} \equiv 
-\frac{\kappa^2 H^2}{192 \pi^2} \sum_{i=1}^{21}
\! \int \! d^4x' \, \widehat{T}^i(a,a',\partial) \times \mathcal{F}^i_1 
\times \delta^4(x \!-\! x') \times \Psi_0(t',r') } 
\\
& & \hspace{0.05cm}
= -\frac{\kappa^2 H^2 a^2}{192 \pi^2} \! \int \! d^4x' \,
\Bigl\{ \ln(a a') \Bigl[ 
-a {a'}^2 H \partial_0 \!+\! 12 a^2 {a'}^2 H^2 \Bigr] 
\nonumber \\
& & \hspace{1.1cm} 
-\Bigl[ 8 a^2 a' \!+\! 40 a {a'}^2 \Bigr] H \partial_0 
+ 44 a^2 {a'}^2 H^2  \Bigr\} \, \delta^4(x \!-\! x') 
\times \Psi_0(t',r') 
\;\; , \qquad \label{S1T}
\end{eqnarray}
\begin{eqnarray}
\lefteqn{\mathcal{S}_{1T_{A}} \equiv 
\frac{\kappa^2 H^2}{384 \pi^3} \sum_{i=1}^{21} 
\! \int \! d^4x' \, \widehat{T}^i_{A}(a,a',\partial) \times \mathcal{F}^i_1 
\times \theta(\Delta \eta \!-\! \Delta r) 
\times \Psi_0(t',r') } 
\qquad \\
& & \hspace{-0.65cm} 
= \frac{\kappa^2 H^2}{384 \pi^3} \int \! d^4x' \Bigl\{
\Bigl[ a a' \partial^2 + (a^2 a' \!-\! 4 a {a'}^2) H \partial_0 
\!+\! a^2 {a'}^2 H^2 \Bigr] \partial^4 
\nonumber \\
& & \hspace{0.3cm} 
+ \Bigl[ -3 a a' \partial^4 \!+\! 2 a a' \nabla^2 \partial^2 
\!-\! (2 a^2 a' \!-\! 4 a {a'}^2) H \partial_0 \partial^2 
\!+\! 2 a^2 {a'}^2 H^2 \partial^2
\qquad \nonumber \\
& & \hspace{2.1cm}
- 4 a^2 {a'}^2 H^2 \nabla^2 \Bigr] \nabla^2 \Bigr\}
\, \theta(\Delta \eta \!-\! \Delta r) 
\times \Psi_0(t',r') 
\;\; , \qquad \label{S1TA}
\end{eqnarray}
\begin{eqnarray}
\lefteqn{\mathcal{S}_{1T_{B}} \equiv 
\frac{\kappa^2 H^2}{384 \pi^3} \sum_{i=1}^{21} 
\! \int \! d^4x' \, \widehat{T}^i_{B}(a,a',\partial) \times \mathcal{F}^i_1 
\times f_B(x;x') \times \Psi_0(t',r') } 
\nonumber \\
& & \hspace{0.2cm} 
= \frac{\kappa^2 H^2}{384 \pi^3} \! \int \! d^4x' \Bigl\{
\Bigl[ -\frac{5 a}{4 H} \!+\! \frac{a'}{2 H} \Bigr] 
\partial_0 \partial^2 
\!+\! \Bigl[ \frac54 a^2 \!-\! 4 a a' + \frac12 {a'}^2 \Bigr] 
\partial^2 
\nonumber \\
& & \hspace{-0.5cm} 
- \Bigl[ 4 a^2 a' \!-\! a {a'}^2 \Bigr] H \partial_0
+ \Bigl[ \frac{\partial^2}{4 H^2} \!-\! \frac{\nabla^2}{4 H^2} 
\!+\! \Bigl( \frac{a}{2 H} \!-\! \frac{a}{H} \Bigr) \partial_0 
\!-\! \frac12 a^2 \!+\! 3 a a' \!-\! {a'}^2 \Bigr] \nabla^2 \Bigr\} 
\nonumber \\
& & \hspace{0.3cm}
\times f_B(x;x') \times \Psi_0(t',r')
\;\; . \label{S1TB}
\end{eqnarray}
We next simplify using the identities (\ref{f2tof1}) and 
(\ref{f3tof2}), express the result in terms of the generic 
integrals evaluated in the Appendix:
\begin{eqnarray}
\lefteqn{\mathcal{S}_{1T} \!+\! \mathcal{S}_{1T_{A}} 
\!+\! \mathcal{S}_{1T_{B}} = }
\nonumber \\
& &
\frac{\kappa^2 H^2}{384 \pi^3} \Bigl\{ 
- a^2 H \partial_0 \partial^2 \!\times\! I_{A\delta}^{1}
+ \Bigl[ 2 a H \partial_0 \partial^2 
\!+\! 2 a^2 H^2 \partial^2 \!-\! 4 a^2 H^2 \nabla^2 \Bigr] 
\!\times\! I_{A\delta}^{2}
\qquad \nonumber \\
& &
- \frac{3 a \partial_0 \partial^2}{4 H} \!\times\! I_{B}^{0} 
- \Bigl[\frac34 a \partial^2 \!+\! \frac32 a^2 H \partial_0 \Bigr] 
\!\times\! I_{B}^{1} 
\nonumber \\
& &
+ \Bigl[ \frac{\partial^2}{4 H^2} \!-\! \frac{\nabla^2}{5 H^2} 
\!-\! \frac{a \partial_0}{2 H} \Bigr] \!\times\! I_{B\delta}^{0} 
- \frac12 a \!\times\! I_{B\delta}^{1} \Bigr\} 
\;\; , \label{S1ints}
\end{eqnarray}
and conclude:
\begin{equation}
\mathcal{S}_{1T} \!+\! \mathcal{S}_{1T_{A}} 
\!+\! \mathcal{S}_{1T_{B}} 
= a^4 H^2 \Psi_0 \,
\Bigl\{-\tfrac{\kappa^2 H^2}{80 \pi^2 a^4 H^4 r^4} \Bigr\} 
\;\; . \qquad \label{S1final}
\end{equation}
\noindent
{\bf -} {\it Corrections to the potentials $\Phi$ and $\Psi$:} 
The 3rd and final term in expression (\ref{S3final}) is a spatial
constant and would therefore be annihilated by the prefactor of 
$\overline{\partial}^{\mu} \overline{\partial}^{\nu}$ which 
$\mathcal{S}_3$ carries as seen from (\ref{scalarS}) . If we 
drop this term and include the Weyl contribution (\ref{largerS3})
to $\mathcal{S}_3$, the $\mathcal{E}_3 = \mathcal{S}_3$ equation 
reads:
\begin{equation}
-a^2 (\Psi_1 + \Phi_1) = 
a^2 \Psi_0 \Bigl\{ -\tfrac{\kappa^2}{240 \pi^2 a^2 r^2}
- \tfrac{3 \kappa^2 H^2}{160 \pi^2} \Bigr\} 
\;\; . \label{E3=S3}
\end{equation}
Solving for $\Psi_1$ gives:
\begin{equation}
\Psi_1 = -\Phi_1 + \Psi_0 \Bigl\{ 
\tfrac{\kappa^2}{240 \pi^2 a^2 r^2}
+ \tfrac{3 \kappa^2 H^2}{160 \pi^2} \Bigr\}
\;\; . \label{Psi1initial}
\end{equation}
Substituting (\ref{Psi1initial}) in the $\mathcal{E}_2 
= \mathcal{S}_2$ equation, and making some simple 
manipulations implies:
\begin{equation}
\partial_0 (a \Phi_1) = a^2 H \Psi_0 \Bigl\{ 
- \tfrac{\kappa^2}{480 \pi^2 a^2 r^2}
- \tfrac{3 \kappa^2 H^2}{160 \pi^2} \Bigr\} 
\;\; . \label{E2=S2}
\end{equation}
The solution, up to a function $f(r)$, takes the form:
\begin{equation}
\Phi_1 = \Psi_0 \Bigl\{ 
\tfrac{\kappa^2}{960 \pi^2 a^2 r^2} 
+ \tfrac{3 \kappa^2 H^2}{160 \pi^2} \!\times\! \ln(a) \Bigr\} 
+ \frac{f(r)}{a} 
\;\; . \label{Phi1initial}
\end{equation}
We can determine the function $f(r)$ by combining the 
$\mathcal{E}_1 = \mathcal{S}_1$ and $\mathcal{E}_2 = 
\mathcal{S}_2$ equations to find:
\begin{equation}
\nabla^2 \Phi_1 = \Psi_0 \Bigl\{ 
\tfrac{\kappa^2}{160 \pi^2 a^2 r^4}
- \tfrac{3 \kappa^2 H^2}{160 \pi^2 r^2} \Bigr\} 
\;\; . \label{E23=S23}
\end{equation}
If we choose the integration constant so that $\Psi_1$ 
agrees with the flat space limit at the initial time, 
the two 1-loop potentials are:
\begin{eqnarray}
\Psi_1 & = & \Psi_0 \Bigl\{ 
\tfrac{\kappa^2}{320 \pi^2 a^2 r^2}
- \tfrac{3 \kappa^2 H^2}{160 \pi^2} 
\!\times\! \ln(a H r) \Bigr\} 
\;\; , \label{Psi1} \\
\Phi_1 & = & \Psi_0 \Bigl\{ 
\tfrac{\kappa^2}{960 \pi^2 a^2 r^2}
+ \tfrac{3 \kappa^2 H^2}{160 \pi^2}
\Bigl[ \ln(a H r) \!+\! 1 \Bigr] \Bigr\} 
\;\; . \label{Phi1} 
\end{eqnarray}

\section{Epilogue}

This re-computation of the MMC scalar loop contribution to the
graviton self-energy was prompted by the recent discovery that
a finite renormalization of the cosmological constant is needed
to make $-i [\mbox{}^{\mu\nu} \Sigma^{\rho\sigma}](x;x')$
conserved \cite{Glavan:2024elz}. The absence of this 
renormalization in the original
computation \cite{Park:2011ww} was compounded by the decision 
to express the result as a sum of conserved tensor differential 
operators acting on structure functions \cite{Leonard:2014zua}.
While it was possible that the two mistakes might have canceled
one another, it was obviously necessary to check. In addition 
to including the missing renormalization we have expressed the
fully renormalized, in-in result (\ref{SigmaSK}) without any 
preconceptions about its tensor structure. Our solution of the
effective field equations in Section 4 confirms the original 
finding of no secular 1-loop corrections to graviton mode
function \cite{Park:2011kg,Leonard:2014zua}, however, our 
results (\ref{Psi1}-\ref{Phi1}) for the response to a point mass
differ from the previous calculation \cite{Park:2015kua} in two 
ways:
\begin{enumerate}
\item{The coefficient of the $\ln(a)$ correction changed to 
agree with that of the old $\ln(Hr)$ correction so that the
two add to give $\ln(a H r)$; and}
\item{We concluded that the large gravitational slip originally
reported \cite{Park:2015kua} is not present.}
\end{enumerate}
We also pushed the calculation of the graviton mode function to
include falling corrections which nonetheless cause the electric
components of the Weyl tensor (\ref{radiation}) to experience secular
growth compared to the classical result.

It is significant that all three of our secular 
1-loop corrections (\ref{radiation}) and (\ref{Psi1}-\ref{Phi1}) have 
the same coefficient of $-\frac{3 \kappa^2 H^2}{160 \pi^2}$. This 
is reminiscent of the secular 1-graviton loop corrections to the
electric field strength of plane wave photons \cite{Wang:2014tza} 
and to the Coulomb potential \cite{Glavan:2013jca}. Those effects
both had a renormalization group explanation \cite{Glavan:2024elz}
and we will show, in a follow-up work \cite{Miao:2024mmc2}, that 
the same applies to (\ref{radiation}) and (\ref{Psi1}-\ref{Phi1}). 
Support for this comes from the observation that all three of 
these secular corrections derive from the factors of $\ln(a)$
which were induced by the incomplete cancellation of primitive 
divergences and counterterms:
\begin{equation}
\frac{(2 H)^{D-4}}{D \!-\! 4} - \frac{(\mu a)^{D-4}}{D \!-\! 4}
= -\ln\Bigl(\frac{\mu a}{2 H}\Bigr) + O(D\!-\!4) \; .
\end{equation}
for the Einstein counterterm (\ref{DL1b}) and for the $\partial_0^2$
part of the Weyl counterterm.

We should also comment on the sign of the three secular 1-loop 
effects. In each case they reduce the classical result. We believe
this might arise from the inflationary production of scalars 
sucking energy from the gravitational sector. Additional support for
this supposition comes from the fact that the scalar loop induces a 
{\it negative} cosmological constant, which is what required a 
positive renormalization of the cosmological constant 
\cite{Tsamis:2023fri}, both in order to make the graviton 
self-energy conserved and so that the constant ``$H$'' corresponds 
to the true Hubble parameter.

Finally, a major reason for making this computation was to
facilitate the analysis of graviton loops. It seems inevitable 
that a finite renormalization of the cosmological constant is
also necessary for them. The fact that we have seen here that 
this matters for the scalar loop contribution points to the need
for re-computing the effects of gravitons \cite{Tan:2021lza,
Tan:2022xpn}.

\vskip 0.5cm

\centerline{\bf Acknowledgements}

This work was partially supported by Taiwan NSTC grants 
111-2112-M-006-038 and 112-2112-M-006-017, by NSF grant 
PHY-2207514 and by the Institute for Fundamental Theory 
at the University of Florida.

\vskip 1cm

\section{Appendix: Figures \& Tables}

\vskip 0.5cm

\begin{center} 
\begin{picture}(300,100)(0,0)
\Arc(150,50)(40,-90,270)
\Photon(110,50)(50,50){2}{5}
\Vertex(110,50){2}
\Text(105,40)[b]{$x$}
\Photon(250,50)(190,50){2}{5}
\Vertex(190,50){2}
\Text(197,40)[b]{$x'$}
\end{picture} 
\end{center}
\vskip -0.5cm
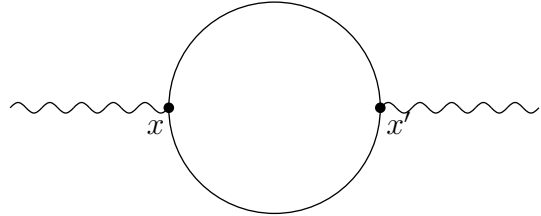
{\bf Fig.~1:} {\footnotesize Contribution from the two 3-point vertices;
graviton lines are wavy, scalar lines are solid.} 

\vskip 1cm

\begin{center}
\begin{picture}(300,100)(0,0)
\Photon(150,50)(90,50){2}{5}
\Photon(210,50)(150,50){2}{5}
\Vertex(150,52){2}
\Text(150,40)[b]{$x$}
\Arc(150,75)(23,-90,270)
\end{picture}
\end{center}
\vskip -1.7cm
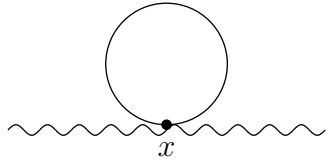
{\bf Fig.~2:} {\footnotesize Contribution from the 4-point vertex;
graviton lines are wavy, scalar lines are solid.}

\vskip -0.5cm

\begin{center}
\begin{picture}(300,100)(0,0)
\Photon(150,50)(90,50){2}{5}
\Photon(210,50)(150,50){2}{5}
\Vertex(150,50){2}
\Text(150,50)[]{\LARGE $\times$}
\Text(150,38)[b]{$x$}
\end{picture}
\end{center}
\vskip -1.7cm
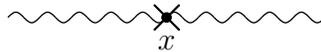
{\bf Fig.~3:} {\footnotesize Contribution from the counterterms;
graviton lines are wavy.}

\newpage

\begin{table}[H]
\setlength{\tabcolsep}{8pt}
\def\arraystretch{1.5}
\centering
\begin{tabular}{|@{\hskip 1mm }c@{\hskip 1mm }||c|}
\hline
$i$ & $V^i$ \\
\hline\hline
1 & $\frac{\eta^{\mu \rho}}{\Delta x^D}$ \\
\hline
2 & $- \frac{D \Delta x^{\mu} \Delta x^{\rho}}{\Delta x^{D+2}}$ \\
\hline
3 & $\frac{(D\!-\!2) [a H \delta^{\mu}_{~0} \Delta x^{\rho} 
- \Delta x^{\mu} a' H \delta^{\rho}_{~0}]}{2 \Delta x^D}$ \\
\hline
4 & $\frac{(D\!-\!2) a a' H^2 \delta^{\mu}_{~0} \delta^{\rho}_{~0}}
{4 \Delta x^{D-2}}$ \\
\hline
5 & $\frac{D a a' H^2 \eta^{\mu\rho}}{8 \Delta x^{D-2}}$ \\
\hline
6 & $-\frac{(D\!-\!2) D a a' H^2\Delta x^{\mu} \Delta x^{\rho}}
{8 \Delta x^{D}}$ \\
\hline
7 & $\frac{(D\!-\!4) D a a' H^2 [a H \delta^{\mu}_{~0} \Delta x^{\rho} 
- \Delta x^{\mu} a' H \delta^{\rho}_{~0}]}{16 \Delta x^{D-2}}$ \\
\hline
8 & $\frac{(D\!-\!4) D a^2 {a'}^2 H^4 \delta^{\mu}_{~0} \delta^{\rho}_{~0}}
{32 \Delta x^{D-4}}$ \\
\hline
\end{tabular}
\caption{\footnotesize
Contributing terms from the doubly-differentiated propagator.}
\label{Vi}
\end{table}

\begin{table}[H]
\setlength{\tabcolsep}{8pt}
\def\arraystretch{1.5}
\centering
\begin{tabular}{|@{\hskip 1mm }c@{\hskip 1mm }||c|}
\hline
$i$ & $T^i$ \\
\hline\hline
3 & $- \frac{(D\!-\!2)}{2} \mathcal{T}^3
- \frac12 \mathcal{T}^9
+ \frac12 \mathcal{T}^{13}
- \frac12 \mathcal{T}^{15} \partial_0
- \frac12 \mathcal{T}^{17} \partial^2$ \\
\hline
4 & $- \frac{(D\!-\!2)}{2} \mathcal{T}^4
- \frac12 \mathcal{T}^9
+ \frac12 \mathcal{T}^{13}
- \frac12 \mathcal{T}^{14} \partial_0
- \frac12 \mathcal{T}^{16} \partial^2$ \\
\hline
5 & $- \frac{(D\!-\!2)}{2} \mathcal{T}^5 
- \frac12 (\mathcal{T}^{10} + \mathcal{T}^{11}) 
+ \frac12 \mathcal{T}^{14} 
- \frac12 \mathcal{T}^{18} \partial_0 
- \frac12 \mathcal{T}^{20} \partial^2$ \\
\hline
6 & $- \frac{(D\!-\!2)}{2} \mathcal{T}^6 
- \frac12 (\mathcal{T}^{10} + \mathcal{T}^{11}) 
+ \frac12 \mathcal{T}^{15} 
- \frac12 \mathcal{T}^{18} \partial_0 
- \frac12 \mathcal{T}^{19} \partial^2$ \\
\hline
7 & $- \frac{(D\!-\!2)}{2} \mathcal{T}^7 
- \frac12 \mathcal{T}^{12} 
+ \frac12 \mathcal{T}^{16} 
- \frac12 \mathcal{T}^{19} \partial_0 
- \frac12 \mathcal{T}^{21} \partial^2$ \\
\hline
8 & $- \frac{(D\!-\!2)}{2} \mathcal{T}^8 
- \frac12 \mathcal{T}^{12} 
+ \frac12 \mathcal{T}^{17} 
- \frac12 \mathcal{T}^{20} \partial_0 
- \frac12 \mathcal{T}^{21} \partial^2$ \\
\hline
\end{tabular}
\caption{\footnotesize
The coefficients $T^i$ as linear combinations of the
$\mathcal{T}^{i\,}$'s.}
\label{Ti-to-Ti}
\end{table}

\newpage

\begin{table}[H]
\setlength{\tabcolsep}{8pt}
\def\arraystretch{1.5}
\centering
\begin{tabular}{|@{\hskip 1mm }c@{\hskip 1mm }||c|}
\hline
$i$ & $\mathcal{T}^i(x;x')$ \\
\hline\hline
1 & $\frac{\partial^4}{64 (D + 1) (D - 1) (D - 2)} 
+ \frac{D a a' H^2 \partial^2}{128 (D-1)} 
+ \frac{D^2 (D-2)^2 a^2 {a'}^2 H^4}{1024 (D-1)}$ \\
\hline
2 & $\frac{\partial^4}{32 (D + 1) (D - 1) (D - 2)} 
+ \frac{D a a' H^2 \partial^2}{64 (D-1)} 
+ \frac{D^2 (D-2)^2 a^2 {a'}^2 H^4}{512 (D-1)}$ \\
\hline
3 & $\frac{(D-2) {a'}^2 H^2 \partial^2}{64 (D-1)}$ \\
\hline
4 & $\frac{(D-2) a^2 H^2 \partial^2}{64 (D-1)}$ \\
\hline
5 & $-\frac{a' H \partial^2}{32 (D-1)} 
- \frac{D (D-2)^2 a {a'}^2 H^3}{128 (D-1)}$ \\
\hline
6 & $\frac{a H \partial^2}{32 (D-1)} 
+ \frac{D (D-2)^2 a^2 a' H^3}{128 (D-1)}$ \\
\hline
7 & $\frac{D \partial^2}{64 (D+1) (D-1) (D-2)} 
+ \frac{D (D-2) a a' H^2}{128 (D-1)}$ \\
\hline
8 & $\frac{D \partial^2}{64 (D+1) (D-1) (D-2)} 
+ \frac{D (D-2) a a' H^2}{128 (D-1)}$ \\
\hline
9 & $\frac{D (D-2)^2 a^2 {a'}^2 H^4}{128}$ \\
\hline
10 & $-\frac{D (D-2) a^2 a' H^3}{64 (D-1)}$ \\
\hline 
11 & $\frac{D (D-2) a {a'}^2 H^3}{64 (D-1)}$ \\
\hline
12 & $-\frac{\partial^2}{16 (D+1) (D-1) (D-2)} 
- \frac{D a a' H^2}{32 (D-1)}$ \\
\hline
13 & $\frac{(D-2)^3 a^2 {a'}^2 H^4}{64}$ \\
\hline
14 & $-\frac{(D-2)^2 a^2 a' H^3}{32}$ \\
\hline
15 & $\frac{(D-2)^2 a {a'}^2 H^3}{32}$ \\
\hline
16 & $\frac{(D-2)^2 a^2 H^2}{64 (D-1)}$ \\
\hline
17 & $\frac{(D-2)^2 {a'}^2 H^2}{64 (D-1)}$ \\
\hline
18 & $-\frac{(D-2) a a' H^2}{16}$ \\
\hline
19 & $\frac{(D-2) a H}{32 (D-1)}$ \\
\hline
20 & $-\frac{(D-2) a' H}{32 (D-1)}$ \\
\hline
21 & $\frac{D}{64 (D+1) (D-1)}$ \\
\hline
\end{tabular}
\caption{\footnotesize 
Primitive divergences before including the three trace terms.
\newline $\mbox{}\hspace{1.5cm}$
Act each $\mathcal{T}^i$ on 
$\, -\mathcal{K} \times [\mbox{}^{\mu\nu} D_i^{\rho\sigma}] 
\times i\delta^D(x-x')$.} 
\label{Tinitial}
\end{table}

\newpage

\begin{table}[H]
\setlength{\tabcolsep}{8pt}
\def\arraystretch{1.5}
\centering
\begin{tabular}{|@{\hskip 1mm }c@{\hskip 1mm }||c|}
\hline
$i$ & $T^i(x;x')$ \\
\hline\hline
3 & $-\frac{(D-2)^2 {a'}^2 H^2 \partial^2}{64 (D-1)} 
- \frac{(D-2)^2 a {a'}^2 H^3 \partial_0}{64} 
+ \frac{(D-2)^2 (D-4) a^2 {a'}^2 H^4}{256}$ \\
\hline
4 & $-\frac{(D-2)^2 a^2 H^2 \partial^2}{64 (D-1)} 
+ \frac{(D-2)^2 a^2 a' H^3 \partial_0}{64} 
+ \frac{(D-2)^2 (D-4) a^2 {a'}^2 H^4}{256}$ \\
\hline
5 & $\!\!\frac{(D-2) a' H \partial^2}{32 (D-1)} 
\!+\! \frac{(D-2) a a' H^2 \partial_0}{32} 
- \frac{(D-2)^2 a^2 a' H^3}{64} 
+ \frac{D (D-2)^3 a {a'}^2 H^3}{256 (D-1)} 
\!+\! \frac{D (D-2) a^2 {a'}^2 H^4 \Delta \eta}{128 (D-1)}\!\!$ \\
\hline
6 & $\!\!\!-\frac{(D-2) a H \partial^2}{32 (D-1)} 
\!+\! \frac{(D-2) a a' H^2 \partial_0}{32} 
\!+\! \frac{(D-2)^2 a {a'}^2 H^3}{64} 
\!-\! \frac{D (D-2)^3 a^2 a' H^3}{256 (D-1)} 
\!+\! \frac{D (D-2) a^2 {a'}^2 H^4 \Delta \eta}{128 (D-1)} \!\!\!$ \\
\hline
7 & $-\frac{(D^2 - 2 D - 2) \partial^2}{64 (D+1) (D-1) (D-2)} 
- \frac{(D-2) a H \partial_0}{64 (D-1)} 
- \frac{D^2 (D-4) a a' H^2}{256 (D-1)}
+ \frac{(D-2)^2 a^2 H^2}{128 (D-1)}$ \\
\hline
8 & $-\frac{(D^2 - 2 D - 2) \partial^2}{64 (D+1) (D-1) (D-2)} 
+ \frac{(D-2) a' H \partial_0}{64 (D-1)} 
- \frac{D^2 (D-4) a a' H^2}{256 (D-1)}
+ \frac{(D-2)^2 {a'}^2 H^2}{128 (D-1)}$ \\
\hline
\end{tabular}
\caption{\footnotesize 
Primitive divergences after including the three trace terms. 
\newline $\mbox{}\hspace{1.5cm}$
Act each $T^i$ on 
$\, -\mathcal{K} \times [\mbox{}^{\mu\nu} D_i^{\rho\sigma}]
\times i\delta^D(x - x')$.}
\label{Tfinal}
\end{table}

\begin{table}[H]
\setlength{\tabcolsep}{8pt}
\def\arraystretch{1.5}
\centering
\begin{tabular}{|@{\hskip 1mm }c@{\hskip 1mm }||c|}
\hline
$i$ & $\Delta T^i_{2}(x;x')$ \\
\hline\hline
1 & $\frac{\partial^4}{32 (D + 1) (D - 1)^2 (D - 2)}$ \\
\hline
2 & $-\frac{\partial^4}{32 (D + 1) (D - 1) (D - 2)}$ \\
\hline
7 & $-\frac{\partial^2}{32 (D+1) (D-1)^2 (D-2)}$ \\
\hline
8 & $-\frac{\partial^2}{32 (D+1) (D-1)^2 (D-2)}$ \\
\hline
12 & $\frac{\partial^2}{16 (D+1) (D-1) (D-2)}$ \\
\hline
21 & $-\frac{1}{32 (D+1) (D-1)^2}$ \\
\hline
\end{tabular}
\caption{\footnotesize 
Divergent contributions from the Weyl counterterm (\ref{S2contB}). 
\newline $\mbox{}\hspace{1.5cm}$
Act each $\Delta T^i_2$ on $-\mathcal{K} \times 
[\mbox{}^{\mu\nu} D_i^{\rho\sigma}] \times i\delta^D(x -x')$.}
\label{WeylDiv}
\end{table} 

\newpage

\begin{table}[H]
\setlength{\tabcolsep}{8pt}
\def\arraystretch{1.5}
\centering
\begin{tabular}{|@{\hskip 1mm }c@{\hskip 1mm }||c|}
\hline
$i$ & $\Delta T^i_{1b}(x;x')$ \\
\hline\hline
1 & $\frac{D (D-2) a a' H^2 \partial^2}{128 (D - 1)}
+ \frac{D^2 (D-2)^2 a^2 {a'}^2 H^4}{512 (D-1)}$ \\
\hline
2 & $-\frac{D (D-2) a a' H^2 \partial^2}{128 (D - 1)}
- \frac{D^2 (D-2)^2 a^2 {a'}^2 H^4}{512 (D-1)}$ \\
\hline
3 & $-\frac{D^2 (D-2)^2 a^2 {a'}^2 H^4}{512 (D-1)}$ \\
\hline
4 & $-\frac{D^2 (D-2)^2 a^2 {a'}^2 H^4}{512 (D-1)}$ \\
\hline
5 & $\frac{D (D-2)^2 a^2 a' H^3}{128 (D-1)}$ \\
\hline
6 & $-\frac{D (D-2)^2 a {a'}^2 H^3}{128 (D-1)}$ \\
\hline
7 & $-\frac{D (D-2) a a' H^2}{128 (D-1)}$ \\
\hline
8 & $-\frac{D (D-2) a a' H^2}{128 (D-1)}$ \\
\hline
9 & $-\frac{D (D-2)^3 a^2 {a'}^2 H^4}{256 (D-1)}$ \\
\hline
10 & $\frac{D (D-2)^2 a {a'}^2 H^3}{128 (D-1)}$ \\
\hline
11 & $-\frac{D (D-2)^2 a^2 a' H^3}{128 (D-1)}$ \\
\hline
12 & $\frac{D (D-2) a a' H^2}{64 (D-1)}$ \\
\hline
\end{tabular}
\caption{\footnotesize 
Divergent contributions from the Einstein counterterm
(\ref{S1bcont}). 
\newline $\mbox{}\hspace{1.5cm}$
Act each $\Delta T^i_{1b}$ on $-\mathcal{K} \times
[\mbox{}^{\mu\nu} D_i^{\rho\sigma}] \times i \delta^D(x-x')$.}
\label{EinsteinDiv}
\end{table}

\newpage

\begin{table}[H]
\setlength{\tabcolsep}{8pt}
\def\arraystretch{1.5}
\centering
\begin{tabular}{|@{\hskip 1mm }c@{\hskip 1mm }||c|}
\hline
$i$ & $\Delta T^i_{1a}(x;x')$ \\
\hline\hline
1 & $-\frac{(D-2) \mathcal{D}_1 \mathcal{D}_1'}{64 (D - 1)^2}$ \\
\hline
3 & $\frac{(D-2)^2 {a'}^2 H^2 \mathcal{D}_1}{64 (D-1)}$ \\
\hline
4 & $\frac{(D-2)^2 a^2 H^2 \mathcal{D}_1'}{64 (D-1)}$ \\
\hline
5 & $-\frac{(D-2) a' H \mathcal{D}_1}{32 (D-1)}$ \\
\hline
6 & $\frac{(D-2) a H \mathcal{D}_1'}{32 (D-1)}$ \\
\hline
7 & $\frac{(D-2) \mathcal{D}_1}{64 (D-1)^2}$ \\
\hline
8 & $\frac{(D-2) \mathcal{D}_1'}{64 (D-1)^2}$ \\
\hline
13 & $-\frac{(D-2)^3 a^2 {a'}^2 H^4}{64}$ \\
\hline
14 & $\frac{(D-2)^2 a^2 a' H^3}{32}$ \\
\hline
15 & $-\frac{(D-2)^2 a {a'}^2 H^3}{32}$ \\
\hline
16 & $-\frac{(D-2)^2 a^2 H^2}{64 (D-1)}$ \\
\hline
17 & $-\frac{(D-2)^2 {a'}^2 H^2}{64 (D-1)}$ \\
\hline
18 & $\frac{(D-2) a a' H^2}{16}$ \\
\hline
19 & $-\frac{(D-2) a H}{32 (D-1)}$ \\
\hline
20 & $\frac{(D-2) a' H}{32 (D-1)}$ \\
\hline
21 & $-\frac{(D-2)}{64 (D-1)^2}$ \\
\hline
\end{tabular}
\caption{\footnotesize 
Divergent contributions from the Eddington counterterm 
(\ref{S1avariation}). 
\newline $\mbox{}\hspace{1.5cm}$
Act each $\Delta T^i_{1a}$ on $-\mathcal{K} \times
[\mbox{}^{\mu\nu} D_i^{\rho\sigma}] \times i\delta^D(x-x')$.} 
\label{EddingtonDiv}
\end{table} 

\newpage

\begin{table}[H]
\setlength{\tabcolsep}{8pt}
\def\arraystretch{1.5}
\centering
\begin{tabular}{|@{\hskip 1mm }c@{\hskip 1mm }||c|}
\hline
$i$ & $T^i(x;x')$ \\
\hline\hline
1 & $\frac{(D^2 - 2D - 2) \partial^4}{32 (D + 1) (D - 1) (D - 2)} 
+ \frac{(3 D^3 - 18 D^2 + 24 D - 16) a a' H^2 \partial^2}{512 (D-1)}$ \\
& $- \frac{(D-2) (D-3) a a' H^2 \partial_0^2}{64 (D-1)} 
+ \frac{(D-2) (D - 4) (D^4 - 48 D + 64) a^2 {a'}^2 H^4}{4096 (D-1)}$ \\
\hline
2 & $\frac{\partial^4}{32 (D + 1) (D - 1) (D - 2)} 
+ \frac{D a a' H^2 \partial^2}{64 (D-1)} 
+ \frac{D^2 (D-2)^2 a^2 {a'}^2 H^4}{512 (D-1)}$ \\
\hline
3 & $-\frac{(D-2)^2 {a'}^2 H^2 \partial^2}{64 (D-1)} 
- \frac{(D-2)^2 a {a'}^2 H^3 \partial_0}{64} 
+ \frac{(D-2)^2 (D-4) a^2 {a'}^2 H^4}{256}$ \\
\hline
4 & $-\frac{(D-2)^2 a^2 H^2 \partial^2}{64 (D-1)} 
+ \frac{(D-2)^2 a^2 a' H^3 \partial_0}{64} 
+ \frac{(D-2)^2 (D-4) a^2 {a'}^2 H^4}{256}$ \\
\hline
5 & $\!\!\frac{(D-2) a' H \partial^2}{32 (D-1)} 
\!+\! \frac{(D-2) a a' H^2 \partial_0}{32} 
\!-\! \frac{(D-2)^2 a^2 a' H^3}{64} 
\!+\! \frac{D (D-2)^3 a {a'}^2 H^3}{256 (D-1)} 
\!+\! \frac{D (D-2) a^2 {a'}^2 H^4 \Delta \eta}{128 (D-1)}\!\!$ \\
\hline
6 & $\!\!\!-\frac{(D-2) a H \partial^2}{32 (D-1)} 
\!+\! \frac{(D-2) a a' H^2 \partial_0}{32} 
\!+\! \frac{(D-2)^2 a {a'}^2 H^3}{64} 
\!-\! \frac{D (D-2)^3 a^2 a' H^3}{256 (D-1)} 
\!+\! \frac{D (D-2) a^2 {a'}^2 H^4 \Delta \eta}{128 (D-1)} \!\!\!$ \\
\hline
7 & $-\frac{(D^2 - 2 D - 2) \partial^2}{64 (D+1) (D-1) (D-2)} 
- \frac{(D-2) a H \partial_0}{64 (D-1)} 
- \frac{D^2 (D-4) a a' H^2}{256 (D-1)}
+ \frac{(D-2)^2 a^2 H^2}{128 (D-1)}$ \\
\hline
8 & $-\frac{(D^2 - 2 D - 2) \partial^2}{64 (D+1) (D-1) (D-2)} 
+ \frac{(D-2) a' H \partial_0}{64 (D-1)} 
- \frac{D^2 (D-4) a a' H^2}{256 (D-1)}
+ \frac{(D-2)^2 {a'}^2 H^2}{128 (D-1)}$ \\
\hline
9 & $\frac{D (D-2)^2 a^2 {a'}^2 H^4}{128}$ \\
\hline
10 & $-\frac{D (D-2) a^2 a' H^3}{64 (D-1)}$ \\
\hline
11 & $\frac{D (D-2) a {a'}^2 H^3}{64 (D-1)}$ \\
\hline
12 & $-\frac{\partial^2}{16 (D+1) (D-1) (D-2)} 
- \frac{D a a' H^2}{32 (D-1)}$ \\
\hline
13 & $\frac{(D-2)^3 a^2 {a'}^2 H^4}{64}$ \\
\hline
14 & $-\frac{(D-2)^2 a^2 a' H^3}{32}$ \\
\hline
15 & $\frac{(D-2)^2 a {a'}^2 H^3}{32}$ \\
\hline
16 & $\frac{(D-2)^2 a^2 H^2}{64 (D-1)}$ \\
\hline
17 & $\frac{(D-2)^2 {a'}^2 H^2}{64 (D-1)}$ \\
\hline
18 & $-\frac{(D-2) a a' H^2}{16}$ \\
\hline
19 & $\frac{(D-2) a H}{32 (D-1)}$ \\
\hline
20 & $-\frac{(D-2) a' H}{32 (D-1)}$ \\
\hline
21 & $\frac{D}{64 (D+1) (D-1)}$ \\
\hline
\end{tabular}
\caption{\footnotesize Coefficients $T^i$ of primitive 
divergences in (\ref{genform1}). Each term acts on 
\newline $\mbox{}\hspace{1.5cm}$
$-\mathcal{K} \times [\mbox{}^{\mu\nu} D_i^{\rho\sigma}] 
\times i \delta^D(x-x')$.}
\label{Tprimitive}
\end{table}

\newpage

\begin{table}[H]
\setlength{\tabcolsep}{8pt}
\def\arraystretch{1.5}
\centering
\begin{tabular}{|@{\hskip 1mm }c@{\hskip 1mm }||c|}
\hline
$i$ & $\Delta T^i(x;x')$ \\
\hline\hline
1 & $\!\!\!-\frac{(D^2 - 2 D -2) \partial^4}{(D+1) (D-1) (D-2)} 
\!+\! \frac{(D-2)^2 \!a a'\! H^2 \!\partial^2}{128 (D-1)} 
\!+\! \frac{(D-2) (D-3) a a'\! H^2 \! \partial_0^2}{64 (D-1)} 
\!+\! \frac{(D+4) (D-2)^2\! (D-4) a^2\! {a'}^2 \!H^4}{
512 (D - 1)} \!\!\!$ \\
\hline
2 & $-\frac{\partial^4}{32 (D+1) (D-1) (D-2)} 
- \frac{D (D-2) a a' H^2 \partial^2}{128 (D-1)} 
- \frac{D^2 (D-2)^2 a^2 {a'}^2 H^4}{512 (D-1)}$ \\
\hline
3 & $\frac{(D-2)^2 {a'}^2 H^2 \partial^2}{64 (D-1)} 
+ \frac{(D-2)^2 a {a'}^2 H^3 \partial_0}{64} 
+ \frac{(D-2)^2 a^2 {a'}^2 H^4}{64} 
- \frac{D^2 (D-2)^2 a^2 {a'}^2 H^4}{512 (D-1)}$ \\
\hline
4 & $\frac{(D-2)^2 a^2 H^2 \partial^2}{64 (D-1)} 
- \frac{(D-2)^2 a^2 a' H^3 \partial_0}{64} 
+ \frac{(D-2)^2 a^2 {a'}^2 H^4}{64} 
- \frac{D^2 (D-2)^2 a^2 {a'}^2 H^4}{512 (D-1)}$ \\
\hline
5 & $-\frac{(D-2) a' H \partial^2}{32 (D-1)} 
- \frac{(D-2) a a' H^2 \partial_0}{32}
- \frac{(D-2) a^2 a' H^3}{32} 
+ \frac{D (D-2)^2 a^2 a' H^3}{128 (D-1)}$ \\
\hline
6 & $\frac{(D-2) a H \partial^2}{32 (D-1)} 
- \frac{(D-2) a a' H^2 \partial_0}{32}
+ \frac{(D-2) a {a'}^2 H^3}{32} 
- \frac{D (D-2)^2 a {a'}^2 H^3}{128 (D-1)}$ \\
\hline
7 &  $\frac{(D^2 - 2 D -2) \partial^2}{64 (D+1) (D-1) (D-2)} 
+ \frac{(D-2) a H \partial_0}{64 (D-1)} 
+ \frac{(D-2) a^2 H^2}{64 (D-1)} 
- \frac{D (D-2) a a' H^2}{128 (D-1)}$\\
\hline
8 & $\frac{(D^2 - 2 D -2) \partial^2}{64 (D+1) (D-1) (D-2)} 
- \frac{(D-2) a' H \partial_0}{64 (D-1)} 
+ \frac{(D-2) {a'}^2 H^2}{64 (D-1)} 
- \frac{D (D-2) a a' H^2}{128 (D-1)}$ \\
\hline
9 & $-\frac{D (D-2)^3 a^2 {a'}^2 H^4}{256 (D-1)}$ \\
\hline
10 & $\frac{D (D-2)^2 a {a'}^2 H^3}{128 (D-1)}$ \\
\hline
11 & $-\frac{D (D-2)^2 a^2 a' H^3}{128 (D-1)}$ \\
\hline
12 & $\frac{\partial^2}{16 (D+1) (D-1) (D-2)} 
+ \frac{D (D-2) a a' H^2}{64 (D-1)}$ \\
\hline
13 & $-\frac{(D-2)^3 a^2 {a'}^2 H^4}{64}$ \\
\hline
14 & $\frac{(D-2)^2 a^2 a' H^3}{32}$ \\
\hline
15 & $-\frac{(D-2)^2 a {a'}^2 H^3}{32}$ \\
\hline
16 & $-\frac{(D-2)^2 a^2 H^2}{64 (D-1)}$ \\
\hline
17 & $-\frac{(D-2)^2 {a'}^2 H^2}{64 (D-1)}$ \\
\hline
18 & $\frac{(D-2) a a' H^2}{16}$ \\
\hline
19 & $-\frac{(D-2) a H}{32 (D-1)}$ \\
\hline
20 & $\frac{(D-2) a' H}{32 (D-1)}$ \\
\hline
21 & $-\frac{D}{64 D+1) (D-1)}$ \\
\hline
\end{tabular}
\caption{\footnotesize Divergent coefficients $\Delta T^i$ 
of the counterterms in (\ref{genform2}). Each term acts on
\newline $\mbox{}\hspace{1.5cm}$
$-\mathcal{K} \times [\mbox{}^{\mu\nu} D_i^{\rho\sigma}] 
\times i\delta^D(x-x')$.}
\label{cterms}
\end{table}

\newpage

\begin{table}[H]
\setlength{\tabcolsep}{8pt}
\def\arraystretch{1.5}
\centering
\begin{tabular}{|@{\hskip 1mm }c@{\hskip 1mm }||c|}
\hline
$i$ & $T^i(x;x') + \Delta T^i(x;x')$ \\
\hline\hline
1 & $\frac{D (D-4) (3D-2) a a' H^2 \partial^2}{512 (D-1)} 
+ \frac{D (D-2) (D-4) (D^3 + 8 D - 32) a^2 {a'}^2 H^4}{4096 (D - 1)}$ \\
\hline
2 & $- \frac{D (D-4) a a' H^2 \partial^2}{128 (D-1)}$ \\
\hline
3 & $\frac{(D-2)^2 (D-4) a^2 {a'}^2 H^4}{256} 
+ \frac{(D-2)^2 a^2 {a'}^2 H^4}{64} 
- \frac{D^2 (D-2)^2 a^2 {a'}^2 H^4}{512 (D-1)}$ \\
\hline
4 & $\frac{(D-2)^2 (D-4) a^2 {a'}^2 H^4}{256} 
+ \frac{(D-2)^2 a^2 {a'}^2 H^4}{64}
- \frac{D^2 (D-2)^2 a^2 {a'}^2 H^4}{512 (D-1)}$ \\
\hline
5 & $-\frac{D^2 (D-2) a^2 a' H^3}{128 (D-1)} 
+ \frac{D (D-2)^3 a {a'}^2 H^3}{256 (D-1)} 
+ \frac{D (D-2) a^2 {a'}^2 H^4 \Delta \eta}{128 (D-1)}$ \\
\hline
6 & $\frac{D^2 (D-2) a {a'}^2 H^3}{128 (D-1)} 
- \frac{D (D-2)^3 a^2 a' H^3}{256 (D-1)} 
+ \frac{D (D-2) a^2 {a'}^2 H^4 \Delta \eta}{128 (D-1)}$ \\
\hline
7 & $-\frac{D^2 (D-4) a a' H^2}{256 (D-1)} 
+ \frac{D (D-2) a^2 a' H^3 \Delta \eta}{128 (D-1)}$\\
\hline
8 & $-\frac{D^2 (D-4) a a' H^2}{256 (D-1)} 
- \frac{D (D-2) a {a'}^2 H^3 \Delta \eta}{128 (D-1)}$ \\
\hline
9 & $\frac{D^2 (D-2)^2 a^2 {a'}^2 H^4}{256 (D-1)}$ \\
\hline
10 & $\frac{D (D-2) (D-4) a {a'}^2 H^3}{128 (D-1)}
-\frac{D (D-2) a^2 {a'}^2 H^4 \Delta \eta}{64 (D-1)}$ \\
\hline
11 & $-\frac{D (D-2) (D-4) a^2 a' H^3}{128 (D-1)}
-\frac{D (D-2) a^2 {a'}^2 H^4 \Delta \eta}{64 (D-1)}$ \\
\hline
12 & $\frac{D (D-4) a a' H^2}{64 (D-1)}$ \\
\hline
13 & $0$ \\
\hline
14 & $0$ \\
\hline
15 & $0$ \\
\hline
16 & $0$ \\
\hline
17 & $0$ \\
\hline
18 & $0$ \\
\hline
19 & $0$ \\
\hline
20 & $0$ \\
\hline
21 & $0$ \\
\hline
\end{tabular}
\caption{\footnotesize Sum of Tables~\ref{Tprimitive} and \ref{cterms}.
Each term multiplies 
$\, -\mathcal{K} \times [\mbox{}^{\mu\nu} D_i^{\rho\sigma}]
\times i\delta^D(x - x')$.}
\label{residual}
\end{table}

\newpage

\begin{table}[H]
\setlength{\tabcolsep}{8pt}
\def\arraystretch{1.5}
\centering
\begin{tabular}{|@{\hskip 1mm }c@{\hskip 1mm }||c|}
\hline
$i$ & $T^i(x;x') + \Delta T^i(x;x')$ \\
\hline\hline
1 & $\frac{D (D-4) (3D-2) a a' H^2 \partial^2}{512 (D-1)} 
+ \frac{D (D-2) (D-4) (D^3 + 8 D - 32) a^2 {a'}^2 H^4}{4096 (D - 1)}$ \\
\hline
2 & $- \frac{D (D-4) a a' H^2 \partial^2}{128 (D-1)}$ \\
\hline
3 & $\frac{D (D-2) (D-4)^2 a^2 {a'}^2 H^4}{512 (D-1)}$ \\
\hline
4 & $\frac{D (D-2) (D-4)^2 a^2 {a'}^2 H^4}{512 (D-1)}$ \\
\hline
5 & $\frac{D (D-2)^2 (D-4) a {a'}^2 H^3}{256 (D-1)}$ \\
\hline
6 & $-\frac{D (D-2)^2 (D-4) a^2 a' H^3}{256 (D-1)}$ \\
\hline
7 & $-\frac{D^2 (D-4) a a' H^2}{256 (D-1)}$\\
\hline
8 & $-\frac{D^2 (D-4) a a' H^2}{256 (D-1)}$ \\
\hline
9 & $\frac{(D+2) D (D-2) (D-4) a^2 {a'}^2 H^4}{256 (D-1)}$ \\
\hline
10 & $\frac{D (D-2) (D-4) a {a'}^2 H^3}{128 (D-1)}$ \\
\hline
11 & $-\frac{D (D-2) (D-4) a^2 a' H^3}{128 (D-1)}$ \\
\hline
12 & $\frac{D (D-4) a a' H^2}{64 (D-1)}$ \\
\hline
13 & $0$ \\
\hline
14 & $0$ \\
\hline
15 & $0$ \\
\hline
16 & $0$ \\
\hline
17 & $0$ \\
\hline
18 & $0$ \\
\hline
19 & $0$ \\
\hline
20 & $0$ \\
\hline
21 & $0$ \\
\hline
\end{tabular}
\caption{\footnotesize Reduction of Table~\ref{residual} 
using $\, \Delta \eta \partial^{\mu} = 
\partial^{\mu} \Delta \eta + \delta^{\mu}_{~0}$.}
\label{finite}
\end{table}

\newpage

\begin{table}[H]
\setlength{\tabcolsep}{8pt}
\def\arraystretch{1.5}
\centering
\begin{tabular}{|@{\hskip 1mm }c@{\hskip 1mm }||c|}
\hline
$i$ & $\lim_{D=4} \frac1{D-4} [T^i(x;x') + \Delta T^i(x;x')]$ \\
\hline\hline
1 & $\frac{5 a a' H^2 \partial^2}{192} 
+ \frac{a^2 {a'}^2 H^4}{24}$ \\
\hline
2 & $-\frac{a a' H^2 \partial^2}{96}$ \\
\hline
3 & $0$ \\
\hline
4 & $0$ \\
\hline
5 & $\frac{a {a'}^2 H^3}{48}$ \\
\hline
6 & $-\frac{a^2 a' H^3}{48}$ \\
\hline
7 & $-\frac{a a' H^2}{48}$\\
\hline
8 & $-\frac{a a' H^2}{48}$ \\
\hline
9 & $\frac{a^2 {a'}^2 H^4}{16}$ \\
\hline
10 & $\frac{a {a'}^2 H^3}{48}$ \\
\hline
11 & $-\frac{a^2 a' H^3}{48}$ \\
\hline
12 & $\frac{a a' H^2}{48}$ \\
\hline
13 & $0$ \\
\hline
14 & $0$ \\
\hline
15 & $0$ \\
\hline
16 & $0$ \\
\hline
17 & $0$ \\
\hline
18 & $0$ \\
\hline
19 & $0$ \\
\hline
20 & $0$ \\
\hline
21 & $0$ \\
\hline
\end{tabular}
\caption{\footnotesize Final finite residuals. Act each factor on
$\, -\frac{\kappa^2}{2\pi^2} \times [\mbox{}^{\mu\nu} D_i^{\rho\sigma}]
\times i \delta^4(x - x')$.}
\label{D4limit}
\end{table}

\newpage

\begin{table}[H]
\setlength{\tabcolsep}{8pt}
\def\arraystretch{1.5}
\centering
\begin{tabular}{|@{\hskip 1mm }c@{\hskip 1mm }||c|}
\hline
$i$ & $\mathcal{T}_A^i(a,a',\partial)$ \\
\hline\hline
3 & $\frac{a {a'}^3 H^4 \partial^4}{64}$ \\
\hline
4 & $\frac{a^3 a' H^4 \partial^4}{64}$ \\
\hline
5 & $-\frac{a {a'}^2 H^3 \partial^4}{96}$ \\
\hline
6 & $\frac{a^2 a' H^3 \partial^4}{96}$ \\
\hline
7 & $\frac{a^2 {a'}^2 H^4 \partial^2}{192}$ \\
\hline
8 & $\frac{a^2 {a'}^2 H^4 \partial^2}{192}$ \\
\hline
9 & $-\frac{a^2 {a'}^2 H^4 \partial^4}{32}$ \\
\hline
10 & $\frac{a^2 a' H^3 \partial^4}{192}$ \\
\hline 
11 & $-\frac{a {a'}^2 H^3 \partial^4}{192}$ \\
\hline
12 & $-\frac{a^2 {a'}^2 H^4 \partial^2}{96}$ \\
\hline
18 & $-\frac{a^2 {a'}^2 H^4 \partial^2}{32}$ \\
\hline
19 & $\frac{a^2 a' H^3 \partial^2}{96}$ \\
\hline
20 & $-\frac{a {a'}^2 H^3 \partial^2}{96}$ \\
\hline
21 & $\frac{a a' H^2 \partial^2}{192} 
- \frac{a^2 {a'}^2 H^4}{96}$ \\
\hline
\end{tabular}
\caption{\footnotesize 
Non-local contributions acting on $\, \ln(\mu^2 \Delta x^2)$ 
before including the trace terms.} 
\label{TauA}
\end{table}

\newpage

\begin{table}[H]
\setlength{\tabcolsep}{8pt}
\def\arraystretch{1.5}
\centering
\begin{tabular}{|@{\hskip 1mm }c@{\hskip 1mm }||c|}
\hline
$i$ & $T_A^i(a,a',\partial)$ \\
\hline\hline
1 & $\frac{a a' H^2 \partial^6}{768} 
- \frac{a^2 {a'}^2 H^4 \partial^4}{128}
- \frac{a^2 {a'}^2 H^4 \partial_0^2 \partial^2}{128} 
+ \frac{7 a^2 {a'}^2 H^4 \Delta \eta \partial_0 \partial^4}{768} 
- \frac{a^3 {a'}^3 H^6 \Delta \eta^2 \partial^4}{128}$ \\
\hline
3 & $\frac{a^2 {a'}^3 H^5 \Delta \eta \partial^4}{64}$ \\
\hline
4 & $-\frac{a^3 {a'}^2 H^5 \Delta \eta \partial^4}{64}$ \\
\hline
5 & $\frac{a {a'}^2 H^3 \partial^4}{64} 
- \frac{a^2 {a'}^2 H^4 \Delta \eta \partial^4}{384} 
+ \frac{a^2 {a'}^2 H^4 \partial_0 \partial^2}{64}$ \\
\hline
6 & $-\frac{a^2 a' H^3 \partial^4}{64} 
- \frac{a^2 {a'}^2 H^4 \Delta \eta \partial^4}{384} 
+ \frac{a^2 {a'}^2 H^4 \partial_0 \partial^2}{64}$ \\
\hline
7 & $-\frac{a a' H^2 \partial^4}{384} 
- \frac{a^2 a' H^3 \partial_0 \partial^2}{192} 
+ \frac{a^2 {a'}^2 H^4 \partial^2}{192}$ \\
\hline
8 & $-\frac{a a' H^2 \partial^4}{384} 
+ \frac{a {a'}^2 H^3 \partial_0 \partial^2}{192} 
+ \frac{a^2 {a'}^2 H^4 \partial^2}{192}$ \\
\hline
9 & $-\frac{a^2 {a'}^2 H^4 \partial^4}{32}$ \\
\hline
10 & $\frac{a^2 a' H^3 \partial^4}{192}$ \\
\hline 
11 & $-\frac{a {a'}^2 H^3 \partial^4}{192}$ \\
\hline
12 & $-\frac{a^2 {a'}^2 H^4 \partial^2}{96}$ \\
\hline
18 & $-\frac{a^2 {a'}^2 H^4 \partial^2}{32}$ \\
\hline
19 & $\frac{a^2 a' H^3 \partial^2}{96}$ \\
\hline
20 & $-\frac{a {a'}^2 H^3 \partial^2}{96}$ \\
\hline
21 & $\frac{a a' H^2 \partial^2}{192} 
- \frac{a^2 {a'}^2 H^4}{96}$ \\
\hline
\end{tabular}
\caption{\footnotesize 
Non-local contributions acting on $\, \ln(\mu^2 \Delta x^2)$ 
in expression (\ref{genform3}).} 
\label{TA}
\end{table}

\newpage

\begin{table}[H]
\setlength{\tabcolsep}{8pt}
\def\arraystretch{1.5}
\centering
\begin{tabular}{|@{\hskip 1mm }c@{\hskip 1mm }||c|}
\hline
$i$ & $\mathcal{T}_B^i(a,a',\partial)$ \\
\hline\hline
1 & $-\frac{\partial^4}{3840} 
- \frac{a a' H^2 \partial^2}{192}
- \frac{a^2 {a'}^2 H^4}{96}$ \\
\hline
2 & $-\frac{\partial^4}{1920} 
- \frac{a a' H^2 \partial^2}{96}
- \frac{a^2 {a'}^2 H^4}{48}$ \\
\hline
3 & $-\frac{{a'}^2 H^2 \partial^2}{192}$ \\
\hline
4 & $-\frac{a^2 H^2 \partial^2}{192}$ \\
\hline
5 & $\frac{a' H \partial^2}{192} 
+ \frac{a {a'}^2 H^3}{48}$ \\
\hline
6 & $-\frac{a H \partial^2}{192} 
- \frac{a^2 a' H^3}{48}$ \\
\hline
7 & $-\frac{\partial^2}{960} 
- \frac{a a' H^2}{96}$ \\
\hline
8 & $-\frac{\partial^2}{960} 
- \frac{a a' H^2}{96}$ \\
\hline
9 & $-\frac{a^2 {a'}^2 H^4}{16}$ \\
\hline
10 & $\frac{a^2 a' H^3}{48}$ \\
\hline 
11 & $-\frac{a {a'}^2 H^3}{48}$ \\
\hline
12 & $\frac{\partial^2}{960} 
+ \frac{a a' H^2}{48}$ \\
\hline
13 & $-\frac{a^2 {a'}^2 H^4}{16}$ \\
\hline 
14 & $\frac{a^2 a' H^3}{16}$ \\
\hline
15 & $-\frac{a {a'}^2 H^3}{16}$ \\
\hline
16 & $-\frac{a^2 H^2}{96}$ \\
\hline
17 & $-\frac{{a'}^2 H^2}{96}$ \\
\hline
18 & $\frac{a a' H^2}{16}$ \\
\hline
19 & $-\frac{a H}{96}$ \\
\hline 
20 & $\frac{a' H}{96}$ \\
\hline
21 & $-\frac1{480}$ \\
\hline
\end{tabular}
\caption{\footnotesize 
Nonlocal contributions acting on 
$\partial^2 [\frac{\ln(\mu^2 \Delta x^2)}{\Delta x^2}]$ 
before including the trace terms.} 
\label{TauB}
\end{table}

\newpage

\begin{table}[H]
\setlength{\tabcolsep}{8pt}
\def\arraystretch{1.5}
\centering
\begin{tabular}{|@{\hskip 1mm }c@{\hskip 1mm }||c|}
\hline
$i$ & $T_B^i(a,a',\partial)$ \\
\hline\hline
1 & $-\frac{\partial^4}{640} 
+ \frac{a a' H^2 \partial_0^2}{64} 
- \frac{a a' H^2 \Delta \eta \partial_0 \partial^2}{192}
+ \frac{a^2 {a'}^2 H^4 \Delta \eta^2 \partial^2}{192} 
- \frac{a^2 {a'}^2 H^4 \Delta \eta \partial_0}{48} 
- \frac{a^2 {a'}^2 H^4}{96}$ \\
\hline
2 & $-\frac{\partial^4}{1920} 
- \frac{a a' H^2 \partial^2}{96}
- \frac{a^2 {a'}^2 H^4}{48}$ \\
\hline
3 & $\frac{{a'}^2 H^2 \partial^2}{96} 
+ \frac{a {a'}^2 H^3 \partial_0}{32}$ \\
\hline
4 & $\frac{a^2 H^2 \partial^2}{96} 
- \frac{a^2 a' H^3 \partial_0}{32}$ \\
\hline
5 & $-\frac{a' H \partial^2}{96} 
- \frac{a a' H^2 \partial_0}{32}
+ \frac{a^2 a' H^3}{48} 
- \frac{a {a'}^2 H^3}{96}$ \\
\hline
6 & $\frac{a H \partial^2}{96} 
- \frac{a a' H^2 \partial_0}{32}
+ \frac{a^2 a' H^3}{96} 
- \frac{a {a'}^2 H^3}{48}$ \\
\hline
7 & $\frac{\partial^2}{640} 
+ \frac{a H \partial_0}{192}
- \frac{a^2 H^2}{192}$ \\
\hline
8 & $\frac{\partial^2}{640} 
- \frac{a' H \partial_0}{192} 
- \frac{{a'}^2 H^2}{192}$ \\
\hline
9 & $-\frac{a^2 {a'}^2 H^4}{16}$ \\
\hline
10 & $\frac{a^2 a' H^3}{48}$ \\
\hline 
11 & $-\frac{a {a'}^2 H^3}{48}$ \\
\hline
12 & $\frac{\partial^2}{960} 
+ \frac{a a' H^2}{48}$ \\
\hline
13 & $-\frac{a^2 {a'}^2 H^4}{16}$ \\
\hline 
14 & $\frac{a^2 a' H^3}{16}$ \\
\hline
15 & $-\frac{a {a'}^2 H^3}{16}$ \\
\hline
16 & $-\frac{a^2 H^2}{96}$ \\
\hline
17 & $-\frac{{a'}^2 H^2}{96}$ \\
\hline
18 & $\frac{a a' H^2}{16}$ \\
\hline
19 & $-\frac{a H}{96}$ \\
\hline 
20 & $\frac{a' H}{96}$ \\
\hline
21 & $-\frac1{480}$ \\
\hline
\end{tabular}
\caption{\footnotesize 
Non-local contributions acting on 
$\, \partial^2 [\frac{\ln(\mu^2 \Delta x^2)}{\Delta x^2}]$ 
in expression (\ref{genform3}).} 
\label{TB}
\end{table}

\newpage

\section{Appendix: Integrals for the Potential}

The purpose of this section is to evaluate the integrations 
needed for Section \ref{response}. These have the generic 
form:
\begin{eqnarray}
I_{A}^{J}(t,r) &\!\!\! \equiv \!\!\!& 
\int \! d^4x' \, \theta(\Delta \eta \!-\! \Delta r) 
\times {a'}^J \, \Psi_0(t',r') 
\;\; , \label{IAJ} \\
I_{B}^{J}(t,r) &\!\!\! \equiv \!\!\!& 
\partial^4 \!\! \int \! d^4x' \, \Bigl\{ 
\theta(\Delta \eta \!-\! \Delta r) 
\Bigl( \ln[\mu^2 (\Delta \eta^2 \!-\! \Delta r^2)] 
\!-\! 1 \Bigr) \Bigr\}
\nonumber \\
& & \hspace{7cm} \times {a'}^J \, \Psi_0(t',r') 
\;\; . \qquad \label{IBJ}
\end{eqnarray}
Relations (\ref{f2tof1}) and (\ref{f2tozero}) 
constrain $I_A^J$:
\begin{equation}
\partial^4 I_{A}^{J} = 8\pi a^J \Psi_0(t,r) 
\qquad , \qquad
\Bigl( \frac{\partial^2}{a} - 2 H \partial_0 
\Bigr) I_{A}^{J} =
\partial^2 I_{A}^{J-1}
\;\; . \label{Aconstraints}
\end{equation}
Similarly, expression (\ref{f3tof2}) implies a 
relation between $I_{A}^{J}$ and $I_{B}^{J}$:
\begin{equation}
-2 H \partial_0 \partial^2 I_{A}^{J} = 
I_{B}^{J-1} - \tfrac1{a} I_{B}^{J}
\; . \label{IAtoIB}
\end{equation}

Both the $A$-Type and $B$-Type integrand factors:
\begin{equation}
\theta(\Delta \eta \!-\! \Delta r)
\qquad , \qquad 
\partial^4 \Bigl\{ \theta(\Delta \eta \!-\! \Delta r) 
\Bigl( \ln[\mu^2 (\Delta \eta^2 \!-\! \Delta r^2)] 
\!-\! 1\Bigr) \Bigr\}  
\;\; . 
\end{equation}
depend only on the conformal coordinate difference 
$(x - x')^{\mu}$, so derivatives can be reflected 
$\partial_{\mu} \rightarrow -\partial'_{\mu}$.
Partially integrating with respect to time produces 
surface terms which we will always avoid. On the 
other hand, causality compels the integrands to 
vanish at spatial infinity, so we will partially 
integrate factors of $\nabla^2$ and take advantage 
of the simplification:
\begin{equation}
{\nabla'}^2 \Psi_0(t',r') = 
-\frac{4\pi G M \delta^3(\vec{x}')}{a'} \; .
\end{equation}
This implies two additional generic integrations:
\begin{eqnarray}
I_{A\delta}^{J} &\!\!\! \equiv \!\!\!& 
\int \! d^4x' \, \theta(\Delta \eta \!-\! \Delta r) 
\times\! {a'}^J \, {\nabla'}^2 \Psi_0(t',r') \; , 
\label{IAJdelta} \\
I_{B\delta}^{J} &\!\!\! \equiv \!\!\!& 
\partial^4 \!\! \int \! d^4x' 
\Bigl\{ \theta(\Delta \eta \!-\! \Delta r) 
\Bigl( \ln[\mu^2 (\Delta \eta^2 \!-\! \Delta r^2)] 
\!-\! 1\Bigr) \Bigr\} 
\nonumber \\
& & \hspace{7cm} \times {a'}^J \, {\nabla'}^2
\Psi_0(t',r') 
\;\; , \qquad \label{IBJdelta}
\end{eqnarray}
obviously related to (\ref{IAJ}-\ref{IBJ}):
\begin{equation}
I_{A\delta}^{J}(t,r) = \nabla^2 I_{A}^{J}(t,r) 
\qquad , \qquad 
I_{B\delta}^{J}(t,r) = \nabla^2 I_{B}^{J}(t,r) 
\;\; . \label{Ideltanabla}
\end{equation}

The temporal integrations begin at $\eta_i = -H^{-1}$, 
and we will assume that the coordinate radius from 
source to observer obeys:
\begin{equation}
H r < 1 - \frac{1}{a} 
\;\; . \label{radius}
\end{equation}
\noindent
{\bf -} {\it Evaluation of the $I_{A}^{J}$ integral:} \\
We begin by making the change of variable $\vec{x}' = 
\vec{x} - \vec{y}$ and then performing the angular 
integrations:
\begin{eqnarray}
I_{A}^{J} &\!\!\! = \!\!\!& 
G M \! \int_{\eta_i}^{\eta} \!\! d\eta' \, {a'}^{J-1} 
\! \int \! d^3y \; \frac{\theta(\Delta \eta \!-\! y)}
{\Vert \vec{x} \!-\! \vec{y}\Vert} 
\qquad \\
&\!\!\! = \!\!\!&
\frac{2 \pi G M}{r} \! \int_{\eta_i}^{\eta} \!\! d\eta' 
\, {a'}^{J-1} \! \int_{0}^{\Delta \eta} \! dy \, y 
\Bigl[ r \!+\! y - \vert r \!-\! y\vert \Bigr] 
\;\; . \qquad \label{step1}
\end{eqnarray}
We now make use of (\ref{radius}) to decompose the 
$\eta'$ and $y$ integrations of (\ref{step1}) into 
regions for which the absolute value $\vert r - y\vert$ 
is either $r-y$ or $y-r$:
\begin{eqnarray}
I_{A}^{J} &\!\!\! = \!\!\!& 
\frac{4\pi G M}{r} \Biggl\{ \! \int_{\eta_i}^{\eta-r}
\!\!\!\! d\eta' \, {a'}^{J-1} \Bigl[ \int_{0}^{r} \!\!\!\! dy \, y^2 
\!+\!\! \int_{r}^{\Delta \eta} \!\!\!\! dy \, y r \Bigr] 
\!+\!\! \int_{\eta - r}^{\eta} \!\!\!\! d\eta' \, {a'}^{J-1} 
\!\! \int_{0}^{\Delta \eta} \!\!\!\! dy \, y^2
\Biggr\} \qquad \\
&\!\!\! = \!\!\!& 
\frac{4\pi G M}{r} \Biggl\{ 
\int_{\eta_i}^{\eta-r} \!\!\!\! d\eta' \, {a'}^{J-1} 
\Bigl[ \tfrac12 r \, \Delta \eta^2 - \tfrac16 r^3 \Bigr] 
+ \int_{\eta - r}^{\eta} \!\!\!\! d\eta' \, {a'}^{J-1} \, 
\tfrac13 \Delta\eta^3 \Biggr\}
\;\; . \label{step2}
\end{eqnarray}
In acting $\partial^2$ it is useful to exploit the identity:
\begin{equation}
\partial^2 f(\eta,r) = 
\frac1{r} (\partial_r \!-\! \partial_0) 
(\partial_r \!+\! \partial_0) [r f(\eta,r)] 
\;\; . \label{partial}
\end{equation}
It follows that:
\begin{equation}
\partial^2 I_{A}^{J} = 
-\frac{8\pi G M}{r} \Biggl\{ 
r \! \int_{\eta_i}^{\eta-r}
\!\!\!\! d\eta' \, {a'}^{J-1} 
+ \int_{\eta - r}^{\eta} \!\!\!\! d\eta' \,
{a'}^{J-1} \Delta \eta \Biggr\} 
\;\; . \label{d2IAJ}
\end{equation}
Proceeding similarly, it is straightforward to check 
relations (\ref{Aconstraints}).

\newpage

\noindent
{\bf -} {\it Evaluation of the $I_{B}^{J}$ integral:} \\
The initial reduction of $I_{B}^{J}(t,r)$ is the same 
as that of $I_{A}^{J}$, except for the logarithm and 
the external derivatives:
\begin{eqnarray}
\lefteqn{ I_{B}^{J} = \frac{4\pi G M}{r} 
(\partial_r^2 \!-\! \partial_0^2)^2 \, \times
\Biggl\{ \int_{\eta_i}^{\eta - r} \!\!\!\! d\eta' \, {a'}^{J-1} 
\Bigl[ \int_{0}^{r} \!\! dy \, y^2 
\Bigl( \ln[\mu^2 (\Delta \eta^2 \!-\! y^2)] \!-\! 1 \Bigr) }
\qquad \nonumber \\
& & \hspace{1.9cm}
\!+\! \int_{r}^{\Delta \eta} \!\!\!\! dy \, y r
\times \Bigl( \ln[\mu^2 (\Delta \eta^2 \!-\! y^2)] \!-\! 1 \Bigr) 
\Bigr] 
\nonumber \\
& & \hspace{1.9cm}
+ \int_{\eta - r}^{\eta} \!\!\!\! d\eta' \, {a'}^{J-1} 
\! \int_{0}^{\Delta \eta} \!\!\!\! dy \, y^2 
\Bigl( \ln[\mu^2 (\Delta \eta^2 \!-\! y^2)] \!-\! 1 \Bigr) \Biggr\} 
\;\; , \qquad
\end{eqnarray}
so that we get:
\begin{eqnarray}
\lefteqn{ I_{B}^{J} = \frac{4\pi G M}{r} 
(\partial_r^2 \!-\! \partial_0^2)^2 \, \times
\Biggl\{ \int_{\eta_i}^{\eta - r} \!\!\!\! d\eta' \, {a'}^{J-1} 
\Bigl[ -\tfrac13 (\Delta \eta^3 \!-\! r^3) 
\ln[\mu (\Delta \eta \!-\! r)] }
\qquad \nonumber \\
& & \hspace{1.5cm}
+ \tfrac13 (\Delta \eta^3 \!+\! r^3) \ln[\mu (\Delta \eta \!+\! r)]
- \tfrac23 \Delta \eta^2 r - \tfrac59 r^3 \Bigr] 
\nonumber \\
& & \hspace{1.5cm}
+ \int_{\eta_i}^{\eta - r} \!\!\!\! d\eta' \, {a'}^{J-1} 
r(\Delta \eta^2 \!-\! r^2)
\Bigl( \tfrac12 \ln[\mu^2 (\Delta \eta^2 \!-\! r^2)] 
\!-\! 1 \Bigr)
\qquad \nonumber \\
& & \hspace{1.5cm}
+ \! \int_{\eta - r}^{\eta} \!\!\!\! d\eta' \, 
{a'}^{J-1} \Delta \eta^3 
\Bigl[ \tfrac23 \ln(2\mu \Delta \eta) \!-\! \tfrac{11}{9} \Bigr] 
\Biggr\} 
\;\; . \qquad 
\end{eqnarray}
As a result, $I_{B}^{J}$ equals:
\begin{equation}
I_{B}^{J} = -\frac{16 \pi G M}{r} \Bigl\{ 
-2 a_r^{J-1} \ln(2 \mu r) + (\partial_r \!-\! \partial_0) 
\! \int_{\eta - r}^{\eta} \!\!\!\! d\eta' \, {a'}^{J-1} 
\ln(2\mu \Delta \eta) \Bigr\} 
\;\; . 
\end{equation}
Here $a_r = (Hr + \tfrac1a)^{-1}$ is the scale factor 
evaluated at $\; \eta' = \eta - r$. 

\newpage

\noindent
{\bf -} {\it Evaluation of the $I_{A\delta}^{J}$ and 
$I_{B\delta}^{J}$ integrals:} \\
The $\delta$-function integrals (\ref{IAJdelta}-\ref{IBJdelta}) 
are rather simple. For the $A$-type generic integral we have:
\begin{equation}
I_{A\delta}^{J} = 
-4\pi G M \! \int_{\eta_i}^{\eta-r} 
\!\!\!\! d\eta' \, {a'}^{J-1}
\;\; , 
\end{equation}
while the generic $B$-type integral is:
\begin{eqnarray}
I_{B\delta}^{J} &\!\!\! = \!\!\!& 
-\frac{4\pi G M}{r} (\partial_r^2 \!-\! \partial_0^2)^2 \,
\Biggl\{ r \! \int_{\eta_i}^{\eta-r} \!\!\!\! d\eta' \, {a'}^{J-1} 
\Bigl( \ln[\mu^2 (\Delta \eta^2 \!-\! r^2)] \!-\! 1 \Bigr) \Biggr\} 
\qquad \\
&\!\!\! = \!\!\!& -\frac{16\pi G M}{r} \Biggl\{
\frac{a_r^{J-1}}{r^2} + \frac{(J\!-\!1) H a_r^{J}}{r} \Biggr\} 
\;\; .
\end{eqnarray} 
\noindent
{\bf -} {\it Particular integrals required:} \\
We display here the list of the specific integrals 
needed for the computation of the gravitationally
induced potentials in Section \ref{response}.

{\it (i)} For $I_{A}^{J}$ we require only $J=2$, acted 
on by either $\partial^2$ or $\partial_0 \partial^2$:
\begin{eqnarray}
\partial^2 I_{A}^{2} 
&\!\!\! = \!\!\!& 
-\frac{8\pi G M}{r} \Bigl\{ 
\tfrac{r}{H} \ln(a_r) + \tfrac{r}{H} 
- \tfrac{1}{H^2 a} \ln(\tfrac{a}{a_r}) \Bigr\} 
\;\; , \label{d2IA2} \\
\partial_0 \, \partial^2 I_{A}^{2} 
&\!\!\! = \!\!\!& 
-\frac{8\pi G M}{r} \Bigl\{
\tfrac1{H} \ln(\tfrac{a}{a_r}) \Bigr\} 
\;\; . \label{d0d2IA2}
\end{eqnarray}

{\it (ii)} For $I_{B}^{J}$ we only need two values of $J$:
\begin{equation}
I_{B}^{0} = \frac{16 \pi G M}{r} \Bigl\{ 
H r + \tfrac1{a} \ln(2 \mu r) \Bigr\} 
\quad , \quad 
I_{B}^{1} = \frac{16 \pi G M}{r} \Bigl\{ 
\ln(2 \mu r) \Bigr\}
\; . \label{IBcases}
\end{equation}

{\it (iii)} For $I_{A\delta}^{J}$ the only cases we need 
are $J=1$ and $J=2$:
\begin{equation}
I_{A\delta}^{1} = 
-\frac{4\pi G M}{r} \!\times\! 
\frac{r}{H} \Bigl[ 1 \!-\! \frac1{a_r} \Bigr] 
\qquad , \qquad 
I_{A\delta}^{2} = 
-\frac{4\pi G M}{r} \!\times\! 
\frac{r}{H} \ln(a_r) 
\;\; . \label{IAdeltacases}
\end{equation}

{\it (iv)} For $I_{B\delta}^{J}$ only the cases of $J=0$ 
and $J=1$ are required:
\begin{equation}
I_{B\delta}^{0} = 
-\frac{16 \pi G M}{r} \!\times\! \frac1{a r^2} 
\qquad , \qquad 
I_{B\delta}^{1} = 
-\frac{16 \pi G M}{r} \!\times\! \frac1{r^2} 
\;\; . \label{IBdeltacases}
\end{equation}

\end{document}